 \documentclass[12pt]{article}

\usepackage{amsmath}
\usepackage{mathtools}
\usepackage{geometry} 
\usepackage[parfill]{parskip} 
\usepackage{graphicx} \usepackage{amssymb} \usepackage{epstopdf}
\usepackage[sort&compress]{natbib}
\usepackage{verbatim} \usepackage{lscape}
\usepackage{rotating}
\usepackage{setspace}
\usepackage{fancyhdr}
\usepackage{subfigure}
\usepackage{rotating}
\usepackage{pdflscape}

\usepackage{floatpag}
\floatpagestyle{plain}

\usepackage{caption}

\usepackage{hyperref}

\usepackage{pdfpages}

\usepackage{placeins}

\usepackage{theorem} 

\usepackage{thrmappendix}

\makeatletter
\newcommand\footnoteref[1]{\protected@xdef\@thefnmark{\ref{#1}}\@footnotemark}
\makeatother

\long\def\rptstmtonlyproof#1{
	\pf \cs{#1proof} \endpf
}

\newtype{result}{statement}{Result} 
\newtype{assumption}{assumption}{Assumption}
\newtype{property}{statement}{Property}
\newtype{cor}{statement}{Corollary}
\def\endpf{\hfill $\blacksquare$}

\numberwithin{statement}{section}

\newcommand{\q}[2]{ {}_{#1}q_{#2}}

\newcommand{\ffqf}{ \q{45}{15} }

\newcommand{\tfqf}{ \q{35}{15} }

\newcommand{\E}{ \mathbb{E} }
\newcommand{\wprime}{w^{\prime}}
\newcommand{\wdiff}{\epsilon}
\newcommand{\piprime}{\pi^{\prime}}

\newcommand{\correl}{\text{cor}}
\newcommand{\cv}{\text{cv}}

\graphicspath{{figures/manual/},{figures/from_code/}}

\title{The network survival method for estimating adult mortality:
Evidence from a survey experiment in Rwanda\footnotemark[1]}

\author{Dennis M. Feehan\footnotemark[2]~, Mary Mahy\footnotemark[3]~, and Matthew J. Salganik\footnotemark[4]}

\footnotetext[1]{%
    {\scriptsize
  The study ``Estimating the Size of Populations through a Household Survey'' was
  conducted in Rwanda from June 2011 to August 2011 by the School of Public
  Health, University of Rwanda, in collaboration with the Rwanda Biomedical
  Center/Institute of HIV/AIDS, Disease Prevention and Control Department, with
  the assistance of the National Institute of Statistics of Rwanda.  Technical
  assistance and funding for the project was provided by the Joint United
  Nations Programme on HIV/AIDS and ICF International (Calverton, Maryland)
  through the MEASURE Demographic and Health Surveys Program, a US Agency for
  International Development (USAID)-funded project providing support and
  technical assistance in the implementation of population and health surveys
  in countries worldwide. Additional assistance was provided by the US Centers
  for Disease Control and Prevention, Princeton University, and the University
  of Florida. The government of Japan, the United Nations Joint Team on HIV of
  Kigali, Rwanda, and USAID provided additional funding. The research was also
  supported by grants to M.J.S. from the US National Science Foundation (grant
  CNS-0905086) and the US National Institutes of Health (grants R01-HD062366
  and R24-HD047879).   Some of this work was performed while M.J.S. was an
  employee of Microsoft Research (New York, New York).  The opinions expressed
  in this paper are those of the authors and do not necessarily reflect the
  views of the government of Rwanda, the funding agencies, or the collaborating
  organizations.
\\
  We thank Bernard Barrere, Patrick Ndimubanzi, Aline Umubyeyi, and Wolfgang
  Hladik for helping to design and conduct the study, and we thank Noreen
  Goldman, Georges Reniers, Bruno Masquelier, Doug Massey, Scott Lynch, 
  Brandon Stewart, and three anonymous reviewers for helpful comments. 
\\
Our dataset is freely available from the DHS website and replication code is available
on the Harvard Dataverse (\texttt{http://dx.doi.org/10.7910/DVN/GSKQIZ}).
  }
  }
\footnotetext[2]{Department of Demography, University of California, Berkeley}
\footnotetext[3]{Joint United Nations Programme on HIV/AIDS (UNAIDS), Geneva, Switzerland}
\footnotetext[4]{Department of Sociology and Office of Population Research, Princeton University}

\date{\today}

\begin{document}
\maketitle

\newpage


\begin{abstract}
Adult death rates are a critical indicator of population health and wellbeing.  Wealthy countries have high-quality vital registration systems, but poor countries lack this infrastructure and must rely on estimates that are often problematic.  In this paper, we introduce the \emph{network survival method}, a new approach for estimating adult death rates.  We derive the precise conditions under which it produces estimates that are consistent and unbiased.  Further, we develop an analytical framework for sensitivity analysis.  To assess the performance of the network survival method in a realistic setting, we conducted a nationally-representative survey experiment in Rwanda $(n=4,669)$.  Network survival estimates were similar to estimates from other methods, even though the network survival estimates were made with substantially smaller samples and are based entirely on data from Rwanda, with no need for model life tables or pooling of data from other countries.  Our analytic results demonstrate that the network survival method has attractive properties, and our empirical results show that it can be used in countries where reliable estimates of adult death rates are sorely needed.
\end{abstract}

\newpage


\newpage


\section{Introduction}
\label{sec:intro}

Adult death rates are a critical indicator of population health and wellbeing.
In developed countries, a variety of legal, medical, and financial systems
ensure that virtually every death is recorded in a vital registration system.
These vital registration systems enable researchers to produce high quality
estimates of adult death rates by age and sex. Most developing countries, on
the other hand, are victims of the \emph{scandal of invisibility}: they do not
have administrative systems that reliably produce death certificates, meaning
that most adults die without ever having their deaths formally recorded
\citep{setel_scandal_2007,mikkelsen_global_2015, abouzahr_civil_2015}.  The scandal of invisibility
is, unfortunately, vast: \citet{mikkelsen_global_2015} estimates that two-thirds
of worldwide deaths are never formally recorded. 

The long-term solution to the scandal of invisibility is for all countries to
develop effective vital registration systems.  Progress on this front, however,
has been very slow: \citet{mikkelsen_global_2015} estimates that between 2000
and 2012, the proportion of deaths registered worldwide went from 36\% to 38\%.
Because of the absence of high-quality vital registration data in developing
countries, researchers have worked on the problem of estimating adult death
rates for decades. Unfortunately, this problem has proven to be extremely
difficult. In the meantime, critical questions about
science and policy in the world's poorest countries continue to go unanswered. 

This paper helps address the scandal of invisibility by developing and testing
the \emph{network survival method}, a new survey-based method for estimating
adult mortality.  
Roughly, this new method generalizes the sibling survival
method, which is the survey-based approach that is most widely used today.
Whereas the sibling survival method collects information about the deaths of
siblings of respondents, the network survival method collects information about
deaths in a wider social network around each respondent.  
The generalization dramatically increases the amount of information that is
collected from each respondent, but it also introduces a variety of
complexities that our methodology addresses.  
Because the network survival method uses data that could be collected in a
standard household survey---the kind of surveys that are routinely fielded in
most developing countries---it could potentially be deployed in developing
countries around the world.

The remainder of this paper is divided into six sections.
In Section 2, we review survey-based adult mortality estimation, paying
particular attention to the current state of the art: the direct sibling
survival estimator. 
In Section 3, we introduce and derive the network survival method. 
In Section 4, we describe the results of a nationally-representative survey
experiment in Rwanda $(n = 4,669)$ that we conducted to test---under realistic
field conditions---two different versions of the network survival method.  
We find that both arms of our survey experiment produced similar estimates.
In Section 5, we compare both network survival estimates to estimates from the
sibling survival method and estimates from three organizations
(e.g., the World Health Organization).  
We find that the network survival estimates were similar to estimates from
other methods, even though the network survival estimates were made with
substantially smaller samples and are based entirely on data from Rwanda, with
no need for model life tables or pooling of data from other countries. 
In Section 6, we introduce and derive a sensitivity analysis framework that
enables researchers to quantitatively assess the sensitivity of network
survival estimates to all of the conditions they rely upon, and we use this
framework to assess our estimates from Rwanda.  
Finally, in Section 7 we conclude with an outline of possible next steps.
Online appendices A - I contain proofs, additional technical details, and additional empirical results.


\section{Background}
\label{sec:background}

\subsection{Estimating death rates}

The death rate is the number of deaths that occur in a group, relative
to the group's exposure to the possibility of dying. Mathematically, for a
demographic group $\alpha$ (for example, women aged 45-49 in 2011), the death
rate can be written
\begin{align}
    \label{eqn:asdr-defn}
    M_{\alpha} &= \frac{D_{\alpha}}{N_{\alpha}},
\end{align}
\noindent where $D_{\alpha}$ is the number of deaths and $N_{\alpha}$ is the
amount of exposure to demographic group $\alpha$.    
Death rates are a type of occurrence-exposure rate.

Adult death rates are difficult to estimate from a survey for two main reasons
\citep{timaeus_measurement_1991}. First, surveys typically ask respondents to
report about themselves; for example, a survey might ask respondents to report
their age, education, or income. 
This approach is not possible for deaths, since people who have died cannot be
interviewed. 
Second, adult deaths are quite rare, even in poor countires. Rare events are difficult to
study using standard survey techniques because they require very large
samples to yield estimates that are precise enough to be useful
\citep{kalton_sampling_1986}. Any survey-based approach to estimating adult
death rates will have to overcome these two primary obstacles.

If death rates are difficult to estimate from surveys, why focus on
survey-based approaches at all?  
We believe that surveys offer the best hope for immediate, global, and
sustained progress, as has been illustrated by the progress that has been made
using surveys to estimate other critical demographic quantities, such as
fertility and child mortality.
In countries that lack good vital registration systems, fertility rates and
child mortality were once as poorly understood as adult mortality is now. 
But today, even the world's poorest countries have high-quality estimates of
fertility and child mortality rates. 
This change did not happen automatically.  
Rather, researchers had to develop new methods to estimate these quantities
from household surveys~\citep{hill_adult_2004, timaeus_measurement_1991}, and
these methods had to be tested and refined in realistic field conditions until
they were able to be deployed at a global scale, first with the World Fertility
Survey Program, and now through the massive, internationally coordinated,
Demographic and Health Survey program and the Multiple Indicator Cluster Survey program \citep{hill_interim_2007,
corsi_demographic_2012, fabic_systematic_2012, hancioglu_measuring_2013}. 
In fact, because of these earlier efforts, high-quality household surveys are
already being regularly conducted in countries without vital registration systems.  
This survey infrastructure can be harnessed to estimate adult mortality. 

\subsection{Sibling survival method}
\label{sec:background-sibling}

Previous research on adult mortality estimation has considered many different
strategies for collecting information about deaths, including surveys,
prospective or cohort designs, incomplete sources of death certificates, one or
many censuses, and historical records.
Other authors have provided more complete overviews of mortality estimation
    \citep[see, for example,][]{
        hill_manual_1983, 
        timaeus_measurement_1991,
        hill_methods_2000, 
        hill_adult_2003, gakidou_adult_2004, hill_unconventional_2005,
        bradshaw_levels_2006,
    hill_interim_2007, reniers_adult_2011}.
In this paper, we focus on survey-based techniques because they are most
relevant to our new estimator.   There are many survey-based approaches to estimating death rates, but the most common is the direct sibling survival
method~\citep{rutenberg_direct_1991}.\footnote{Another survey-based approach
focuses on collecting information about deaths in the household 
\citep{el_arifeen_maternal_2014, koenig_maternal_2007, hill_how_2006}.}
The direct sibling survival method requires collecting \emph{sibling histories}: each
respondent is asked to enumerate her or his siblings and then to provide each
sibling's birthday, survival status, and date of death (when applicable). 

The direct sibling survival method seems like a promising way to overcome the
two fundamental challenges in estimating death rates from surveys: since
respondents report about their siblings, it is possible to learn about people
who have died; and, since respondents typically have multiple siblings, each
interview produces information about more than one person, increasing the
effective size of the sample. 
As a part of the Demographic and Health Survey (DHS) program, 
sibling histories have been collected in over 150 surveys from
dozens of countries across the developing world \citep{corsi_demographic_2012, fabic_systematic_2012}. 
Nonetheless, relatively few researchers have made use of these DHS sibling histories to study adult mortality~\citep{reniers_adult_2011,gakidou_adult_2004}. 
For example, despite the fact that very little is known about
adult mortality in Sub-Saharan Africa
\citep{setel_scandal_2007}, only a handful of studies have
tried to use the DHS sibling histories to construct estimates of recent trends in
adult mortality
~\citep{timaeus_adult_2004,obermeyer_measuring_2010,rajaratnam_worldwide_2010,
reniers_adult_2011, wang_age-specific_2013, masquelier_divergences_2014}.  

There are two reasons why the DHS sibling histories may have been relatively
under-used. 
First, surveys with typical DHS sample sizes---between 5,000 and 30,000 
respondents \citep{corsi_demographic_2012}---cannot be used to produce
timely direct estimates of age- and sex-specific death rates because the
sampling variation from the direct sibling survival estimator is too large
\citep{stanton_assessment_2000, timaeus_adult_2004, hill_how_2006}.
Instead, researchers have had to resort to a combination of pooling data across
countries and across time, smoothing regressions, and model life tables to
estimate adult mortality from DHS sibling histories
\citep{timaeus_adult_2004,obermeyer_measuring_2010,rajaratnam_worldwide_2010,
reniers_adult_2011, wang_age-specific_2013, masquelier_divergences_2014}.  
This need to smooth the raw data requires researchers to make several
difficult-to-verify assumptions, reducing the appeal of producing estimates
based on sampled data~\citep{masquelier_adult_2013}.

The second reason that DHS sibling histories may be relatively under-used is that there
is methodological uncertainty about how sibling histories should be analyzed.
Several common methodological concerns have emerged from research about the sibling histories: 
(i) there is no way to learn about
sibships (sets of people who are siblings) that have no survivors left to be
sampled by the survey; (ii)  more generally, sibships with more survivors are
more likely to be sampled by the survey, potentially biasing estimates if
sibship size and mortality are correlated~%
\citep{gakidou_death_2006,trussell_note_1990,graham_estimating_1989,
masquelier_adult_2013,reniers_adult_2011,gakidou_adult_2004};
(iii) there are many ways that respondents' reports about their siblings may not
be accurate; for example, respondents may omit some
siblings from their survey reports, and if the tendency to omit a sibling is
correlated with the chances that the sibling is alive, then this may introduce
bias into the resulting estimates~%
\citep{helleringer_reporting_2014,helleringer_improving_2014,
helleringer_misclassification_2013,merdad_improving_2013,masquelier_sibship_2014};
(iv) the respondent is, by definition, alive, making it unclear whether the
respondent's experience should be included or omitted from the death rate
estimates~%
\citep{reniers_adult_2011,masquelier_adult_2013}.

Uncertainty about these methodological issues has not been resolved. For
example, \citet{gakidou_death_2006} proposed a solution to address the
potential correlation between sibship size and mortality, but it has proven to
be controversial in practice~\citep{masquelier_adult_2013}. Subsequent studies
have therefore been divided: one group has applied the Gakidou-King selection
bias adjustments \citep{kassebaum2014global, wang_age-specific_2013,
rajaratnam_worldwide_2010} while another has not \citep{reniers_adult_2011,
moultrie_tools_2013, masquelier_divergences_2014}.

To conclude, the direct sibling survival method is a promising approach to
overcoming the two main challenges that must be faced to estimate death rates
from a survey: it enables researchers to learn about people who died, and it
enables researchers to learn about more than one person from each interview.
Unfortunately, in practice, the direct sibling survival method has two big
disadvantages: first, it cannot typically be used to produce direct estimates
of death rates because the sampling variation of direct estimates is too large;
and, second, the sibling survival method is clouded by several
potential sources of bias. It is not clear precisely what effect these
potential biases might have on sibling survival estimates, or how these
potential biases might interact with one another. 


\section{The network survival method}
\label{sec:network_survival}

In this paper we introduce the network survival method, which can be seen as a
generalization of the direct sibling survival method.  Whereas the direct
sibling survey method collects information about mortality in sibling networks,
the network survival method collected information about mortality in \emph{any}
type of network in which respondents are embedded.   

The network survival methods collects two types of information about survey
respondents' personal networks: first, respondents are asked about their
connections to people who died (e.g., ``How many people do you know who died in
the previous 12 months?'', where ``know'' could be replaced with other types of
social relationships, as we discuss below). Similar to a sibling history,
respondents are asked to enumerate each person who died and to provide
additional information about each one (for example, the deceased's age and
sex). Second, unlike the sibling survival method, respondents are also asked about their
connections to several different groups whose total size is known (e.g., ``How
many policemen do you know?'' where the number of policemen is available from
administrative records or estimated from a survey). This information about
connections to groups of known size is used to estimate the total size of
respondents' personal networks, an approach that has been used previously as
part of the network scale-up method~\citep{killworth_estimation_1998,
bernard_counting_2010, feehan_generalizing_2016}.

Asking survey respondents to report about the members of their personal
networks helps resolve both of the major difficulties in estimating death
rates from a survey: since respondents report about others, it is possible to
learn about people who have died, even though the people who died cannot be
interviewed directly. And, since respondents are asked to report about all of
the people in their personal networks, researchers get information about much
more than just one person from each interview, increasing the effective
sample size. 

In the remainder of this section, we turn to a more detailed description of how
the network survival method estimates death rates. 
Our focus will be on describing the main ideas behind the new estimator; Online
Appendices~\ref{ap:ns-kp} through \ref{ap:ns-instrument} have proofs and
further technical details.

\subsection{Estimating the number of deaths, $D_\alpha$}
\label{sec:estimating_deaths}

\begin{figure}
  \centering
  \subfigure[]{%
     \label{fig:reporting-network-panel1} 
     \includegraphics[width=0.4\textwidth]{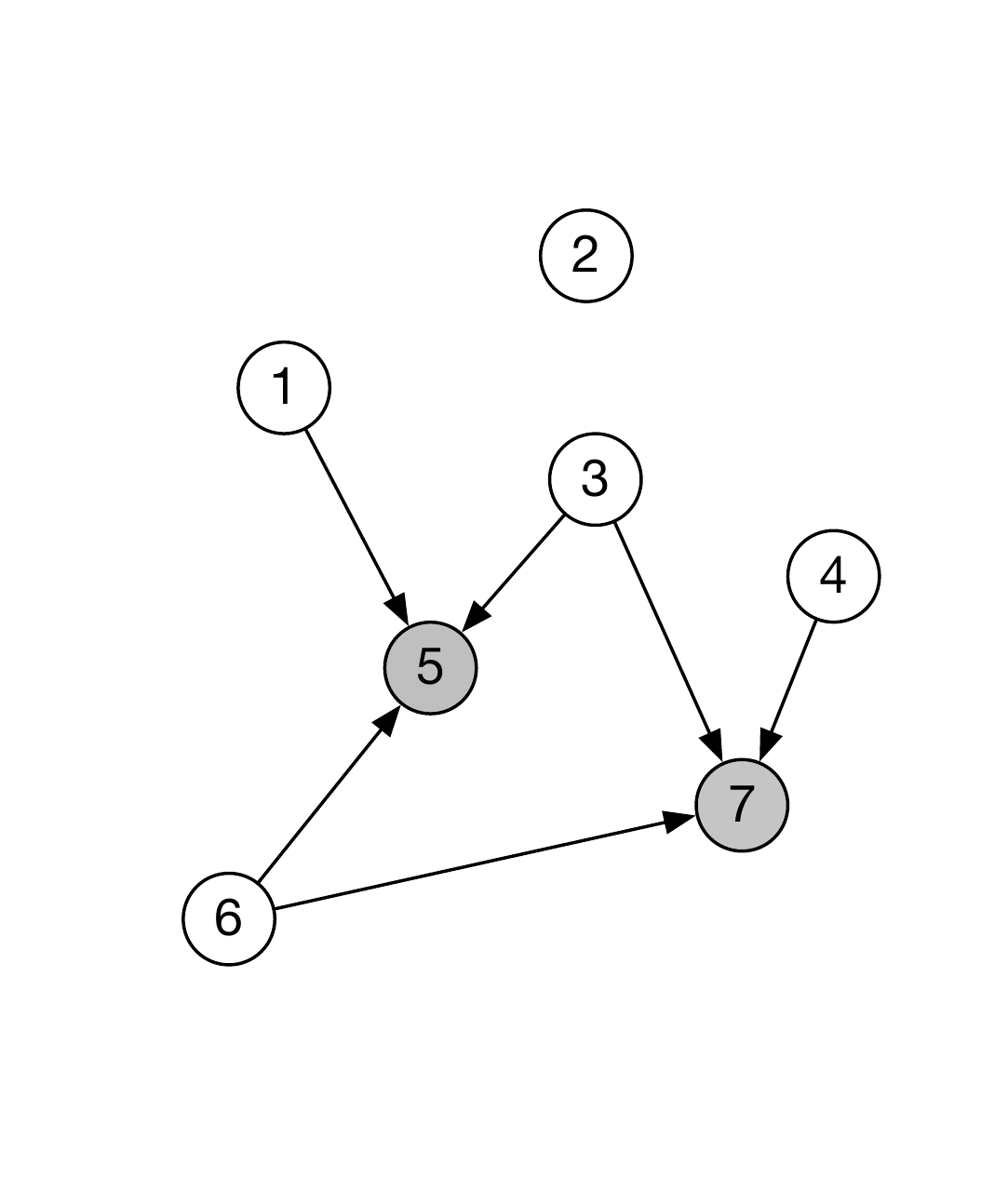}}
  \hspace{0.0in}
  \subfigure[]{%
     \label{fig:reporting-network-panel2} 
     \includegraphics[width=0.4\textwidth]{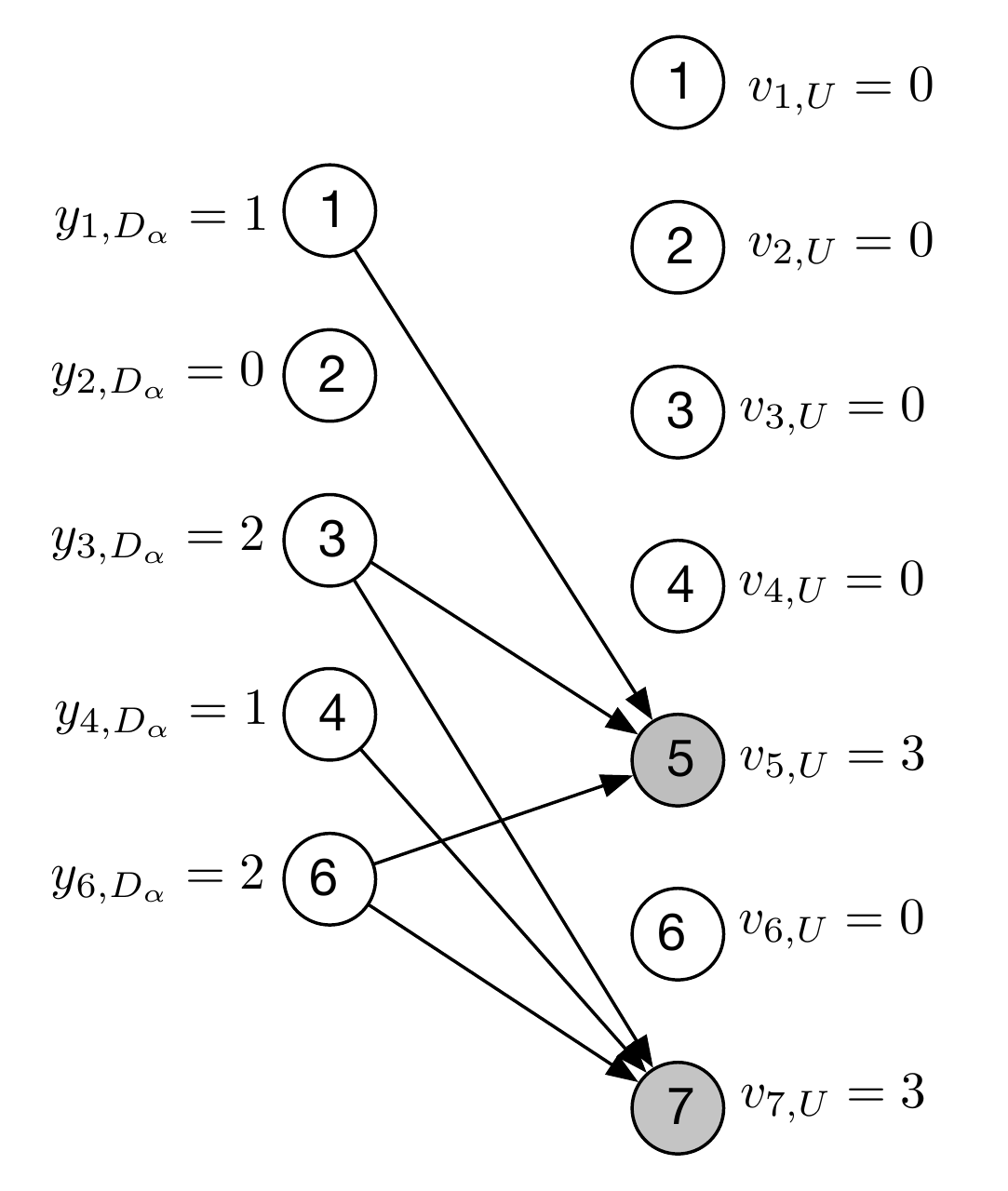}}
  \hspace{0.0in}
     \caption{
         Panel (a) shows a population of 7 people, 2 of whom have died
         (shown in grey).  
         A directed edge $i \rightarrow j$
         indicates that $i$ counts $j$ as having died when answering the
         question ``How many people do you know who have died in the past 12 months?'' 
         Panel (b) shows the same population, but redrawn so that each person
         now appears twice: as someone who reports, on the left, and as a
         someone who could be reported about, on the right.
         People who have died cannot report, since they cannot be
         interviewed.
         Note that this figure depicts detailed individual
         reports $i \rightarrow j$, but in practice reports are not typically collected
         at that level of detail (i.e., we typically would know that person $i$ reports
         one death, but not that the death was specifically person $j$). 
         Fortunately, the identity in Eq.~\ref{eqn:text-id} requires estimates of
         aggregate quantities, so this level of detail is not required.
     }
     \label{fig:reporting-network} 
\end{figure}

The numerator of a death rate is the number of deaths in demographic group
$\alpha$ ($D_\alpha$)\footnote{%
  To avoid over-complicating our notation, we use $D_{\alpha}$ to
  represent both the number of deaths and also the set of people who
  have died; the intended meaning should be clear from context.
}.  
Estimating this quantity from network reports is complex because each individual death
could be reported multiple times (or not at all).  We must therefore 
convert respondents' \emph{reports} about deaths into an
estimate for the \emph{number} of deaths in the population. 
To make this conversion, we use
the network reporting framework~\citep{feehan_generalizing_2016, feehan_network_2015-1}, which is illustrated in Figure~\ref{fig:reporting-network}. 
Figure~\ref{fig:reporting-network-panel1} depicts individuals in a
population who have been asked to report which of their personal
network members have died in the past 12 months (of course, only living people
can be interviewed). 
Each directed arrow $i \rightarrow j$ indicates that $i$ reports that $j$ has
died. 
Figure~\ref{fig:reporting-network-panel2} presents the
same information, but rearranged so that the people who
report are on the left hand side, and the people who could be reported about
are on the right-hand side (note that living people can both report and be
reported about, since it can happen that a living person is erroneously
reported as dead). 
Using this framework, we can create a reporting identity:
\begin{align}
    \label{eqn:text-id0}
    \text{total number of reports about deaths} &=
    \text{number of deaths} \times \text{average reports per death}.  
\end{align}
Rearranging Eq.~\ref{eqn:text-id0} yields
\begin{align}
    \label{eqn:text-id}
    \text{number of deaths} &= 
    \frac{\text{total number of reports about deaths}}{\text{average reports per death}}.
\end{align}
The identity in Eq.~\ref{eqn:text-id} reveals that we can estimate the
number of deaths from respondents' reports by estimating (i) the total
number of reports about deaths that would be collected if we interviewed everyone,
and (ii) the average number of reports per death.  A helpful way to think about the identity in Eq.~\ref{eqn:text-id} is that it clarifies the appropriate way to adjust reports of deaths to avoid overcounting the same death multiple times.

Mathematically, the identity in Eq.~\ref{eqn:text-id} can be written
\begin{align}
    \label{eqn:nontext-intermsofu-id}
    D_{\alpha} &= \frac{y_{F, D_{\alpha}}}{v_{U, F}/D_\alpha},
\end{align}
where
$U$ is the entire population; $F$ is the frame population (the set of people on the sampling frame;  in many cases, this will be all
adults); 
$y_{F, D_\alpha} = \sum_{i \in F} y_{i, D_\alpha}$ is the number of
deaths in demographic group $\alpha$ that would be reported if everyone in the
frame population $F$ was interviewed (i.e., in a census); and 
$v_{U, F} = \sum_{j \in U} v_{j, F}$ is the 
total visibility of all deaths (i.e., the number deaths in the entire population that would be reported if everyone in the frame population was
interviewed). 

There turns out to be a practical problem with trying to develop an estimator
from the identity in Eq.~\ref{eqn:nontext-intermsofu-id}: $v_{U,F}$ is the
number of times anyone in the population would be reported as dead, but it is
much more feasible to estimate the number of times that anyone who actually
died would be reported as dead. Therefore, we make the assumption that
respondents do not incorrectly report that someone died when in fact she did
not. In this case, we say that there are no \emph{false positive} reports.  (In
Section~\ref{sec:ns-adjustment-factors} we develop a full framework for
sensitivity analysis that shows exactly how estimates can be impact by
violations of this assumption).

If there are no false positive reports, then $v_{j,F} = 0$ for all people $j$ who are
alive and therefore $v_{U,F}~=~v_{D_\alpha,F}$.
We can then re-write Eq.~\ref{eqn:nontext-intermsofu-id} as
\begin{align}
    \label{eqn:nontext-id}
    D_{\alpha} &= \frac{y_{F, D_{\alpha}}}{\bar{v}_{D_\alpha, F}},
\end{align}
where $\bar{v}_{D_\alpha, F} = v_{D_\alpha,F}/D_\alpha$.
$\bar{v}_{D_\alpha,F}$ is
the \emph{visibility} of deaths: the average number of times that each death in
group $\alpha$ would be reported if everyone in the frame population was
interviewed.

The network survival estimate for the number of deaths in demographic group
$\alpha$ ($D_\alpha$) is based on 
Eq.~\ref{eqn:nontext-id}.  
The numerator of Eq.~\ref{eqn:nontext-id}, $y_{F, D_\alpha}$,
is the total reported connections to deaths.
This quantity can be estimated from the data we collect about respondents'
connections to people who have died using a standard Horvitz-Thompson
approach:
\begin{align}
    \label{eqn:yhat-f-oalpha}
    \widehat{y}_{F,D_\alpha} = \sum_{i \in s} y_{i, D_\alpha} / \pi_i,
\end{align}
where $\pi_i$ is the probability that respondent $i$ was included in our
sample. $\pi_i$ is typically known from the survey's sampling design.
See Result~\ref{res:y-f-oalpha} for a formal statement and proof.

The denominator of Eq.~\ref{eqn:nontext-id} is the visibility of deaths,
$\bar{v}_{D_\alpha, F}$.  This quantity is more difficult to estimate. There
are many possible approaches, but we propose using the estimated average
personal network size of \emph{survey respondents} in demographic group
$\alpha$ to estimate the visibility of deaths in demographic group $\alpha$.
(We will describe how to estimate personal network sizes below.) For example,
our approach is to assume that the visibility of deaths among women aged 45-54
(i.e., the number of times each of these deaths could be reported) is the same
as the personal network size of women in the frame population aged 45-54. Using
respondents' average personal network size to estimate the visibility of deaths
will be exactly correct if (1) people who die in group $\alpha$ have personal
networks that are the same size, on average, as survey respondents in group
$\alpha$ (the \emph{decedent network assumption}); and, (2), survey respondents
are perfectly aware of and report all of the deaths in their personal networks
(the \emph{accurate reporting assumption}) (see Result~\ref{res:vbar-oalpha-f}
for a formal statement and proof). These are both strong assumptions; for
example, people who die might have smaller personal networks if they experience
an illness that reduces the size of their personal networks in the time leading
up to death.  Again, in Section~\ref{sec:ns-adjustment-factors}, we develop a
full framework for sensitivity analysis that shows exactly how estimates are
impacted by violations of these assumptions.

\subsubsection{Estimating the average personal network size of group $\alpha$, $\widehat{\bar{d}}_{F_\alpha, F}$}

In order to estimate the average personal network size of respondents in
demographic group $\alpha$, we adapt the known population
method~\citep{killworth_social_1998}.
The known population method asks respondents questions about their connections
to groups of known size (e.g., ``How many policemen do you know?'');
intuitively, the more connections a respondent reports to policemen, the bigger
we estimate her personal network to be. 
Respondents are typically asked about their connections to about 20 different
groups of known size, and the results are combined using the known population
estimator~\citep{killworth_social_1998, bernard_counting_2010,
feehan_generalizing_2016}.

The known population estimator was designed to estimate personal network sizes
for individual respondents.  Fortunately, we have a slightly easier problem:
estimating the average personal network size for a group of people. Therefore,
in Online Appendix~\ref{ap:ns-kp}, we derive an adapted estimator for the
average network size of respondents in a particular demographic group $\alpha$.
The main advantage of our adapted approach is that it requires slightly weaker
conditions than the traditional known population estimator.  The adapted known
population estimator is:
\begin{align}
    \label{eqn:dbar-falpha-f}
    \widehat{\bar{d}}_{F_\alpha, F} &= 
    \frac{\sum_{i \in s_\alpha} \sum_{j} y_{i, A_j} / \pi_i}{\sum_{j} N_{A_j}}~
    \frac{N_F}{N_{F_\alpha}},
\end{align}
where $\bar{d}_{F_\alpha, F} = d_{F_\alpha, F} / N_{F_\alpha}$ is the average
number of network connections between frame population members in demographic
group $\alpha$ ($F_\alpha$) and all the members of the frame population ($F$);
$N_F$ is the size of the frame population; $N_{F_\alpha}$ is the number of
frame population members who are also in demographic group $\alpha$;
$s_\alpha$ is the subset of survey respondents in demographic group $\alpha$;
$j \in \{1, \dots, J\}$ indexes the groups of known size; $y_{i, A_j}$ is the
number of connections that respondent $i$ reports to group of known size $A_j$;
and $N_{A_j}$ is the size of the $j$th group of known size. 
See Result~\ref{res:adapted-kp} for a formal statement and proof.

Combining the estimator for the number of reported deaths in group $\alpha$
(Eq.~\ref{eqn:yhat-f-oalpha})
with the estimator for the personal network size of survey respondents in 
group $\alpha$ (Eq.~\ref{eqn:dbar-falpha-f}) 
yields our estimator for the number of deaths in group $\alpha$:
\begin{equation}
    \label{eqn:dalpha-estimator}
    \widehat{D}_\alpha = 
    \frac{\widehat{y}_{F, D_\alpha}}{\widehat{\bar{d}}_{F_\alpha, F}}.
\end{equation}
See Result~\ref{res:o-alpha} for a formal statement and proof.

\subsection{Estimating the exposure, $N_\alpha$}

In order to convert the estimated total number of deaths into a death rate,
we need to estimate the amount of exposure $N_\alpha$.
If the sampling frame includes all adults, then
\begin{align}
    N_\alpha = N_{F_\alpha}, 
\end{align} 
\noindent and we say the frame population is \emph{complete} for $\alpha$.  
When the frame population is complete for $\alpha$, researchers can use
information from the sample design to estimate 
$N_\alpha$:
\begin{align}
    \label{eqn:exposure-estimator-complete}
    \widehat{N}_\alpha = \sum_{i \in s_\alpha} \frac{1}{\pi_i}.
\end{align}

If the sampling frame is not complete and if high quality estimates for the
exposure $N_\alpha$ are available from other sources, then researchers can use
the alternative approaches described in Result~\ref{res:m-alpha-complete}.

\subsection{Putting it all together to estimate death rates, $\widehat{M}_\alpha$}

Combining the estimator for the number of deaths (Eq.~\ref{eqn:dalpha-estimator}) and the estimator for the
exposure (Eq.~\ref{eqn:exposure-estimator-complete}), and simplifying, leads to the \emph{the network survival estimator} 
for the death rate in group $\alpha$:
\begin{align}
    \label{eqn:ns-all-together-reduced}
    \widehat{M}_\alpha &= 
	\frac{\widehat{y}_{F, D_\alpha}}{\widehat{\bar{d}}_{F_\alpha, F}} ~ 
	   \frac{1}{ \widehat{N}_{F_\alpha}}.
\end{align}
See Result~\ref{res:m-alpha-reduced} in Online Appendix~\ref{ap:ns-estimator} for a formal statement and proof.


\section{The network survival method in Rwanda}
\label{sec:results}

The arguments above and the proofs in the Online Appendices
show that the network survival method has attractive theoretical properties.  
They tell us little, however, about how the method actually works in practice.  
The ideal way to assess any new method is to use it in a situation like
the ones where it will be used in practice \emph{and} where it
can be validated.   
These two conditions, unfortunately, are rarely satisfied together.
Typically, we can test a new method in either a realistic situation \emph{or}
a situation where it can be validated.   
For this paper, we chose to test the network survival method in a realistic
situation: a large household survey in Rwanda, a country without a
high-quality vital registration system.  
This study alone, therefore, cannot be used to fully assess the network
survival method.  
But, neither could a study using the network survival method in the US, a
setting with a high-quality vital registration system but which is unlike
countries where the network survival method will typically be used.  
Ultimately, we think that empirical assessment of the network survival method
must involve both studies in realistic field situations and studies where
estimates can be validated against gold standard measures.

The network survival method can be used to collect reports about people
connected to respondents in almost any way.  Therefore, we had to decide who
we would ask respondents to report about.  In other words, we had to choose
the \emph{tie definition} that would be used in our study; this terminology
comes from the social networks literature, where a connection between nodes in
a network is called a tie.

Since people are embedded in many different personal networks---friendship
networks, family networks, occupational networks, and so forth---the ability to
choose a tie definition makes the network survival method very flexible.
Further, we expect that the choice of tie definition will have implications for both sampling
and non-sampling error because it trades off the quality and quantity of
information collected in each interview~\citep{feehan_quantity_2016}.  
Roughly, we expect that using a weaker tie definition will collect more, noisier
information per interview.  Using a stronger tie definition, on the other hand,
could produce more accurate information but about a small number of other
people.  Obviously, researchers would like to choose a tie definition that would minimize total error (i.e., sampling error + non-sampling error).  Because no network survival data has been collected previously, we had no way to assess this trade-off empirically before embarking.  

Therefore,  we conducted a survey experiment that randomized respondents to
report about one of two different types of personal network: half of our sample
reported a relatively weak tie network---their \emph{acquaintance
network}---while the other half of the sample reported about a relatively
strong tie network---their \emph{meal network} (Table~\ref{tab:tiedefns}).  The
acquaintance tie definition has been used in all previous network
scale-up studies~\citep{bernard_counting_2010}, and our study was the first to
use the meal definition, which we devised and refined in collaborations with
local experts in Rwanda.  We pilot tested both definitions to ensure that they
were appropriate in Rwanda.  Overall, this survey experiment enables us to
better understand this key aspect of the method.

\begin{table}[h]
  \begin{tabular}{p{3in} p{3in}}
\hline
    \multicolumn{2}{c}{\bf \underline{Tie Definitions}}\\
    \multicolumn{1}{c}{\underline{Acquaintance ($n = 2,236$)}} & 
     \multicolumn{1}{c}{\underline{Meal ($n = 2,433$)}}\\
    \begin{itemize}
    \item people of all ages who live in Rwanda
    \item people the respondent knows, by sight AND name, and who also
      know the respondent by sight and name
    \item \emph{people the respondent has had some contact with -- either in
      person, over the phone, or on the computer in the previous 12
      months}
    \end{itemize}
    &
    \begin{itemize}
    \item people of all ages who live in Rwanda
    \item people the respondent knows, by sight AND name, and who also
      know the respondent by sight and name
    \item \emph{people the respondent has shared a meal or drink with in the
      past 12 months, including family members, friends, co-workers,
      or neighbors, as well as meals or drinks taken at any location,
      such as at home, at work, or in a restaurant.}
    \end{itemize}\\
\hline
  \end{tabular}
  \caption{
      The two definitions of a personal network connection (also called a
      tie) that were used in our study.  All of the conditions need to
      be satisfied for the respondent to consider someone a member of
      her network. 
  }
  \label{tab:tiedefns}
\end{table}

\subsection{Data collection}


Our survey used the same interviewers, data entry protocols, training
techniques and sampling procedures as the 2010 Rwanda DHS.
By using the DHS infrastructure, we ensure that our research design can
be used in face-to-face surveys in developing countries across the world.
Our sample--which was a special survey, distinct from the 2010 Rwanda DHS--was
drawn using a stratified, two-stage cluster design, and interviews were
conducted between June and August of 2011.  
The household response rate was 99\% and individual response rate was 97\%.  
The full details of the sampling plan and field procedures are described elsewhere
\citep{rwanda_biomedical_center/institute_of_hiv/aids_estimating_2012}.
Following the guidelines of the DHS program \citep[sec.
1.13.7]{icf_international_demographic_2012}, we de-normalize the sampling
weights by using the UN Population Division estimates for the size of Rwanda's
population aged 15 and above in 2010 \citep{united_nations_world_2013}.  When
quantifying the sampling uncertainty in our estimates we use the rescaled
bootstrap, which accounts for our complex sample
design~\citep{rao_resampling_1988, rao_recent_1992, feehan_generalizing_2016}.

Each sampled household was randomly assigned to one of the two possible
definitions of a network, and balance checks show that the randomization was
successfully implemented \citep{feehan_quantity_2016}.  All adults in each
household were interviewed.  Our choice to interview all adults differs from a
typical DHS, which interviews women up to age 50 and men up to age 60; we discuss 
this difference
and its implication for estimates in greater detail in Online
Appendix~\ref{ap:respondent-age}. Table~\ref{tab:rwanda-known-popns} shows the
known populations that were used to estimate personal network sizes in our
study in Rwanda.  More information about how these particular known populations
were chosen and general advice about choosing known populations can be found
elsewhere \citep{rwanda_biomedical_center/institute_of_hiv/aids_estimating_2012,
feehan_quantity_2016, feehan_generalizing_2016}.

\begin{table}[p]
\begin{tabular}{  l r l }
\hline
	Group name                              & Size      &  Source \\ 
    \hline
	Priests                                 & 1,004     &  Catholic Church \\ 
	Nurses or Doctors                       & 7,807     &  Ministry of Health \\ 
	Twahirwa                                & 10,420    &  ID database \\ 
	Mukandekezi                             & 10,520    &  ID database \\ 
	Nyiraneza                               & 21,705    &  ID database \\ 
	Male Community Health Worker            & 22,000    &  Ministry of Health \\ 
	Ndayambaje                              & 22,724    &  ID database \\ 
	Murekatete                              & 30,531    &  ID database \\ 
	Nsengimana                              & 32,528    &  ID database \\ 
	Mukandayisenga                          & 35,055    &  ID database \\ 
	Widowers                                & 36,147    &  RDHS (05, 07, 10) \\ 
	Ndagijimana                             & 37,375    &  ID database \\ 
	Bizimana                                & 38,497    &  ID database \\ 
	Nyirahabimana                           & 42,727    &  ID database \\ 
	Teachers                                & 47,745    &  Ministry of Educ. \\ 
	Nsabimana                               & 48,560    &  ID database \\ 
	Divorced Men                            & 50,698    &  RDHS (05, 07, 10) \\ 
	Mukamana                                & 51,449    &  ID database \\ 
	Incarcerated people                     & 68,000    &  ICRC 2010 report \\ 
	Women who smoke                         & 119,438   &  RDHS (05) \\ 
	Muslim                                  & 195,449   &  RDHS (05, 07, 10) \\ 
	Women who gave birth in the last 12 mo. & 256,164   &  RDHS (10) \\ 
	~  & ~  \\ 
    \hline
\end{tabular}

\caption{The known populations used to estimate network sizes in the Rwanda study. RDHS denotes the Rwanda Demographic and Health Survey from the years indicated in parentheses,  ID database denotes counts of names from the national identity card database, and ICRC is the International Committee of the Red Cross. }
\label{tab:rwanda-known-popns}
\end{table}

We had to pay careful attention to constructing the wording of the question
that asked respondents to report about deaths. Both tie definitions used in our
study in Rwanda were based on interactions (Table~\ref{tab:tiedefns})---either
contact, for the acquaintance definition, or sharing a meal or drink, for the
meal definition. Of course, people who have died cannot continue to interact
with others.  We therefore expect people who have died in the 12 months before
a survey to have had fewer total interactions than people who did not. This
expected systematic difference is problematic for network survival estimates, which are
based on the assumption that the visibility of deaths can be estimated by the
personal network size of survey respondents (the \emph{decedent network
assumption} in Result~\ref{res:o-alpha}). Thus, we do not want the personal
networks of people who died to be smaller, on average, than people who lived.  We attempted to circumvent this potential problem in our study by asking
respondents to report people who satisfy two conditions: (i) the person died in
the 12 months before the interview; and (ii) the person shared a meal with the
respondent \emph{in the 12 months before death}.   We discuss this choice, its possible impact on estimates, and alternative approaches in Online Appendix~\ref{ap:ns-instrument}.  Online Appendix~\ref{ap:ns-instrument} also includes an excerpt of the English translation of the survey instrument.  All of the survey materials, including the original Kinyarwanda instruments, are freely available from the DHS website
\citep{rwanda_biomedical_center/institute_of_hiv/aids_estimating_2012}.

\subsection{Basic descriptives}

\begin{figure} 
  \centering
  \includegraphics[width=\textwidth,keepaspectratio]{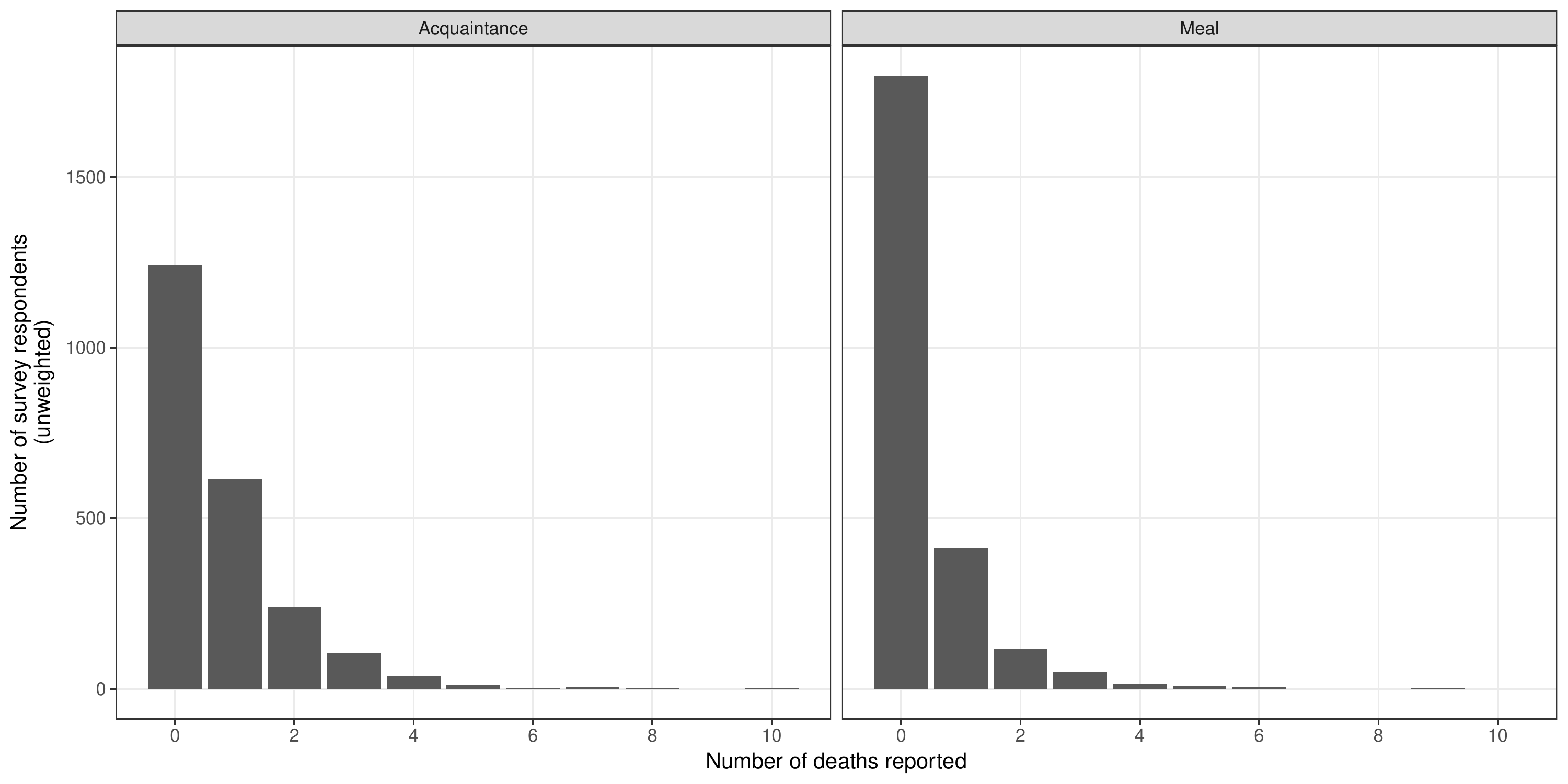}
  \caption{
      Distribution of the number of adult deaths reported by respondents using the acquaintance network
      (left panel) and the meal network (right panel).
  }
  \label{fig:net-num-deaths}
\end{figure}

\begin{figure} 
  \centering
  \includegraphics[width=\textwidth,keepaspectratio]{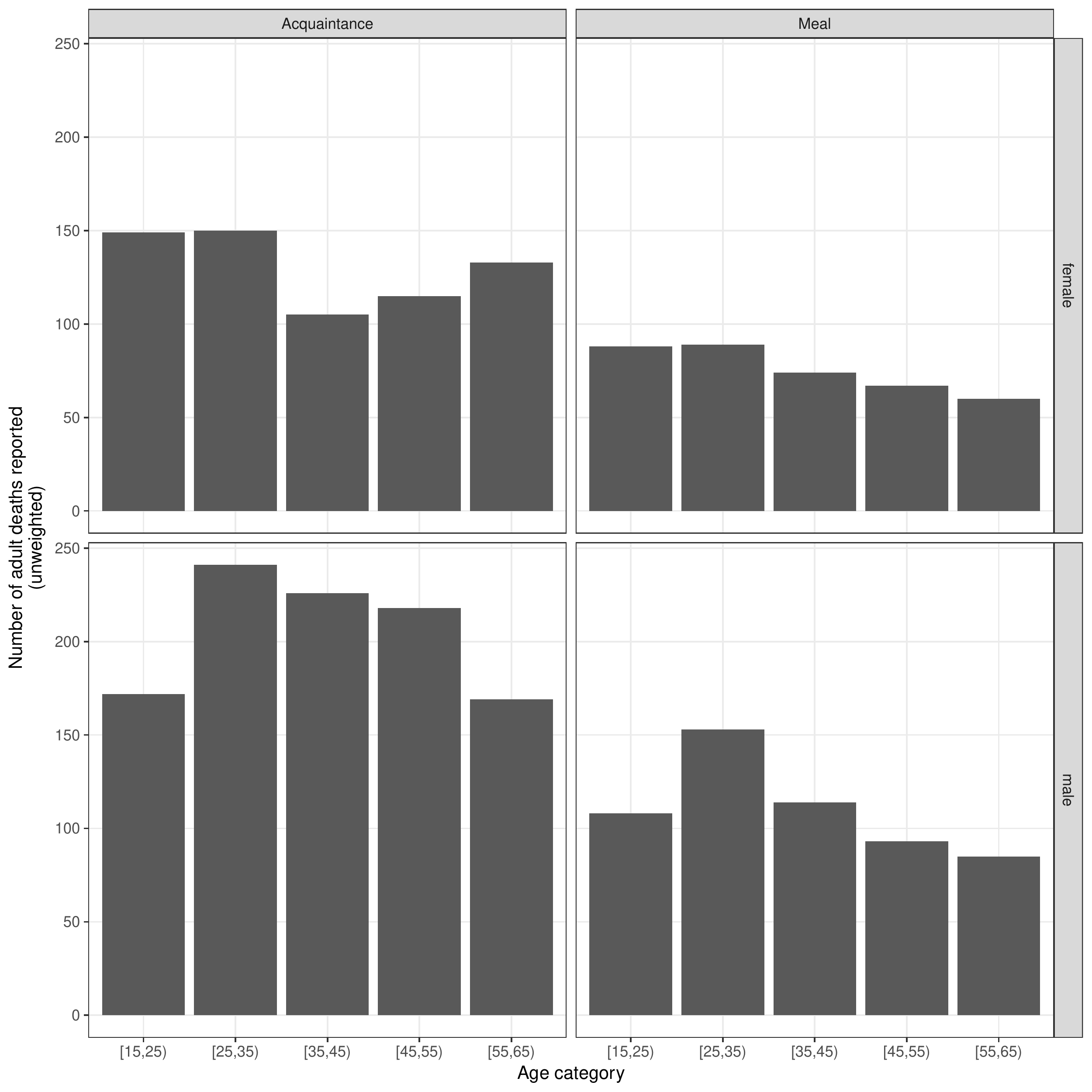}
  \caption{
      Age and sex distribution of adult deaths reported by respondents using the acquaintance network
      (left panels) and the meal network (right panels).
  }
  \label{fig:net-deaths-agesex}
\end{figure}

To provide intuition about the information about deaths that the network
reporting collects, we begin by reporting some basic descriptives.  
Figure~\ref{fig:net-num-deaths}
shows the distribution of the number of deaths per interview in the two arms of
the survey experiment.  
As expected, respondents reported knowing more deaths in the acquaintance
condition (0.7 deaths per interview) than the meal condition (0.4 deaths 
reported per interview) (Table~\ref{tab:deathperint}).  
Further, Figure~\ref{fig:net-deaths-agesex} reports the age-sex distributions
of the reported deaths in the two arms of the survey experiment.\footnote{Out
    of the 3,853 reported deaths, 8 (0.2\%) were missing age, sex, or both.
    These reported deaths are excluded from this analysis.}  
Online Appendix~\ref{ap:quantity-and-quality} has numerous other descriptive plots
including plots about 1) the responses for the groups of known size, 2) heaping
in reported ages of death, and 3) a more detailed comparison between responses
to the questions related to the network reporting method and sibling survival method.

\subsection{Network survival method estimates}

\begin{figure}[p] 
  \centering
  \includegraphics[width=\textwidth,keepaspectratio]{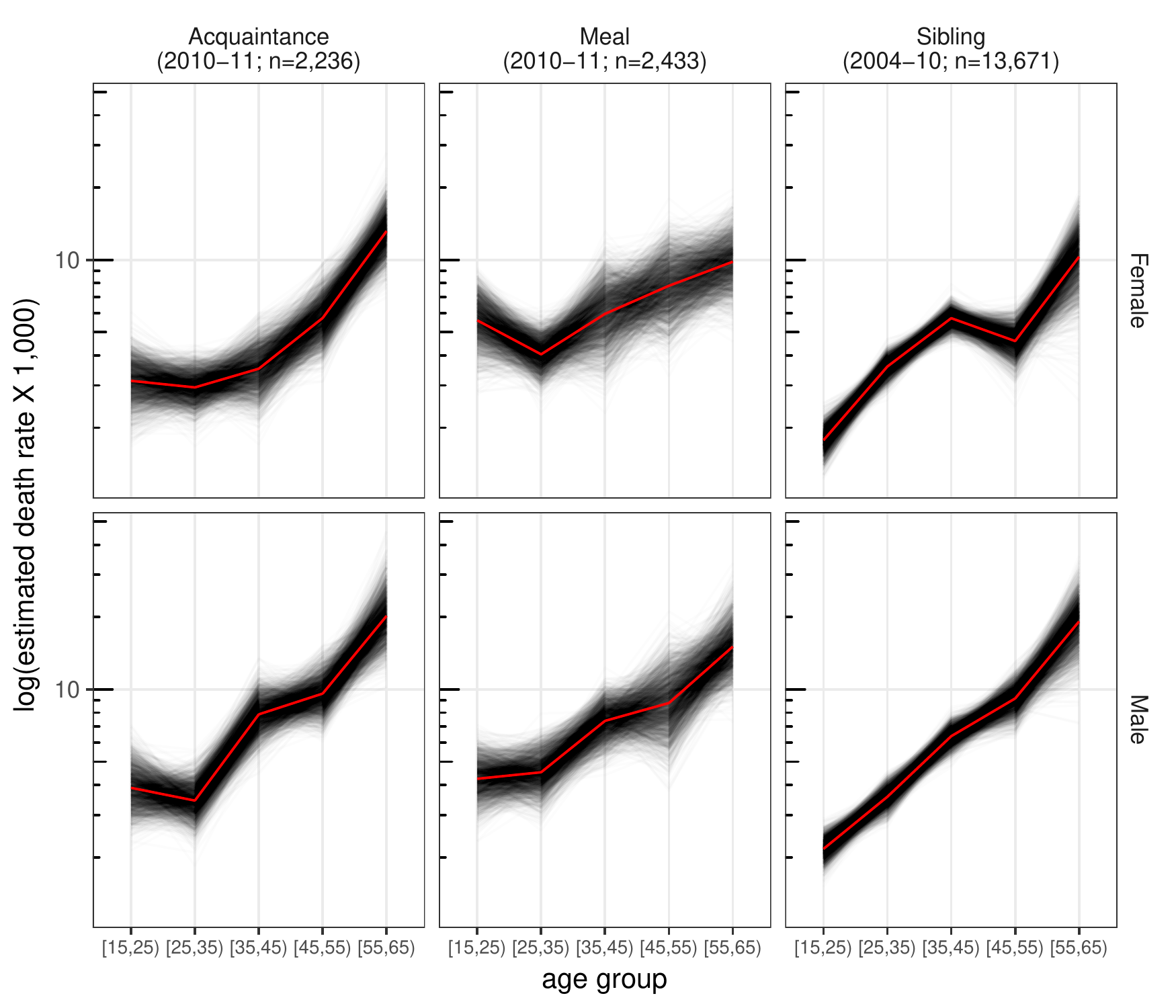}
  \caption{
    Comparison between network survival death rate estimates for two types
    of personal network (left-hand column and middle column), 
    and direct sibling survival death rates estimates from the
    2010 Rwanda Demographic and Health Survey (right-hand column). The top
    row has death rates estimated for females, while the bottom row has death
    rates estimated for males. The network survival estimates are based on
    reported deaths from the 12 months prior to the interview. 
    The sibling estimates are based on reported deaths in the
    84 months prior to the interview because estimates from the 12 months prior
    were too unstable (see Online Appendix~\ref{ap:comparison-estimates}). 
    Each gray line shows the estimate from one bootstrap resample; taken together,
    the set of lines shows the estimated sampling uncertainty of the death rates.
    The thicker black lines show the mean of the bootstrap resamples.
  }
  \label{fig:net-asdr-all}
\end{figure}

\begin{figure} 
  \centering
  \includegraphics[width=.5\textwidth,keepaspectratio]{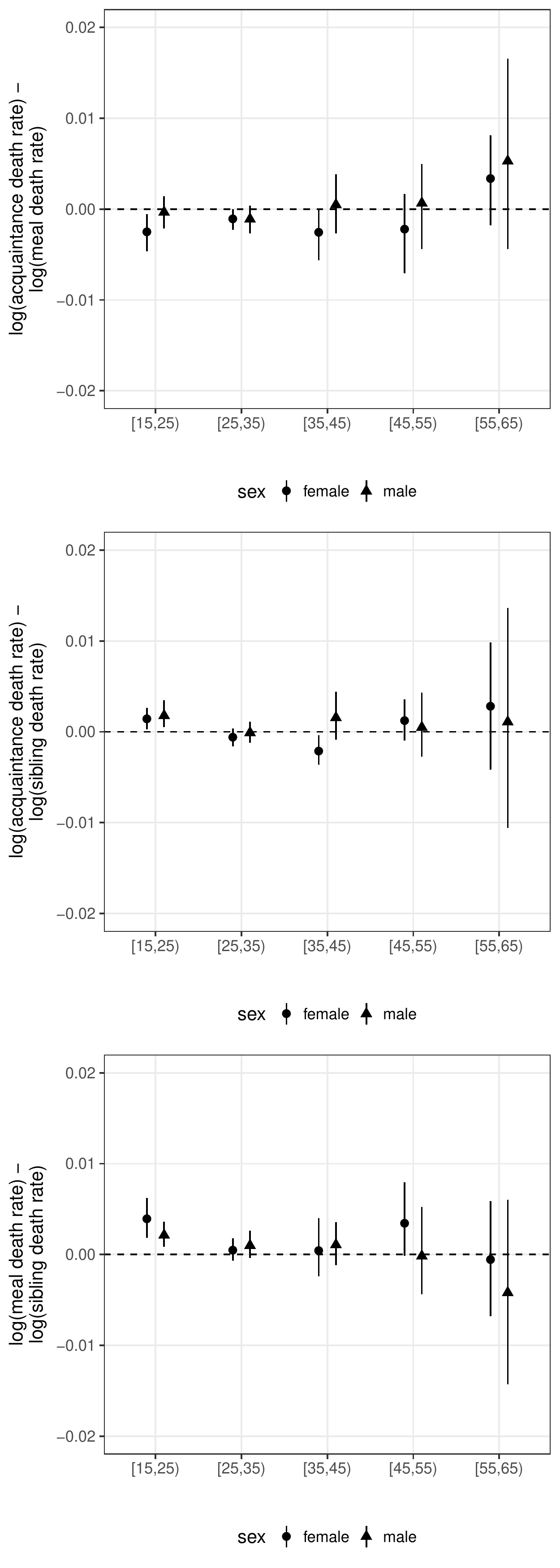}
  \caption{
      Age-specific differences between the estimated log death rate using (i)
      the acquaintance network and the meal network (top panel); (ii) the
      acquaintance network and the sibling histories (middle panel); and (iii)
      the meal network and the sibling histories (bottom panel). Above the
      dotted line, estimated death rates from the meal or acquaintance network
      are higher. These estimates are presented in tabular form
      in Online Appendix~\ref{ap:tabular}.
  }
  \label{fig:all-asdr-age-comparisons}

\end{figure}

Figure~\ref{fig:net-asdr-all} (left and middle columns) reports the estimated age-specific death rates ($\widehat{M}_\alpha$, Eq.~\ref{eqn:ns-all-together-reduced})
across the two tie definitions for males and females\footnote{%
All of our estimates were computed in 
R~\citep{r_core_team_r:_2014} 
using the following packages:
networkreporting~\citep{feehan_networkreporting_2014},
surveybootstrap~\citep{feehan_surveybootstrap:_2016},
plyr~\citep{wickham_split-apply-combine_2011},
dplyr~\citep{wickham_dplyr:_2015},
stringr~\citep{wickham_stringr:_2012},
ggplot2~\citep{wickham_ggplot2:_2009},
devtools~\citep{wickham_devtools:_2013},
stargazer~\citep{hlavac_stargazer:_2014},
car~\citep{fox_r_2011},
and
gridExtra~\citep{auguie_gridExtra:_2012}.  Also, following conventional practice in the network scale-up literature, all network reports about groups of known size were topcoded at 30,
meaning that reported values greater than 30 were treated as 30; this topcoding
affected 0.2 percent of the responses.
}.  
As expected, the estimated death rates generally increase with age (with the exception of young females for the meal definition).  The top panel of Figure~\ref{fig:all-asdr-age-comparisons} directly plots the difference between estimates from the two tie definitions for different age groups, and it shows that there is broad overall agreement between the estimates from each tie definition with the largest differences in the oldest age group.

\section{Comparison to other estimates}
\label{sec:comparisons}

In addition to comparing our network survival estimates to each other, we also
compare them to direct sibling survival estimates produced from the 2010 Rwanda
Demographic and Health Survey (DHS) \citep{nisr_rwanda_2012} and to estimates
produced by three organizations: the World Health
Organization, the United Nations Population Division, and the Institute for
Health Metrics and Evaluation.  
To foreshadow our results, we find that the network survival estimates were
similar to the sibling survival estimates and to estimates from the three
organizations.

\subsection{Comparison to estimates from the sibling survival method}

The 2010 Rwanda DHS finished fieldwork in March 2011, right before our data
collection started. 
As is typical in a Demographic and Health Survey, only
women of reproductive age (aged 15-49) were interviewed using the sibling
survival module. 
Therefore, the sibling survival estimates below are based on the sibling
histories of the 13,671 women between 15 and 49 who were interviewed in the
12,540 households sampled in the DHS. 

Even with 13,671 respondents, however, we found that estimated death rates for
the 12 months before the survey were too imprecise to usefully compare to
network survival estimates (Figure~\ref{fig:asdr-1vs5}). 
Therefore, we follow the recommendations of the
sibling survival literature and pool together information from reports about
84 months (7 years) prior to the survey \citep{stanton_assessment_2000,
timaeus_adult_2004}.  The sibling survival estimates are thus estimated average
death rates over the 84 months before the survey, while the network survival
estimates are estimated death rates for the 12 months prior to the survey. 
(See Online Appendix~\ref{ap:comparison-estimates} for detailed information about how we calculated
sibling survival estimates.)  As with the network survival estimates, we estimate the sampling uncertainty in
the sibling survival estimates using the rescaled bootstrap, which accounts for the complex sample design of the DHS~
\citep{rao_resampling_1988, rao_recent_1992}. 

Figure~\ref{fig:net-asdr-all} shows the age-specific death rates produced from
the network reporting method (left and middle columns) and the ones produced
by the direct sibling survival method (right column).  
Further, Figure~\ref{fig:all-asdr-age-comparisons} directly shows differences
between the acquaintance and sibling estimates (middle panel) and between the
meal and sibling estimates (bottom panel). 
This comparison shows that network survival estimates from both tie definitions
are similar to the sibling survival estimates, even though the network survival
estimates are based on a sample that is roughly one-fifth the size ($n=2,236$
network reporting method (acquaintance); $n=2,433$ network reporting method
(meal); $n=13,671$ sibling survival method).   
One systematic difference between the two methods is that the network survival
estimates are slightly higher than sibling survival estimates for the youngest
age group.

\begin{figure} 
  \centering
  \includegraphics[width=0.5\textwidth,keepaspectratio]{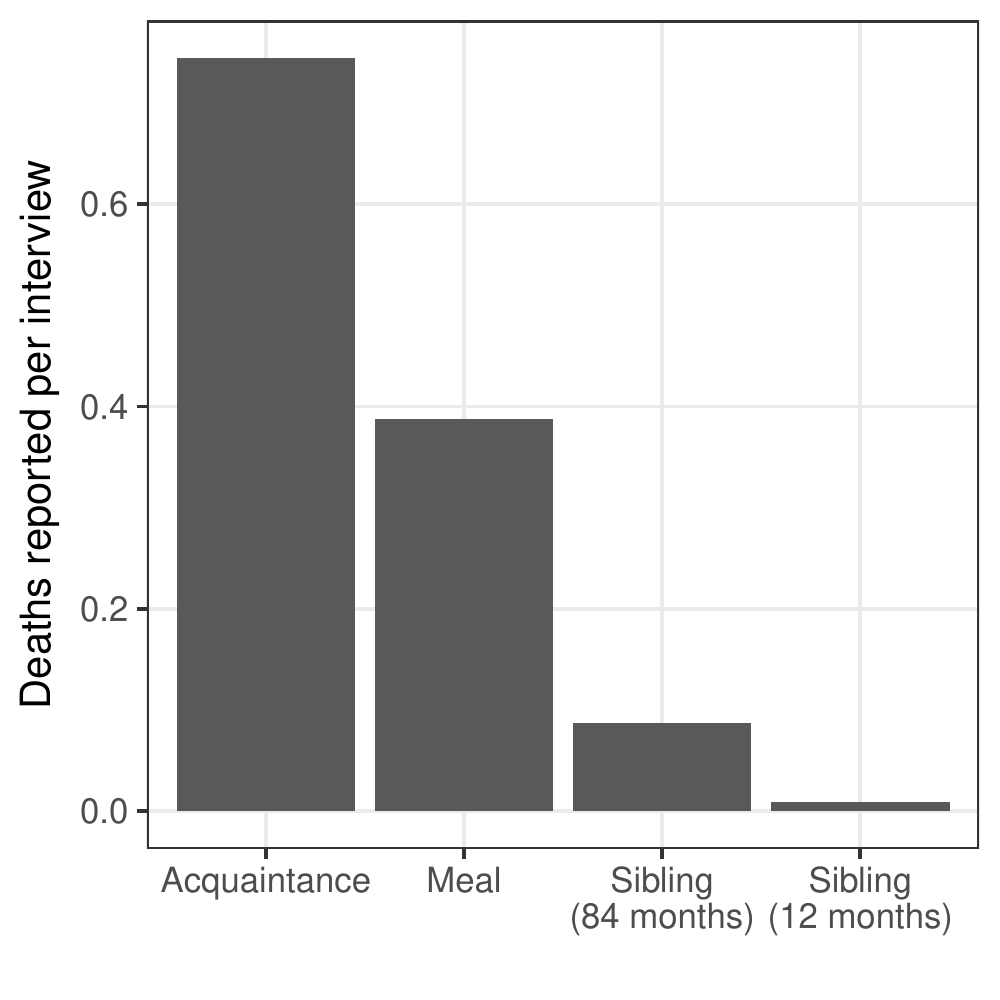}
  \caption{
    Average number of deaths reported from each interview in Rwanda using the
    acquaintance and meal tie definitions from the network survival study, and
    using the sibling history module of the DHS survey. 
    The acquaintance and meal definitions use reported information about deaths
    in the 12 months prior to the survey.
    Compared to sibling reports about 84 months before the survey, 
    network survival respondents reported about
    8 times more deaths using the acquaintance tie definition and
    about 4 times more deaths using the meal tie definition.
    Compared to
    the sibling reports about 12 months before the survey, 
    network survival respondents reported about 82 times more deaths using the
    acquaintance tie definition and about 43 times more deaths using the meal
    tie definition.
  }
  \label{fig:deaths-per-int}
\end{figure}

To clarify how the network survival method was able to produce similar
estimates with substantially smaller samples, Figure~\ref{fig:deaths-per-int}
compares the number of deaths reported per interview for the different
approaches.  Considering a 12 month reporting window, the network survival
method yielded about 40 times (meal) or 80 times (acquaintance) more deaths per
interview than the sibling survival method\footnote{Another way to compare the
amount of information per interview is to compare the number of deaths
reported with the network survival method (12 month reporting window) to the
number of deaths reported with the sibling survival method (84 month reporting
window).  In this case, the network survival method yields 4 times (meal) or 8
times (acquaintance) more deaths per interview than the sibling survival
method.
}.  
Because it yields so many more deaths per interview than the sibling
survival method, the network survival method can produce more granular
estimates in samples of a similar size or can produce similar estimates with
smaller samples.  

\subsection{Comparison to estimates from organizations}

In addition to comparing network survival estimates to sibling survival estimates,
we also compare them to estimated adult mortality
rates produced by three organizations: 
the United Nations Population Division (UNPD)~
\citep{united_nations_population_division_world_2015}%
\footnote{
    UNPD estimates are taken from the 2015 revision of the World Population Prospects:
    \url{http://esa.un.org/unpd/wpp/Download/Standard/ASCII/}
    (accessed March~17,~2016).
};
the World Health Organization~\citep{who_global_2015}%
\footnote{
    WHO estimates are taken from the Global Health Observatory:
    \url{http://apps.who.int/gho/data/node.main.11?lang=en}, and
    \url{http://apps.who.int/gho/data/view.main.61370}
    (accessed March~17,~2016).
};
and, the Institute for Health Metrics and Evaluation~\citep{nagavi_global_2015}%
\footnote{
    IHME estimates are taken from the 2013 Global Burden of Disease study:
    \url{http://ghdx.healthdata.org/global-burden-disease-study-2013-gbd-2013-data-downloads}
    (accessed March~17,~2016).
}.

Researchers typically use estimates from these organizations to compare adult
mortality across countries using an aggregate quantity called  $\ffqf$. 
$\ffqf$ is the conditional probability of dying before age 60 among people who
survive to age 15, and who then face the given age-specific death rates~\citep{preston_demography:_2001-1, wachter_essential_2014}.
For example, a set of age-specific death rates with $\ffqf$ of 0.2 implies that
20\% of people who survive to age 15 and then face those age-specific death rates
will die before age 60.  The estimated $\ffqf$ from each organization is derived from a complex combination of data sources, models, and expert judgment\footnote{
In brief, the methods used to estimate adult mortality for
WHO and the UN Population Division are fairly similar: data from censuses and
household surveys (such as the DHS), are combined with model life tables to
estimate the adult mortality levels.  These estimates, therefore, rely on extrapolating adult mortality from estimates
of child mortality levels \citep[see][for a more detailed discussion]{masquelier_divergences_2014}.
For IHME, a smoothed regression approach is taken that incorporates additional
variables related to health and borrows strength from data from other countries and time periods.  
For a more information about how these organizations produce estimates, see 
\citet{united_nations_population_division_world_2015}, 
\citet{wang_age-specific_2013}, and 
\citet{who_global_2015}.
}.

\begin{figure} 
  \centering
  \includegraphics[keepaspectratio,width=\textwidth]{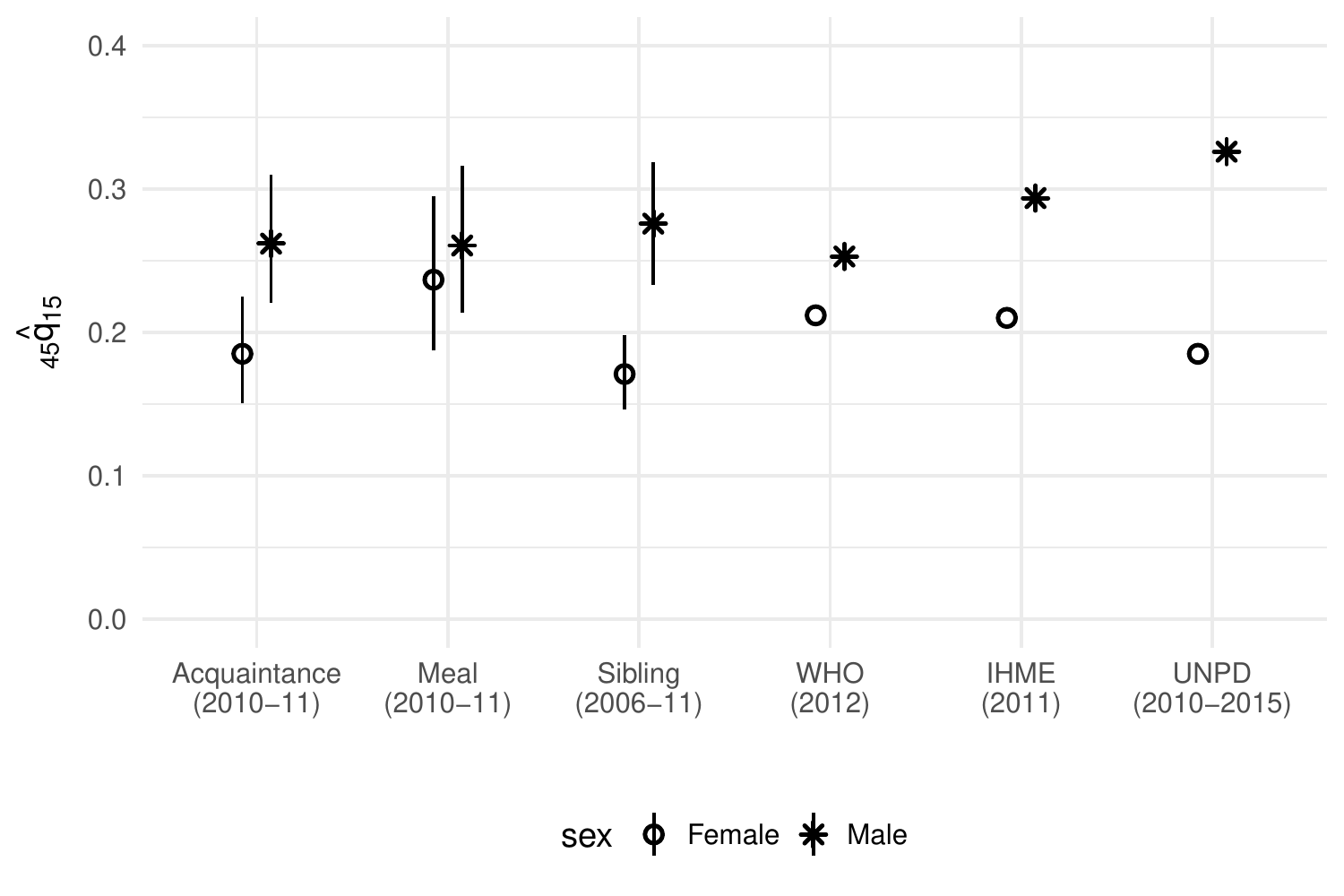}
  \caption{ 
      Estimated $\ffqf$ for Rwanda from six different sources: the acquaintance
      and meal tie definitions from our network survival method; 
      the direct sibling survival method from the 2010 Rwanda Demographic and Health Survey; 
      the United Nations Population Division (UNPD); 
      the World Health Organization (WHO); 
      and the Institute for Health Metrics and Evaluation (IHME). 
      Error bars indicate 95\% sampling uncertainty intervals for the survey-based estimates,
      which were computed using the rescaled bootstrap.
      Note that the estimates are not for exactly the same time periods.
  }
  \label{fig:all-45q15s}
\end{figure}

Figure~\ref{fig:all-45q15s} compares estimated $\ffqf$ for Rwanda from the
network survival method to estimates from three
organizations.  
(No sampling-based uncertainty estimates are available for the estimates from
the organizations.)
Figure~\ref{fig:all-45q15s} shows that estimates from the network survival
method are similar to estimates from the WHO and IHME, and to female estimates
from UNPD (UNPD's male $\ffqf$ estimates are slightly higher than all of the
other estimates).  
Figure~\ref{fig:all-45q15s} also shows that the
difference between male and female mortality appears to be larger for the
acquaintance network than for the meal network, a pattern which was not
as apparent from in Figure~\ref{fig:all-asdr-age-comparisons}.
In Online Appendix~\ref{ap:comparison-estimates}, we extend this comparison to age-specific
death rates and again find that estimates from both arms of our survey
experiment are similar to estimates from WHO, IHME, and UNPD
(Figure~\ref{fig:ref-asdr-all}).  The estimates from the network reporting
method, however, did not require model life tables or other external data from
neighboring countries or time periods.

\section{Framework for sensitivity analysis}
\label{sec:ns-adjustment-factors}

Any approach to estimating adult mortality rates will have to make assumptions.
Unfortunately, it is not clear how the sibling survival method and the methods
used by the organizations are impacted by violations of their
underlying  assumptions.  
Because of the mathematical structure of the network survival
method, however, we were able to derive a complete framework for sensitivity analysis.  
This framework shows analytically how the network survival estimates are
impacted by violations of assumptions, both individually and jointly.  

We develop the full framework in Online
Appendix~\ref{ap:ns-decomposition-framework}, which includes conditions related
to i) respondent reporting behavior; ii) social network structure; iii)
questionnaire construction; and iv) sampling.  Here, we illustrate the
sensitivity framework by focusing on three important conditions, which were
introduced in Section~\ref{sec:estimating_deaths}: the no false positives assumption,
the decedent network condition and
the accurate reporting condition.

The network survival estimator's sensitivity to these three important
conditions is captured by the decomposition in
Eq.~\ref{eqn:ns-adjustmentfactors-main}, which relates the true number of
deaths ($D_\alpha$) to the 
network survival estimand ($y_{F,D_\alpha}/\bar{d}_{F_\alpha,F}$)
and three multiplicative adjustment factors ($\delta_{F,\alpha}$, $\eta_{F,\alpha}$,
and $\tau_{F,\alpha}$): 
\begin{align}
\label{eqn:ns-adjustmentfactors-main}
D_\alpha &= 
\underbrace{%
\left( \frac{y_{F,D_\alpha}}{\bar{d}_{F_\alpha,F}} \right)
}_{\substack{\text{network survival} \\ \text{estimand for } D_\alpha}}
\times 
\underbrace{%
    \left(\frac{1}{\delta_{F, \alpha}}\right) \times 
    \left(\frac{\eta_{F,\alpha}}{\tau_{F,\alpha}}\right).
}_{\mbox{adjustment factors}}
\end{align}
The first adjustment factor---the degree ratio ($\delta_{F, \alpha}$)---is
related to the structure of the underlying social network: it is exactly 1 when
the decedent network assumption is satisfied, less than 1 if survey respondents 
in group $\alpha$ have 
bigger personal networks than people who died, and greater than 1 otherwise.
The other two adjustment factors---the true positive rate ($\tau_{F,\alpha}$)
and the precision ($\eta_{F,\alpha}$)---are related to the accuracy of reporting;
when respondents' reports are perfectly accurate, then both $\tau_{F,\alpha}$
and $\eta_{F,\alpha}$ are 1. If there are false positive reports, then the precision
will be less than 1; and, if respondents do not report all of the deaths that 
actually happen in their personal networks, then the true positive rate will be less
than 1. Online Appendix~\ref{ap:ns-decomposition-framework} has more information,
including precise definitions of each adjustment factor. 

\begin{figure}[p] 
  \centering
  \includegraphics[width=\textwidth,keepaspectratio]{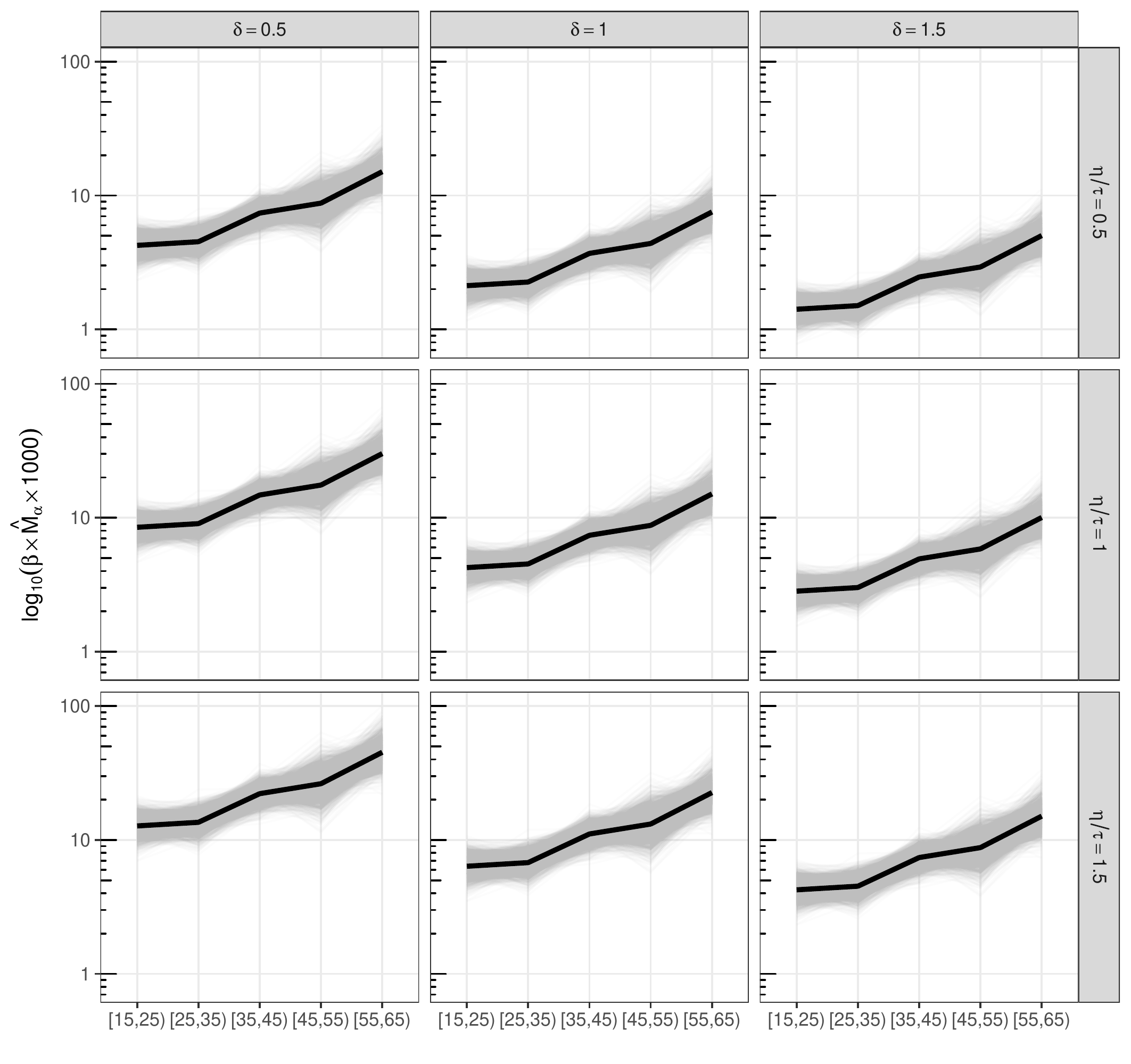}
  \caption{
    Estimated age-specific death rates for Rwandan males using the meal
    definition under violations of reporting and network structure conditions.
    The rows show different types of reporting:
    in the middle row, the accurate reporting condition holds 
    ($\eta_{F,\alpha}/\tau_{F,\alpha}=1$); in the top row, reporting
    tends to omit deaths ($\eta_{F,\alpha} / \tau_{F,\alpha}=0.5$); and in
    the bottom row, reporting tends to erroneously include deaths
    ($\eta_{F,\alpha} / \tau_{F,\alpha} = 1.5$). 
    The columns show different types of personal network structure:
    in the middle column, the decedent network condition holds
    ($\delta_{F, \alpha}=1$);
    in the left column, people who die have smaller personal networks than
    the average frame population member ($\delta_{F, \alpha}=0.5$);
    in the right column, people who die have personal networks
    that are larger than the average frame population member
    ($\delta_{F, \alpha}=1.5$).
    Note that violations of the accurate reporting and decedent network condition
    can work in opposite directions, balancing each other out (top-left and bottom-right panels); 
    or, they can work in the same direction, making estimates
    less accurate (bottom-left and top-right).
  }
  \label{fig:net-asdr-robustness}
\end{figure}

Figure~\ref{fig:net-asdr-robustness} illustrates how the decomposition in
Eq.~\ref{eqn:ns-adjustmentfactors-main} can be used to assess how death rate
estimates are impacted by (1) violations of the decedent network condition
($\delta_{F,\alpha}=1$, columns) and (2) violations of the two reporting
conditions ($\eta_{F,\alpha}/\tau_{F,\alpha}=1$, rows).
Figure~\ref{fig:net-asdr-robustness} shows that violations of these conditions
can work in opposite directions, canceling each other's effects (e.g. the bottom-right panel
of Figure~\ref{fig:net-asdr-robustness}); or they can
work in the same direction, making the estimates less accurate (e.g. the bottom-left panel
of Figure~\ref{fig:net-asdr-robustness}).  This example illustrates a small portion of the sensitivity framework in Online
Appendix~\ref{ap:ns-decomposition-framework}, which can be used to assess how
sensitive death rate estimates are to all of the conditions required by the
network survival estimator, individually and jointly.

\section{Discussion}
\label{sec:discussion}

Understanding adult mortality is critical to a wide range of important research
and policy questions, but estimating adult death rates remains difficult in
countries that lack high-quality vital registration systems.  In this study, we
introduced a promising new method for estimating adult death rates that
overcomes many of the limitations of existing approaches, such as the sibling
survival method. Our approach---the network survival method---uses information
about survey respondents' personal networks to estimate adult death rates. 

In addition to deriving the theoretical properties of the network survival
estimator and developing a framework for sensitivity analysis, we also designed
and conducted a nationally-representative survey experiment to test the method
in Rwanda, a setting where improved methods for estimating adult mortality are
sorely needed.  
We found that two versions of the network reporting method produced estimates
that were similar to those produced by the sibling survival method, even though
the network reporting estimates were based on a sample that was one-fifth the
size.  
Further, the aggregated versions of the network survival estimates were
comparable to the estimates from three organizations that incorporate data from
multiple surveys and model life tables to create smoothed estimates.

Our results---theoretical and empirical---show that the network survival method
can potentially overcome the two fundamental challenges in estimating death rates 
from surveys: it enables researchers to learn about people who died, and 
it can produce estimated death rates by age and sex from survey samples of
moderate size.

The network survival method also has some potential advantages over the sibling
survival method.
First, 
the network survival method collects more information per interview
than the sibling survival method.
In our study in Rwanda, it collected about 80 times more reported deaths using
the acquaintance tie definition and about 40 times more reported deaths using
the meal tie definition (Figure~\ref{fig:deaths-per-int}).  
By collecting more information per interview, the network reporting method was
able to directly estimate adult death rates by age and sex for the 12 months prior to
the survey without any pooling across countries or time.  Because one of the
main goals monitoring adult death rates is to detect---and react to---changes,
the ability to produce direct, local, and timely estimates would be an
improvement over current estimates that are pooled in a variety of different
ways.  
Based on the high number of deaths reported per interview by network survival
respondents in Rwanda, we believe that the network survival estimator could
produce estimates of adult death rates for the past 12 months based only
on data from a survey like the DHS.

Second, the network survival method has a formal framework for sensitivity
analysis which allows researchers to clearly identify and analytically quantify
the impact of structural and reporting errors---and the interaction between
them---on estimates.  
As a result, there is no ambiguity about how potential biases will impact
network survival estimates, and it is straightforward to conduct routine
sensitivity analyses of all estimates.
Such a framework does not yet exist for the sibling survival method,
which has been the subject of methodological
uncertainty about different sources of bias and how they might interact.  

There are many potential directions for future work.
First, we believe that there should be additional studies assessing the quality
of network survival estimates in countries without vital records systems and in
countries where estimates can be compared to gold standard measures.
Second, the flexibility of the network survival method means that the type of network
respondents report about can be customized---and hopefully optimized---for
different settings. For example, in one country it might make sense to ask
about the network of people who attend the same mosque, while in a different
country it would make more sense to ask about people who attend the same
church.  
This choice of tie definition has implications for the size and nature of
reporting errors, structural biases, and sampling uncertainty.  
Therefore, future research should develop methods for choosing the optimal tie
definition for each study.
Third, although we focused on estimating national-level adult death rates as
part of routine household surveys, there is a demand for survey-based
approaches to estimate mortality in a wide range of other settings, including
conflicts, natural disasters, famines, epidemic outbreaks, and other
humanitarian crises~\citep{checchi_documenting_2008,epicentre_wanted:_2007}.
We believe that the network survival method could be tailored to work in some
of these settings as well.
Fourth, our survey interviewed adults of all ages, but some household surveys
restrict the population that they interview by age or sex
 potentially limiting the ability to produce reliable age-specific mortality
rates for age groups other than those of the survey respondents (such as
$\q{60}{20}$). Mortality among older age groups is becoming increasingly important to
measure given the global shift toward monitoring mortality related to
non-communicable diseases which largely occur in the older age groups\footnote{%
See Target 3.4: \url{http://unstats.un.org/sdgs/metadata/}
}. 
We hope that the ideas in Online Appendix~\ref{ap:respondent-age} enable other researchers
to modify our approach for these settings.
Finally, we hope that the network survival method might help inspire
improvements in the sibling survival method, particularly in terms of
sensitivity analysis.

The scandal of invisibility means that almost two-thirds of deaths in the world
are not recorded in a vital registration system~\citep{abouzahr_civil_2015}.
The long-term solution to the scandal of invisibility is develop effective
vital registration systems in every country.
Unfortunately, there has been very little progress improving the systems in developing
countries over the past 15 years~\citep{mikkelsen_global_2015}.  
Other demographic quantities such as fertility and child mortality
were once as poorly understood as adult mortality is now. 
But today, even the
world's poorest countries have high-quality survey-based estimates of fertility
and child mortality rates thanks to the development of appropriate survey-based
methods and a massive, internationally-coordinated, infrastructure to deploy
those methods around the world.  
The same infrastructure could also be
harnessed to estimate adult mortality, and we believe that the network survival
method is a promising step in that direction.

\clearpage

\bibliography{rwanda_mortality}
\bibliographystyle{demography}

\newpage
\part*{Online Appendices}
\appendix

\setcounter{figure}{0} \renewcommand{\thefigure}{A.\arabic{figure}}
\numberwithin{figure}{section}
\setcounter{table}{0} \renewcommand{\thetable}{A.\arabic{table}}
\numberwithin{table}{section}
\setcounter{equation}{0} \renewcommand{\theequation}{A.\arabic{equation}}
\numberwithin{equation}{section}

\clearpage
\pagenumbering{arabic}
\renewcommand*{\thepage}{A\arabic{page}}

\newpage

\section{Estimating personal network size}
\label{ap:ns-kp}

The network survival estimator uses the personal networks of survey respondents
in demographic group $\alpha$ to estimate the visibility of deaths in
demographic group $\alpha$.  This approach requires a method for estimating the
average personal network size of survey respondents in demographic group
$\alpha$, $\bar{d}_{F_\alpha, F}$. 
In this appendix, we adapt an existing personal network size
estimator called the known population method \citep{killworth_social_1998}
so that it can be used to estimate $\bar{d}_{F_\alpha, F}$. 
Most of the contents of this appendix closely parallel the formal
analysis of the known population estimator in \citet[][Online Appendix
B]{feehan_generalizing_2016}.

Before presenting the first result, we first need to introduce some notation
for working with the groups of known size. 
Let $U$ be the entire population (e.g., all of Rwanda), and 
let $F$ be the frame population for the survey (e.g., Rwandan adults).
Suppose that we have several groups $A_1, A_2, \dots, A_J$ with $A_J \subset U$.
These groups are the known populations.
Imagine concatenating all of the people in populations $A_1, A_2, \dots, A_J$
together, repeating each individual once for each population she is in.
The result, which we call the \emph{probe alters} $\mathcal{A}$ is a multiset.
The size of $\mathcal{A}$ is $N_\mathcal{A} = \sum_j N_{A_j}$.
In our notation, we use $\mathcal{A}$ in subscripts like any other set;
for example, $y_{F_\alpha, \mathcal{A}}$ is the reported connections from
frame population members in group $\alpha$ ($F_\alpha$) to the probe alters
($\mathcal{A}$).


\stmt{result}{res:adapted-kp} {
Suppose we have a probability sample $s$ taken from the frame population with
known probabilities of inclusion $\pi_i$. Further, suppose we have a multiset of
probe alters $\mathcal{A}$ that have been chosen
so that two conditions hold:
    \begin{itemize}
        \item $y_{F_\alpha, \mathcal{A}} = d_{F_\alpha, \mathcal{A}}$ (reporting condition)
        \item $\bar{d}_{\mathcal{A}, F_\alpha} = \bar{d}_{F, F_\alpha}$ 
            (probe alter condition).
    \end{itemize}
Then the adapted known population estimator
\begin{align}
\label{eqn:adapted-kp}
\widehat{\bar{d}}_{F_\alpha, F} &= 
\frac{\sum_{i \in s_\alpha} y_{i, \mathcal{A}}/\pi_i}{\sum_j N_{A_j}}~
      \frac{N_F}{N_{F_\alpha}} 
\end{align}
\noindent is consistent and unbiased for $\bar{d}_{F_\alpha, F}$. 
}
\stmtproof{res:adapted-kp}{
	By Property~B.2 of \citet{feehan_generalizing_2016}, 
    $\widehat{y}_{F_\alpha, \mathcal{A}}/N_{\mathcal{A}}$ is consistent and unbiased for
    $y_{F_\alpha, \mathcal{A}}/N_{\mathcal{A}}$.
    By the reporting condition, 
    $y_{F_\alpha, \mathcal{A}}/N_{\mathcal{A}} = d_{F_\alpha, \mathcal{A}}/N_{\mathcal{A}}$.
    Re-writing this quantity, we have
    \begin{align}
        \frac{d_{F_\alpha, \mathcal{A}}}{N_{\mathcal{A}}} &=
        \frac{d_{\mathcal{A}, F_\alpha}}{N_{\mathcal{A}}} =
        \bar{d}_{\mathcal{A}, F_\alpha}.
    \end{align}
    Now, using the probe alter condition,
    \begin{align}
    \bar{d}_{\mathcal{A}, F_\alpha} = \bar{d}_{F, F_\alpha}.
    \end{align}
    So we have shown that, assuming the reporting condition and the
    probe alter condition hold, $\widehat{y}_{F_\alpha, \mathcal{A}}/N_\mathcal{A}$
    is consistent and unbiased for $\bar{d}_{F, F_\alpha}$.
    Now we can re-write $\bar{d}_{F, F_\alpha}$ as
    \begin{align}
        \bar{d}_{F, F_\alpha} &= \frac{d_{F, F_\alpha}}{N_F}
        = \frac{d_{F_\alpha, F}}{N_F}.
    \end{align}
    So we conclude that the estimator is consistent and unbiased for
    \begin{align}
        \frac{d_{F_\alpha, F}}{N_F}~\frac{N_F}{N_{F_\alpha}}
        &= \frac{d_{F_\alpha, F}}{N_{F_\alpha}} = \bar{d}_{F_\alpha, F}.
    \end{align}
}
\rptstmtonlyproof{res:adapted-kp}

\citet[][Online Appendix B]{feehan_generalizing_2016} offers suggestions for how to
choose probe alters for the known population estimator; these suggestions carry
over to the adapted estimator (Result~\ref{res:adapted-kp}) with some modifications
to accommodate the specific reporting condition and probe alter condition
required by the adapted known population estimator.

\section{The network survival estimator}
\label{ap:ns-estimator}

In this appendix, we provide formal results related to the network survival
estimator.  Several of the results in this appendix follow the analysis of the
generalized scale-up estimator found in \citet{feehan_generalizing_2016}.  

\subsection{Estimating the number of deaths, $D_\alpha$}

Equation~\ref{eqn:nontext-id} shows that the two components of the estimated number
of deaths are: (i) the total number of reports about deaths, $y_{F, D_\alpha}$;
and (ii) the average visibility of deaths, $\bar{v}_{D_\alpha, F}$. 
First, we present results about estimators for each of these two components.
Then we show that estimators for these two components can be combined to
estimate $M_\alpha$.

Result~\ref{res:y-f-oalpha}, shows that $y_{F, D_\alpha}$ can be
estimated from survey reports using standard survey techniques.

\stmt{result}{res:y-f-oalpha} {
Suppose we have a probability sample $s$ taken from the frame population with
known probabilities of inclusion $\pi_i$. Then
\begin{align}
\label{eqn:y-f-oalpha-hat}
\widehat{y}_{F, D_\alpha} &= \sum_{i \in s} y_{i, D_\alpha} / \pi_i
\end{align}
\noindent is consistent and unbiased for $y_{F, D_\alpha}$.
}
\stmtproof{res:o-alpha}{
    Equation~\ref{eqn:y-f-oalpha-hat} is a standard Horvitz-Thompson estimator
    \citep[see, eg][chap. 2]{sarndal_model_2003}, so
    it is consistent and unbiased for the total 
    $\sum_{i \in F} y_{i, D_\alpha} = y_{F, D_\alpha}$.
}
\rptstmtonlyproof{res:o-alpha}

Next, Result~\ref{res:vbar-oalpha-f} shows that it is possible to
use information about survey respondents' personal networks to estimate the
visibility of deaths ($\bar{v}_{F, D_\alpha}$) if two additional conditions are
satisfied: the visible deaths condition and the decedent network condition. 

\stmt{result}{res:vbar-oalpha-f} {
    Suppose that $\widehat{\bar{d}}_{F_\alpha, F}$ is a consistent and unbiased
    estimator for $\bar{d}_{F_\alpha, F}$ (such as the one in Result~\ref{res:adapted-kp}).
Furthermore, suppose that the following conditions hold:
    \begin{itemize}
    \item $\bar{v}_{D_\alpha, F} = \bar{d}_{D_\alpha, F}$ (visible deaths condition)
    \item $\bar{d}_{D_\alpha, F} = \bar{d}_{F_\alpha, F}$ (decedent network condition)
    \end{itemize}
    Then $\widehat{\bar{d}}_{F_\alpha, F}$ is a consistent and unbiased estimator for
    $\bar{v}_{D_\alpha, F}$.
}
\stmtproof{res:vbar-oalpha-f}{
    By assumption, $\widehat{\bar{d}}_{F_\alpha, F}$ is consistent and unbiased for
    $\bar{d}_{F_\alpha, F}$. By the decedent network condition,
    $\bar{d}_{F_\alpha, F} = \bar{d}_{D_\alpha, F}$. And, by the
    visible deaths condition, $\bar{d}_{D_\alpha, F} = \bar{v}_{D_\alpha, F}$.
}
\rptstmtonlyproof{res:vbar-oalpha-f}

The \emph{visible deaths condition} says that the average number of times a
death could be reported (the visibility of deaths) is the same as the
average number of network connections people who died have to the frame
population (i.e., $\bar{v}_{D_\alpha, F} = \bar{d}_{D_\alpha, F}$).
Substantively, we would expect this condition to hold when, on average, 
people who are connected to a person who died are aware of that fact and report it on a
survey. 

The \emph{decedent network condition} says that the average size of personal
networks is the same for dead people and for the people who respond to the
survey (i.e.,  $\bar{d}_{D_\alpha, F} = \bar{d}_{F_\alpha, F}$).
For example, suppose that women aged 50-54 who are eligible to be sampled
by our survey have an average personal network size of 100. 
In that case, the decedent network condition is satisfied when women aged 50-54
who died also have an average personal network size of 100.

The visible death condition and the decedent network condition could both
be violated in practice. 
Therefore, in Online Appendix~\ref{ap:ns-decomposition-framework} we develop a
sensitivity analysis framework that enables researchers to understand the impact that violations of 
these two assumptions will have on the accuracy of estimated death rates.

Next, Result~\ref{res:o-alpha} shows how the network survival
method combines Results~\ref{res:y-f-oalpha} and~\ref{res:vbar-oalpha-f}
to form an estimator for the number of deaths ($D_\alpha$). 


\stmt{result}{res:o-alpha} {
    Suppose $\widehat{y}_{F, D_\alpha}$ is a consistent and unbiased estimator for
    $y_{F, D_\alpha}$, and that $\widehat{\bar{v}}_{D_\alpha, F}$ is a consistent
    and unbiased estimator for $\bar{v}_{D_\alpha, F}$. 
    Suppose also that there are no false positive reports, so that
    $v_{i, F} = 0$ for all $i \notin D_\alpha$. Then
\begin{align}
\label{eqn:o-alpha}
\widehat{D}_\alpha &= \frac{\widehat{y}_{F, D_\alpha}}{\widehat{\bar{v}}_{D_\alpha, F}}
\end{align}
\noindent is consistent and essentially unbiased for $D_\alpha$.
}
\stmtproof{res:o-alpha}{
    With  consistent and unbiased estimators for $y_{F, D_\alpha}$ and
    for $\bar{v}_{D_\alpha, F}$, we can form a consistent and essentially unbiased
    estimator for $y_{F, D_\alpha} / \bar{v}_{D_\alpha, F}$ using a standard
    ratio approach~\citep[][chap. 5]{sarndal_model_2003}\footnote{
        Ratio estimator are standard in survey
        research, and a discussion of them can be found in many texts. 
        Ratio estimators are not, strictly speaking, unbiased. 
        However, there is a large literature that confirms that the bias in
        ratio estimators is typically very small when samples are not too small  
        (see, for example, \citet[][chap. 5]{sarndal_model_2003};
        \citet[][Online Appendix E]{feehan_generalizing_2016}; and
        \citet{rao_double_1968}).
        Since ratio estimators are technically biased, but the bias can be expected to
        be very small, we use by the term \emph{essentially unbiased} instead of
        unbiased in several of our results. 
    }. 
    So it remains to show that
    $y_{F, D_\alpha} / \bar{v}_{D_\alpha, F} = D_\alpha$.
    Since in-reports must equal out-reports (see~\citet{feehan_network_2015-1} and
    \citet{feehan_generalizing_2016}), 
    $y_{F, D_\alpha} = v_{U, F}$,
    where $U$ is the set of all of the people who could be reported about,
    living or dead (note that $D_\alpha \subset U$ and $F \subset U$).
    By the no false positives assumption, $v_{i, F} = 0$ 
    for all $i \notin D_\alpha$, which means that 
    \begin{align}
    v_{U, F} = \sum_{i \in U} v_{i, F} = \sum_{i \in D_\alpha} v_{i, F} = v_{D_\alpha, F}.
    \end{align}
    So we conclude that $y_{F, D_\alpha} = v_{D_\alpha, F}$.
    Dividing both sides of this identity by $D_\alpha$ and re-arranging produces
    \begin{align}
        D_\alpha &= \frac{y_{F, D_\alpha}}{v_{D_\alpha, F} / D_\alpha}
        = \frac{y_{F, D_\alpha}}{\bar{v}_{D_\alpha, F}}.
    \end{align}
}
\rptstmtonlyproof{res:o-alpha}

\subsection{Estimator for $M_\alpha$}
\label{ap:combined-estimator}

We now turn to a set of results related to estimating the death rate
$M_\alpha$\footnote{%
Note that, as is typical in demographic research, we use the size of the population
to approximate the exposure in the denominator of the death rate.
This approximation should not be problematic unless (i) the time period
over which death rates are computed is long; or (ii) death rates are extremely
high (much higher than populations typically experience). For the 12-month
death rates we study in Rwanda, we do not expect this approximation to pose
a problem.
}. 
We begin by developing a general expression that can be used to estimate
the death rate $M_\alpha$.
Then we discuss, in detail, the way that we used the general expression to
estimate death rates in our study. 

We begin with a general result.

\stmt{result}{res:m-alpha-complete} {
Suppose we have a probability sample $s$ taken from the frame population with
known probabilities of inclusion $\pi_i$. Suppose also that we have a consistent
and unbiased estimator $\widehat{y}_{F, D_\alpha}$ (eg, Result~\ref{res:y-f-oalpha});
a consistent and unbiased estimator
$\widehat{\bar{v}}_{D_\alpha, F}$ (eg, Result~\ref{res:vbar-oalpha-f});
and a consistent and unbiased estimator $\widehat{N}_\alpha$.
Then 
\begin{align}
\label{eqn:m-alpha-complete}
\widehat{M}_\alpha &= 
\frac{\widehat{y}_{F, D_\alpha}}{\widehat{\bar{v}}_{D_\alpha, F}}
\frac{1}{\widehat{N}_\alpha}
\end{align}
\noindent is consistent and essentially unbiased for $M_\alpha = D_\alpha / N_{\alpha}$.
}
\stmtproof{res:m-alpha-complete}{
    Equation~\ref{eqn:m-alpha-complete} is a compound ratio estimator;
    \citet{rao_double_1968} and \citet[][Online Appendix E]{feehan_generalizing_2016}
    give proofs that compound ratio estimators are consistent and essentially unbiased.
}
\rptstmtonlyproof{res:m-alpha-complete}

Result~\ref{res:m-alpha-complete} is very general in the sense that it can be
used to estimate death rates by combining any consistent and unbiased
estimators of connections to people who died, the visibility of deaths, and the
size of the population.  For our study, we customized this general estimator in
two ways.  First, we used the adapted known population estimator for
$\bar{d}_{F_\alpha, F}$ (Result~\ref{res:adapted-kp}) as an estimator of the
visibility of deaths ($\bar{v}_{D_\alpha, F}$).  Second, we assumed that the
sampling frame was complete $(N_{F_{\alpha}} = N_{\alpha} \mbox{ for all }
\alpha)$\footnote{%
	In our study, we believe that it is reasonable to assume that the sampling frame
was complete (i.e., that all adults could have been selected) because of our
field procedures.  More specifically, our approach was to (1) randomly sample a set of
geographical areas; (2) send a team to visit the geographical areas and produce
a census of dwellings; and then (3) choose a sample of dwellings and interview
all adults who lived in them.  See
\citet{rwanda_biomedical_center/institute_of_hiv/aids_estimating_2012} for more
information about the sampling design.  Researchers concerned about either of
these choices can use the sensitivity framework in Online Appendix
\ref{ap:ns-decomposition-framework} to assess the sensitivity of the estimated
death rates to this assumption.} These two choices lead to a more specific
estimator that we used in this study.

\stmt{result}{res:m-alpha-reduced} {
Suppose we have a probability sample $s$ taken from the frame population with
known probabilities of inclusion $\pi_i$. 
Suppose that we have a set of probe alters $\mathcal{A}$ (also called known populations)
that satisfy the reporting condition 
($y_{F_\alpha, \mathcal{A}} = d_{F_\alpha, \mathcal{A}}$)
and the probe alter condition
($\bar{d}_{\mathcal{A}, F_{\alpha}} = \bar{d}_{F, F_\alpha}$)
from Result~\ref{res:adapted-kp}.
Suppose that the visible deaths condition
($\bar{v}_{D_\alpha, F} = \bar{d}_{D_\alpha, F}$)
and the decedent network condition
($\bar{d}_{D_\alpha, F} = \bar{d}_{F_\alpha, F}$)
from Result~\ref{res:vbar-oalpha-f} are satisfied.
Finally, suppose that the frame population is complete,
($N_{F_\alpha} = N_\alpha$), and that there are no
false positive reports about deaths 
($v_{i, F} = 0$ for all $i \notin D_\alpha$).
Then
\begin{align}
\label{eqn:m-alpha-reduced}
\widehat{M}_\alpha &= 
\frac{\sum_{i \in s} y_{i, D_\alpha} / \pi_i}
     {\sum_{i \in s_\alpha} y_{i, \mathcal{A}}/ \pi_i}~
\frac{N_{\mathcal{A}}}
     {N_F}
=
\frac{\widehat{y}_{F, D_\alpha}}{\widehat{y}_{F_\alpha, \mathcal{A}}}~
\frac{N_{\mathcal{A}}}{N_F}
=
\frac{\widehat{y}_{F, D_\alpha}}{\widehat{\bar{d}}_{F_\alpha,F} \times N_{F_\alpha}}
\end{align}
\noindent is consistent and essentially unbiased for $M_\alpha = D_\alpha / N_{\alpha}$.
}
\stmtproof{res:m-alpha-reduced}{
    First, note that
    \begin{align}
        \frac{\widehat{y}_{F, D_\mathcal{A}}}
             {\widehat{\bar{d}}_{F_\alpha, F} \times N_{F_\alpha}} &=
        \frac{\widehat{y}_{F, D_\alpha}}
             {\widehat{y}_{F_\alpha, \mathcal{A}}} 
        \frac{N_\mathcal{A}}{N_{F_\alpha}}
        \frac{N_{F_\alpha}}{N_F} \\
        &= \frac{\widehat{y}_{F, D_\alpha}}{\widehat{y}_{F_\alpha, \mathcal{A}}}
        \frac{N_\mathcal{A}}{N_F},
        \label{eqn:temp-m-alpha-reduced}
    \end{align}
    where we have plugged in the definition of the adapted known
    population estimator and cancelled the $N_{F_\alpha}$ (Result~\ref{res:adapted-kp}).

    Equation~\ref{eqn:temp-m-alpha-reduced} is a standard ratio estimator,
    so it is consistent and essentially unbiased for the quantity
    \begin{align}
        Q_\alpha = \frac{y_{F, D_\alpha}}{y_{F_\alpha, \mathcal{A}}}~
        \frac{N_\mathcal{A}}{N_F}
    \end{align}
    \citep[see, e.g.][chap. 5]{sarndal_model_2003}. So it remains to show
    that $Q_\alpha = D_\alpha / N_\alpha = M_\alpha$.
    We will do this by working backwards through the discussion above. 
    First, multiply $Q_\alpha$ by $N_{F_\alpha}/N_\alpha$ (which equals 1, by
    the completeness of the frame population), to obtain
    \begin{align}
        Q_\alpha
        = \frac{y_{F, D_\alpha}}{y_{F_\alpha, \mathcal{A}}}~
          \frac{N_\mathcal{A}}{N_F}~
          \frac{N_{F_\alpha}}{N_\alpha}.
    \end{align}
    Now we can use the reporting condition 
    ($y_{F_\alpha, \mathcal{A}} = d_{F_\alpha, \mathcal{A}}$) 
    followed by the probe alter condition 
    ($\bar{d}_{\mathcal{A}, F_\alpha} = \bar{d}_{F, F_\alpha}$)
    to rewrite the expression as
    \begin{align}
        \label{eqn:qalpha-interim}
        Q_\alpha 
        = \frac{y_{F, D_\alpha}}{\bar{d}_{F, F_\alpha}}~
          \frac{1}{N_F}~
          \frac{N_{F_\alpha}}{N_\alpha}.
    \end{align}
    Now, recall that 
    $\bar{d}_{F, F_\alpha}~N_F / N_{F_\alpha} = \bar{d}_{F_\alpha, F}$.
    Applying this relationship to simplify the denominator of
    Eq.~\ref{eqn:qalpha-interim} produces 
    \begin{align}
        Q_\alpha 
        = \frac{y_{F, D_\alpha}}{\bar{d}_{F_\alpha, F}}~
          \frac{1}{N_\alpha}.
    \end{align}
    Finally, applying the decedent network condition 
    ($\bar{d}_{F_\alpha, F} = \bar{d}_{D_\alpha, F}$)
    and the visible deaths condition
    ($\bar{d}_{D_\alpha, F} = \bar{v}_{D_\alpha, F}$),
    we have
    \begin{align}
        Q_\alpha 
        = \frac{y_{F, D_\alpha}}{\bar{v}_{D_\alpha, F}}~
          \frac{1}{N_\alpha}.
    \end{align}
    Now, since there are no false positive reports, we can apply the
    argument in Result~\ref{res:o-alpha} to conclude that
    $y_{F, D_\alpha} / \bar{v}_{D_\alpha, F} = D_\alpha$.
    Therefore,
    \begin{align}
        Q_\alpha 
        = \frac{D_\alpha}{N_\alpha}
        = M_\alpha.
    \end{align}
}
\rptstmtonlyproof{res:m-alpha-reduced}

\section{Sensitivity framework}
\label{ap:ns-decomposition-framework}

The network survival estimator we used in Rwanda relies on several conditions
(Result~\ref{res:m-alpha-reduced}), and
these conditions can be separated into four groups:
(i) reporting (for example, the visible deaths condition);
(ii) network structure (the decedent network connection);
(iii) survey construction (for example, choosing the probe alters for the
adapted known population method);
and (iv) sampling (the requirement that researchers obtain a probability sample).
In practice, we expect that researchers may not be sure that all of the
conditions required by the network survival estimator are exactly satisfied.
Therefore, in this appendix we develop a framework that researchers can use to
quantitatively assess how violating each condition impacts estimated death
rates.
Our framework also identifies precise and well-defined quantities that
future studies may be able to measure. With measurements for these quantities,
network survival estimates could be adjusted and potentially improved%
\footnote{%
Note that this framework is an adapted version of the one introduced for the
scale-up estimator in \citet{feehan_generalizing_2016}, and rigorous proofs for
our sensitivity results can be found there.
Moreover, to keep our derivations as simple as possible, our focus
here will be on the specific estimator we used in Rwanda
(Result~\ref{res:m-alpha-reduced}); 
however, by following the approach in this appendix, researchers can extend our
approach to the more general estimator in Result~\ref{res:m-alpha-complete} as
well. 
}.

In the next section, we focus on the impact of nonsampling errors.
Then, we turn to an analysis of the impact of sampling errors. 
Finally, we combine the results into a unified sensitivity framework for
network survival estimates. 

\subsection{Network survival sensitivity to nonsampling errors}

To understand how different sources of nonsampling error affect
network survival estimates, we will briefly review the network reporting
framework; see \citet{feehan_network_2015-1} and \citet{feehan_generalizing_2016} 
for more detail.
Figure~\ref{fig:reporting-network-panel2} shows an example of a reporting network
that has been rearranged into a
\emph{bipartite reporting graph}. 
The edges in this bipartite reporting graph represent the reports that people
in the frame population make about people who died.
The edges contribute two types of quantities to the vertices in the graph: each
edge adds an \emph{out-report} to the people who do the reporting ($F$, on the left-hand side
of the graph); and each edge also adds an \emph{in-report} to the people who get
reported about ($U$, on the right-hand side of the graph).
We call the sum of all of the out-reports $y_{F, D_\alpha}$, and the sum of all
of the in-reports $v_{U, F}$.

Out-reports can be separated into two groups: (i), \emph{true
positives}, which are reports that correctly lead to people who died; and
(ii) \emph{false positives}, which are reports that incorrectly lead to
people who did not die. We write the true positives as $y_{F, D_\alpha}^{+}$,
and the false positives as $y_{F, D_\alpha}^{-}$.  By definition, all of the
true positive reports lead to $D_\alpha$, meaning that $y_{F, D_\alpha}^{+} =
v_{D_\alpha, F}$. This identity is true in \emph{any} bipartite reporting graph, no
matter how accurate or inaccurate respondents' reports are. 
Starting from $y_{F, D_\alpha}^{+} = v_{D_\alpha, F}$, multiplying both
sides by $D_\alpha$, and then rearranging the terms yields an identity that
is the basis for the network survival estimator:
\begin{align}
    \label{eqn:oalpha-general-id}
    D_\alpha &= \frac{y_{F, D_\alpha}^{+}}{\bar{v}_{D_\alpha, F}}.
\end{align}

Now we will use the network reporting framework to develop an expression for
the sensitivity of network survival estimates for $M_\alpha$, the death rate.
Our approach will be to introduce quantities that capture the extent to which
each required condition is satisfied. We call these quantities \emph{adjustment factors}.

First, we focus on an expression for the sensitivity of the estimator for
$D_\alpha$, the number of deaths.
Estimating the number of deaths requires that three conditions are satisfied: two reporting conditions and one condition related to network structure.
The first condition required to estimate the number of deaths is 
that there are no false positive reports.
To account for this requirement, we introduce a quantity called the \emph{precision}:
\begin{align}
    \eta_{F, \alpha} &=
    \frac{\text{total \# of out-reports from frame popn that correctly lead to deaths}}
         {\text{total \# of out-reports from frame popn}} =
    \frac{y_{F, D_\alpha}^{+}}
         {y_{F, D_\alpha}}.
\end{align}
$\eta_{F, \alpha}$ relates accurate network reports to all network reports; it will range
from 1, when reporting is perfectly accurate, to 0, when none of the out-reports correctly
leads to a death.
Values of $\eta_{F, \alpha}$ other than 1 mean that the no false positives assumption
is violated.

The second condition required to estimate the number of deaths is the
visible deaths condition.
To account for this requirement, we introduce a quantity called the
\emph{true positive rate}:
\begin{align}
    \tau_{F, \alpha} &=
    \frac{\text{avg \# of in-reports from the frame to each death}}
         {\text{avg \# of network connections from a death to the frame population}} =
    \frac{\bar{v}_{D_\alpha, F}}
         {\bar{d}_{D_\alpha, F}}.
\end{align}
$\tau_{F, \alpha}$ relates network degree to network reports; it will range from 1, when reporting is perfectly accurate, to 0, when no network edges leading to deaths are reported.
Values of $\tau_{F, \alpha}$ other than 1 mean that the visible deaths condition is
violated.

The third condition required to estimate the number of deaths is the 
decedent network condition.
To account for this requirement, we introduce a quantity called the
\emph{degree ratio}:
\begin{align}
    \delta_{F, \alpha} &=
    \frac{\text{avg \# edges from a death in $\alpha$ to the frame population}}
         {\text{avg \# edges from a frame pop member in $\alpha$ to the entire frame pop}} =
    \frac{\bar{d}_{D_\alpha, F}}
         {\bar{d}_{F_\alpha, F}}.
\end{align}
$\delta_{F, \alpha}$ will range from 0 to infinity. When it is less than one, people who
die in demographic group $\alpha$ tend to have fewer connections to the frame population than
frame population members in demographic group $\alpha$; when it is greater than one,
people who die in demographic group $\alpha$ tend to have more connections to the frame
population than frame population members in demographic group $\alpha$. Values of
$\delta_{F, \alpha}$ other than 1 mean that the decedent network condition is
violated.

Together, the adjustment factors can be used to propose a decomposition of the
difference between network survival estimand for $D_\alpha$ and the true value
of $D_\alpha$:

\begin{align}
\label{eqn:ns-adjustmentfactors-text}
D_\alpha &= 
\underbrace{%
\left( \frac{y_{F,D_\alpha}}{\bar{d}_{F_\alpha,F}} \right)
	}_{\substack{\text{network} \\ \text{survival} \\ \text{estimand}}}
\times 
\underbrace{%
    \underbrace{\frac{1}{\bar{d}_{D_\alpha,F}/{\bar{d}_{F_\alpha,F}}}}_{\substack{\text{degree ratio} \\ \delta_{F,\alpha}}} \times 
    \underbrace{\frac{1}{\bar{v}_{D_\alpha,F}/{\bar{d}_{D_\alpha,F}}}}_{\substack{\text{true positive rate} \\ \tau_{F,\alpha} }} \times
    \underbrace{\frac{y^{+}_{F, D_\alpha}}{y_{F, D_\alpha}}.}_{\substack{\text{precision} \\ \eta_{F,\alpha} }} 
}_{\mbox{adjustment factors}}
\end{align}

The decomposition in Eq.~\ref{eqn:ns-adjustmentfactors-text}, shows that
the network survival estimand will estimate the true number of deaths if the
three adjustment factors satisfy 
$\eta_{F, \alpha} / (\delta_{F, \alpha} \times \tau_{F,\alpha}) = 1$. 

\subsubsection{Sensitivity of the adapted known population estimator}

We now analyze the sensitivity of the adapted known population estimator
(Result~\ref{res:adapted-kp}) to nonsampling conditions.
The adapted known population estimator is used to estimate the size of survey
respondents' personal networks; it requires three nonsampling conditions:
first, that researchers have accurate information about the size of the known populations ($N_\mathcal{A}$);
second, the probe alter condition ($\bar{d}_{\mathcal{A}, F_\alpha} = \bar{d}_{F, F_\alpha}$);
and third, the reporting condition ($y_{F_\alpha, \mathcal{A}} = d_{F_\alpha, \mathcal{A}}$).
Following the strategy above, 
we introduce a quantitative adjustment factor to capture the extent to which each of these
three conditions is satisfied.
For example, suppose that in a particular study, the reporting condition is not
satisfied, so that $y_{F_\alpha, \mathcal{A}} \neq d_{F_\alpha, \mathcal{A}}$;
in that case, we can write $y_{F_\alpha, \mathcal{A}} = c d_{F_\alpha, \mathcal{A}}$ 
for some constant $c$; when $c=1$, the condition is satisfied. 
The corresponding adjustment factor is then 
$c = \frac{y_{F_\alpha, \mathcal{A}}}{d_{F_\alpha, \mathcal{A}}}$.

By introducing an adjustment factor for each of the three assumptions---
$c_1=\frac{\widehat{N}_\mathcal{A}}{N_\mathcal{A}}$
for the known population totals, 
$c_2=\frac{\bar{d}_{\mathcal{A},F_\alpha}}{\bar{d}_{F, F_\alpha}}$ for the probe alter condition, and 
$c_3=\frac{y_{F_\alpha, \mathcal{A}}}{d_{F_\alpha, \mathcal{A}}}$
for the reporting conditions---the adapted known population estimator
can be decomposed as:
\begin{align}
\label{eqn:adapted-kp-adjustmentfactors}
\bar{d}_{F_\alpha, F} &= 
\underbrace{%
    \left( \frac{\widehat{y}_{F_\alpha,\mathcal{A}}}{\widehat{N}_{\mathcal{A}}} 
    \frac{N_F}{N_{F_\alpha}} 
    \right)
}_{\substack{\text{adapted} \\ \text{known population}}}
\times 
\underbrace{c_1}_{\substack{\text{known} \\ \text{population} \\ \text{totals} }} 
\times 
\underbrace{\frac{1}{c_2}}_{\substack{\text{probe} \\ \text{alter} \\ \text{condition}}}
\times 
\underbrace{\frac{1}{c_3}}_{\substack{\text{reporting} \\ \text{conditions} \\ \text{for known} \\ \text{populations}}}.
\end{align}

\subsubsection{Sensitivity to nonsampling conditions}

We have now developed expressions that illustrate the sensitivity of estimands for 
$y_{F, D_\alpha}$, $D_\alpha$, and $\bar{d}_{F_\alpha, F}$.
The final condition required by the estimator we used in Rwanda (Result~\ref{res:m-alpha-reduced}) is that the
frame population be complete, meaning that $N_{F_\alpha} = N_\alpha$. 
Following the approach in the previous sections, we account for this condition by
introducing the adjustment factor 
$c_4 = \frac{N_{F_\alpha}}{N_\alpha}$. 
With this final adjustment factor, we can combine our analysis of all of the nonsampling
factors to produce

\begin{align}
\label{eqn:m-alpha-reduced-adjustmentfactors}
M_\alpha &= 
\underbrace{%
    \left( 
           \frac{y_{F, D_\alpha}}{y_{F_\alpha,\mathcal{A}}}
           \times
           N_{\mathcal{A}} 
    \right)
}_{\substack{\text{network} \\ \text{survival} \\ \text{estimand} \\ \text{(Result~\ref{res:m-alpha-reduced})}}}
\times 
\underbrace{
    \frac{c_2~c_3}{c_1}
}_{\substack{\text{adapted} \\ \text{known} \\ \text{population} \\ \text{conditions} }} 
\times 
\underbrace{
    \frac{1}{c_4}
}_{\substack{\text{frame} \\ \text{population} \\ \text{is complete}}}
\times 
\underbrace{
    \frac{\eta_{F, \alpha}}{\tau_{F,\alpha} \delta_{F,\alpha}}.
}_{\substack{\text{reporting and} \\ \text{network} \\ \text{structure}}}
\end{align}

To assess the sensitivity of death rate estimates to any of the
nonsampling conditions required by network survival, researchers can
(1) assume values for $c_1$, $c_2$, $c_3$, $c_4$, $\eta_{F, \alpha}$,
$\tau_{F, \alpha}$, and $\delta_{F, \alpha}$ that describe how the conditions
are not satisfied; 
and (2) plug these values into
Equation~\ref{eqn:m-alpha-reduced-adjustmentfactors} to obtain the resulting
death rate.

\subsection{Sensitivity to sampling conditions}

The last type of condition required by the network survival estimator is
that researchers have obtained a probability sample and the associated sampling
weights.

We begin by repeating \citet{feehan_generalizing_2016}'s definition of 
\emph{imperfect sampling weights}, since this concept is critical to
understanding the network survival estimator's sensitivity to sampling error.

\paragraph{Imperfect sampling weights.}
Suppose a researcher obtains a probability sample $s_F$ from the frame population $F$ 
\citep{sarndal_model_2003}.
Let $I_i$ be the random variable that assumes the value $1$ when unit $i \in F$
is included in the sample $s_F$, and $0$ otherwise.
Let $\pi_i = \E[I_i]$ be the true probability of inclusion for unit $i \in F$,
and let $w_i = \frac{1}{\pi_i}$ be the corresponding design weight for
unit $i$.
We say that researchers have \emph{imperfect sampling weights} when
researchers use imperfect estimates of the inclusion probabilities $\piprime_i$
and the corresponding design weights
$\wprime_i = \frac{1}{\piprime_i}$.
Note that we assume that both the true and the imperfect weights satisfy 
$\pi_i > 0$ and $\piprime_i > 0$ for all $i$.

\citet[][Result D.10]{feehan_generalizing_2016} introduces two more quantities
that we will use here. The first quantity, called $\wdiff_i$,
captures the relative error in the imperfect sampling weights for each unit $i$
in the frame population.
It is defined as $\wdiff_i = \frac{\pi_i}{\piprime_i}$.
The second quantity is an index, called $K$, that depends on the quantity being
estimated, as well as on the magnitude of problems with the imperfect sampling weights.
For example, in the case of estimating $y_{F, D_\alpha}$ from imperfect weights,
$K$ is defined as
$K = \cv(\wdiff_i)~\cv(y_{i, D_\alpha})~\correl(\wdiff_i, y_{i, D_\alpha})$,
where $\cv(\cdot)$ is the coefficient of variation (the standard deviation divided 
by the mean), and $\correl(\cdot, \cdot)$ is the correlation coefficient.
$K$ will tend to be large in magnitude when the imperfections in weights
have a lot of variance ($\cv(\wdiff_i)$ is large),
when the quantity being estimated has large variance ($\cv(y_{i, D_\alpha})$ is large),
and when there is a strong relationship between the $\wdiff_i$ and the
quantity being estimated ($\correl(\wdiff_i, y_{i, D_\alpha})$).
When the imperfect weights are exactly correct, $K=0$.

The argument from \citet[][Result D.10]{feehan_generalizing_2016} can now be
used to show that

\begin{align}
\label{eqn:m-alpha-reduced-sensitivity}
\underbrace{
\widehat{M}_\alpha 
}_{
    \substack{
        \text{network} \\ \text{survival} \\ \text{estimator}
    }
}
\leadsto
\underbrace{%
    M_\alpha
}_{\substack{\text{true} \\ \text{death} \\ \text{rate}}}
\times 
\underbrace{
    \frac{c_1}{c_2~c_3}
}_{\substack{\text{adapted} \\ \text{known} \\ \text{population} \\ \text{conditions} }} 
\times 
\underbrace{
    c_4
}_{\substack{\text{frame} \\ \text{population} \\ \text{is complete}}}
\times 
\underbrace{
    \frac{\tau_{F,\alpha} \delta_{F,\alpha}}{\eta_{F, \alpha}}
}_{\substack{\text{reporting and} \\ \text{network} \\ \text{structure}}}
\times 
\underbrace{
    \frac{(1 + K_{F_1})}{(1 + K_{F_2})}
}_{\substack{\text{sampling} \\ \text{conditions}}},
\end{align}
\noindent where 
$\leadsto$ means `is consistent and essentially unbiased for',
$K_{F_1} = \cv(\wdiff_i) \cv(y_{i, D_\alpha}) \correl(\wdiff_i, y_{i, D_\alpha})$ 
is the imperfect sampling index for $y_{F, D_\alpha}$, and 
$K_{F_2} = \cv(\wdiff_i) \cv(y_{i, \mathcal{A}}) \correl(\wdiff_i, y_{i, \mathcal{A}})$ 
is the imperfect sampling index for $y_{F, \mathcal{A}}$.

Researchers who wish to assess how death rates estimated using network survival 
would be impacted by violations of any of the conditions required by the estimator
can use Eq.~\ref{eqn:m-alpha-reduced-sensitivity} to perform a sensitivity analysis by
(i) assuming values or a range of values for $c_1$, $c_2$, $c_3$, $c_4$,
$\tau_{F, \alpha}$, $\delta_{F, \alpha}$, $\eta_{F, \alpha}$, $K_{F_1}$, and
$K_{F_2}$; and then
(ii) using Eq.~\ref{eqn:m-alpha-reduced-sensitivity} to determine the resulting
values of $M_\alpha$.

\paragraph{Worked example.}
For example, in order to create the lower-left panel of Figure~\ref{fig:net-asdr-robustness},
we set 
$\delta_{F, \alpha} = 0.5$ and $\eta_{F,\alpha} / \tau_{F,\alpha} = 1.5$ in
Equation~\ref{eqn:m-alpha-reduced-sensitivity}. All of the other terms are set to
$\frac{c_1}{c_2 c_3} = 1$, $c_4 = 1$, and $\frac{(1 + K_{F_1})}{(1 + K_{F_2})} = 1$.
Rearranging Equation~\ref{eqn:m-alpha-reduced-sensitivity}, we find that in this situation,
the expression
\begin{align}
    \widehat{M}_\alpha \frac{\eta_{F, \alpha}}{\tau_{F, \alpha} \delta_{F, \alpha}} &\leadsto M_\alpha
\end{align}
will be consistent and essentially unbiased for the true death rate $M_\alpha$.
So we multiply the network survival estimates by 
$\frac{\eta_{F, \alpha}}{\tau_{F, \alpha} \delta_{F, \alpha}} = \frac{1.5}{0.5} = 3$.

\section{Tabular versions of results}
\label{ap:tabular}

This appendix provides tabular versions of 
Figure~\ref{fig:net-asdr-all} (in Table~\ref{tab:asdr-ests}), 
Figure~\ref{fig:all-asdr-age-comparisons} 
(in Table~\ref{tab:acqsibasdrcomp} and Table~\ref{tab:mealsibasdrcomp}), 
Figure~\ref{fig:deaths-per-int} (in Table~\ref{tab:deathperint}), and 
Figure~\ref{fig:all-45q15s} (in Table~\ref{tab:ffqf-all}).

\begin{table}[!htbp]  
  \caption{Estimated age-specific death rates using the acquaintance and meal tie definitions from the network survival study, and using the sibling history module of the DHS survey. Estimates are deaths rates per 1,000 person-years.} 
  \label{tab:asdr-ests} 
\small 
\begin{tabular}{@{\extracolsep{5pt}} lccll} 
\\[-1.8ex]\hline \\[-1.8ex] 
Tie definition & Sex & Age group & Estimate & 95\% CI \\ 
\hline \\[-1.8ex] 
Acquaintance (2010-11; n=2,236) & Female & [15,25) & $3.19$ & [2.12, 4.37] \\ 
Acquaintance (2010-11; n=2,236) & Female & [25,35) & $2.97$ & [2.25, 3.82] \\ 
Acquaintance (2010-11; n=2,236) & Female & [35,45) & $3.58$ & [2.43, 5.06] \\ 
Acquaintance (2010-11; n=2,236) & Female & [45,55) & $5.82$ & [4.04, 8.07] \\ 
Acquaintance (2010-11; n=2,236) & Female & [55,65) & $13.40$ & [9.30, 18.80] \\ 
Acquaintance (2010-11; n=2,236) & Male & [15,25) & $3.96$ & [2.75, 5.59] \\ 
Acquaintance (2010-11; n=2,236) & Male & [25,35) & $3.48$ & [2.58, 4.58] \\ 
Acquaintance (2010-11; n=2,236) & Male & [35,45) & $7.97$ & [5.81, 10.54] \\ 
Acquaintance (2010-11; n=2,236) & Male & [45,55) & $9.72$ & [7.17, 13.05] \\ 
Acquaintance (2010-11; n=2,236) & Male & [55,65) & $20.69$ & [13.67, 31.73] \\ 
Meal (2010-11; n=2,433) & Female & [15,25) & $5.71$ & [3.65, 7.93] \\ 
Meal (2010-11; n=2,433) & Female & [25,35) & $4.08$ & [3.07, 5.28] \\ 
Meal (2010-11; n=2,433) & Female & [35,45) & $6.15$ & [3.53, 9.48] \\ 
Meal (2010-11; n=2,433) & Female & [45,55) & $8.03$ & [4.85, 12.42] \\ 
Meal (2010-11; n=2,433) & Female & [55,65) & $10.04$ & [6.48, 14.82] \\ 
Meal (2010-11; n=2,433) & Male & [15,25) & $4.30$ & [3.03, 5.80] \\ 
Meal (2010-11; n=2,433) & Male & [25,35) & $4.57$ & [3.27, 6.12] \\ 
Meal (2010-11; n=2,433) & Male & [35,45) & $7.48$ & [5.46, 9.79] \\ 
Meal (2010-11; n=2,433) & Male & [45,55) & $9.05$ & [5.37, 14.22] \\ 
Meal (2010-11; n=2,433) & Male & [55,65) & $15.40$ & [10.35, 22.93] \\ 
Sibling (2004-10; n=13,671) & Female & [15,25) & $1.78$ & [1.41, 2.18] \\ 
Sibling (2004-10; n=13,671) & Female & [25,35) & $3.61$ & [3.04, 4.18] \\ 
Sibling (2004-10; n=13,671) & Female & [35,45) & $5.73$ & [4.89, 6.67] \\ 
Sibling (2004-10; n=13,671) & Female & [45,55) & $4.63$ & [3.47, 5.88] \\ 
Sibling (2004-10; n=13,671) & Female & [55,65) & $10.62$ & [6.03, 16.03] \\ 
Sibling (2004-10; n=13,671) & Male & [15,25) & $2.18$ & [1.79, 2.58] \\ 
Sibling (2004-10; n=13,671) & Male & [25,35) & $3.58$ & [2.99, 4.20] \\ 
Sibling (2004-10; n=13,671) & Male & [35,45) & $6.41$ & [5.44, 7.45] \\ 
Sibling (2004-10; n=13,671) & Male & [45,55) & $9.23$ & [7.22, 11.39] \\ 
Sibling (2004-10; n=13,671) & Male & [55,65) & $19.60$ & [12.04, 28.19] \\ 
\hline \\[-1.8ex] 
\end{tabular} 
\end{table}

\begin{table}[!htbp] \centering 
  \caption{Comparison between the estimated sampling distribution of the log age-specific death rate (log deaths per person-year) for the acqaintance network and for the sibling histories.} 
  \label{tab:acqsibasdrcomp} 
\small 
\begin{tabular}{@{\extracolsep{5pt}} cccc} 
\\[-1.8ex]\hline \\[-1.8ex] 
Sex & Age group & Mean difference in log(asdr estimate) & 95\% CI \\ 
\hline \\[-1.8ex] 
Female & [15,25) & $0.001$ & [ 0.000, 0.003] \\ 
Female & [25,35) & $$-$0.001$ & [-0.002, 0.000] \\ 
Female & [35,45) & $$-$0.002$ & [-0.004, 0.000] \\ 
Female & [45,55) & $0.001$ & [-0.001, 0.004] \\ 
Female & [55,65) & $0.003$ & [-0.004, 0.010] \\ 
Male & [15,25) & $0.002$ & [ 0.001, 0.003] \\ 
Male & [25,35) & $$-$0.0001$ & [-0.001, 0.001] \\ 
Male & [35,45) & $0.002$ & [-0.001, 0.004] \\ 
Male & [45,55) & $0.0005$ & [-0.003, 0.004] \\ 
Male & [55,65) & $0.001$ & [-0.011, 0.014] \\ 
\hline \\[-1.8ex] 
\end{tabular} 
\end{table}

\begin{table}[!htbp] \centering 
  \caption{Comparison between the estimated sampling distribution of the log age-specific death rate (log deaths per person-year) for the meal network and for the sibling histories.} 
  \label{tab:mealsibasdrcomp} 
\small 
\begin{tabular}{@{\extracolsep{5pt}} cccc} 
\\[-1.8ex]\hline \\[-1.8ex] 
Sex & Age group & Mean difference in log(asdr estimate) & 95\% CI \\ 
\hline \\[-1.8ex] 
Female & [15,25) & $0.004$ & [ 0.002, 0.006] \\ 
Female & [25,35) & $0.0005$ & [-0.001, 0.002] \\ 
Female & [35,45) & $0.0004$ & [-0.002, 0.004] \\ 
Female & [45,55) & $0.003$ & [ 0.000, 0.008] \\ 
Female & [55,65) & $$-$0.001$ & [-0.007, 0.006] \\ 
Male & [15,25) & $0.002$ & [ 0.001, 0.004] \\ 
Male & [25,35) & $0.001$ & [ 0.000, 0.003] \\ 
Male & [35,45) & $0.001$ & [-0.001, 0.004] \\ 
Male & [45,55) & $$-$0.0002$ & [-0.004, 0.005] \\ 
Male & [55,65) & $$-$0.004$ & [-0.014, 0.006] \\ 
\hline \\[-1.8ex] 
\end{tabular} 
\end{table}

\begin{table}[!htbp]  
  \caption{Average number of deaths reported from each interview in Rwanda using the acquaintance and meal tie definitions from the network survival study and the sibling history module of the DHS.} 
  \label{tab:deathperint} 
\small 
\begin{tabular}{@{\extracolsep{5pt}} lccc} 
\\[-1.8ex]\hline \\[-1.8ex] 
Tie definition & Num. Reported deaths & Num. Interviews & Deaths / Interview \\ 
\hline \\[-1.8ex] 
Acquaintance & $1,681$ & $2,259$ & $0.74$ \\ 
Meal & $932$ & $2,404$ & $0.39$ \\ 
Sibling
(12 months) & $124$ & $13,671$ & $0.01$ \\ 
Sibling
(84 months) & $1,197$ & $13,671$ & $0.09$ \\ 
\hline \\[-1.8ex] 
\end{tabular} 
\end{table}

\begin{table}[!htbp] \centering 
  \caption{Estimated $\ffqf$ values, by tie definition and sex. The survey-based estimates have 95\% confidence intervals, which come from the estimated sampling distribution of each estimator.} 
  \label{tab:ffqf-all} 
\small 
\begin{tabular}{@{\extracolsep{5pt}} lccc} 
\\[-1.8ex]\hline \\[-1.8ex] 
Tie definition & Sex & 45q15 & 95\% CI \\ 
\hline \\[-1.8ex] 
Meal
(2010-11) & Female & $0.24$ & [0.19-0.30] \\ 
Sibling
(2006-11) & Female & $0.17$ & [0.15-0.20] \\ 
Acquaintance
(2010-11) & Female & $0.19$ & [0.15-0.23] \\ 
WHO
(2012) & Female & $0.21$ &  \\ 
UNPD
(2010-2015) & Female & $0.19$ &  \\ 
IHME
(2011) & Female & $0.21$ &  \\ 
Meal
(2010-11) & Male & $0.26$ & [0.21-0.32] \\ 
Sibling
(2006-11) & Male & $0.28$ & [0.23-0.32] \\ 
Acquaintance
(2010-11) & Male & $0.26$ & [0.22-0.31] \\ 
WHO
(2012) & Male & $0.25$ &  \\ 
UNPD
(2010-2015) & Male & $0.33$ &  \\ 
IHME
(2011) & Male & $0.29$ &  \\ 
\hline \\[-1.8ex] 
\end{tabular} 
\end{table} 

\FloatBarrier

\section{Network survival results for both sexes and tie definitions}
\label{ap:all-ns-results}

Network survival estimates for adult death rates in Rwanda are shown in the main text
(Figure~\ref{fig:net-asdr-all}).
This appendix has additional plots that provide more detail about how the
network survival death rates were estimated.

\begin{figure}[p] 
  \captionsetup[subfigure]{justification=justified,singlelinecheck=false}
  \centering
  \subfigure[
     $\widehat{y}_{F, D_\alpha}$
  ]{%
     \label{fig:steps-meal-male-dr}
     \includegraphics[keepaspectratio,width=.4\textwidth]{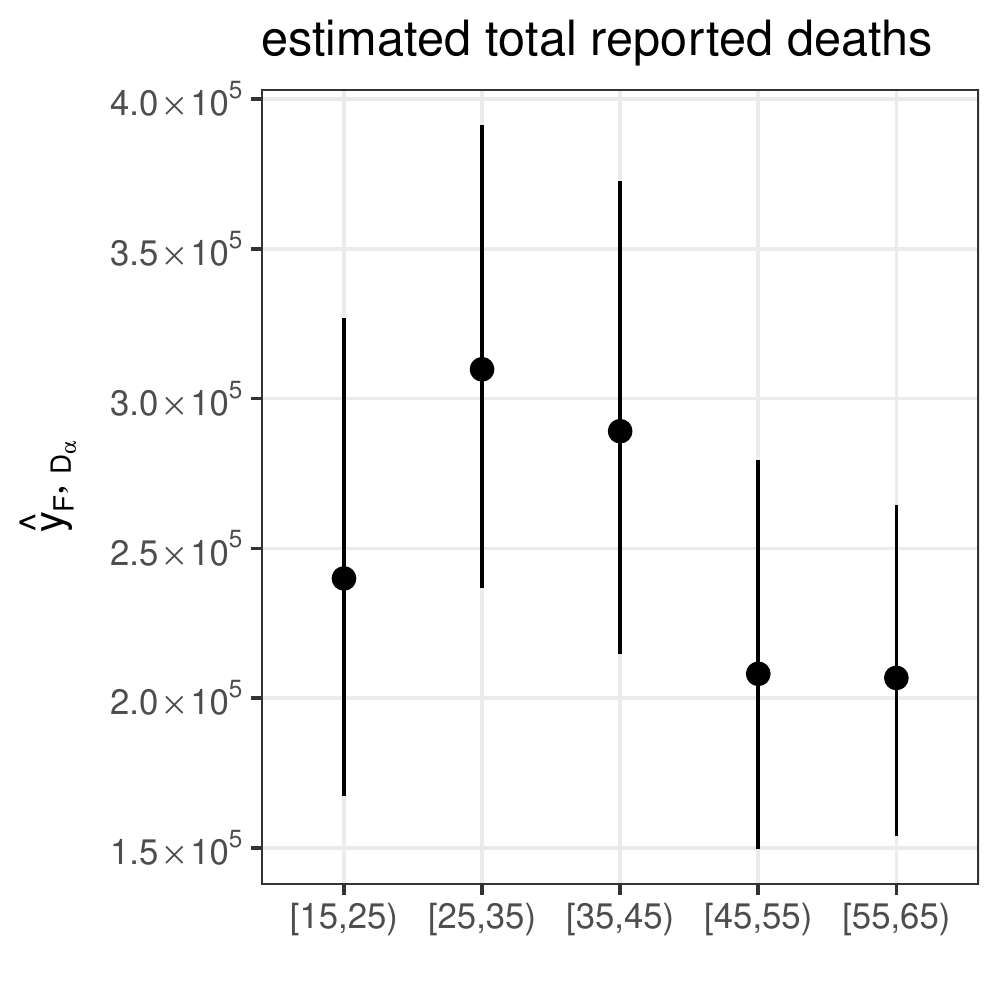}}
  \hspace{0.0in}
  \subfigure[
      $\widehat{\bar{d}}_{F_\alpha, F}$ (an estimate of $\bar{v}_{D_\alpha, F}$)
  ]{%
     \label{fig:steps-meal-male-net}
     \includegraphics[keepaspectratio,width=.4\textwidth]{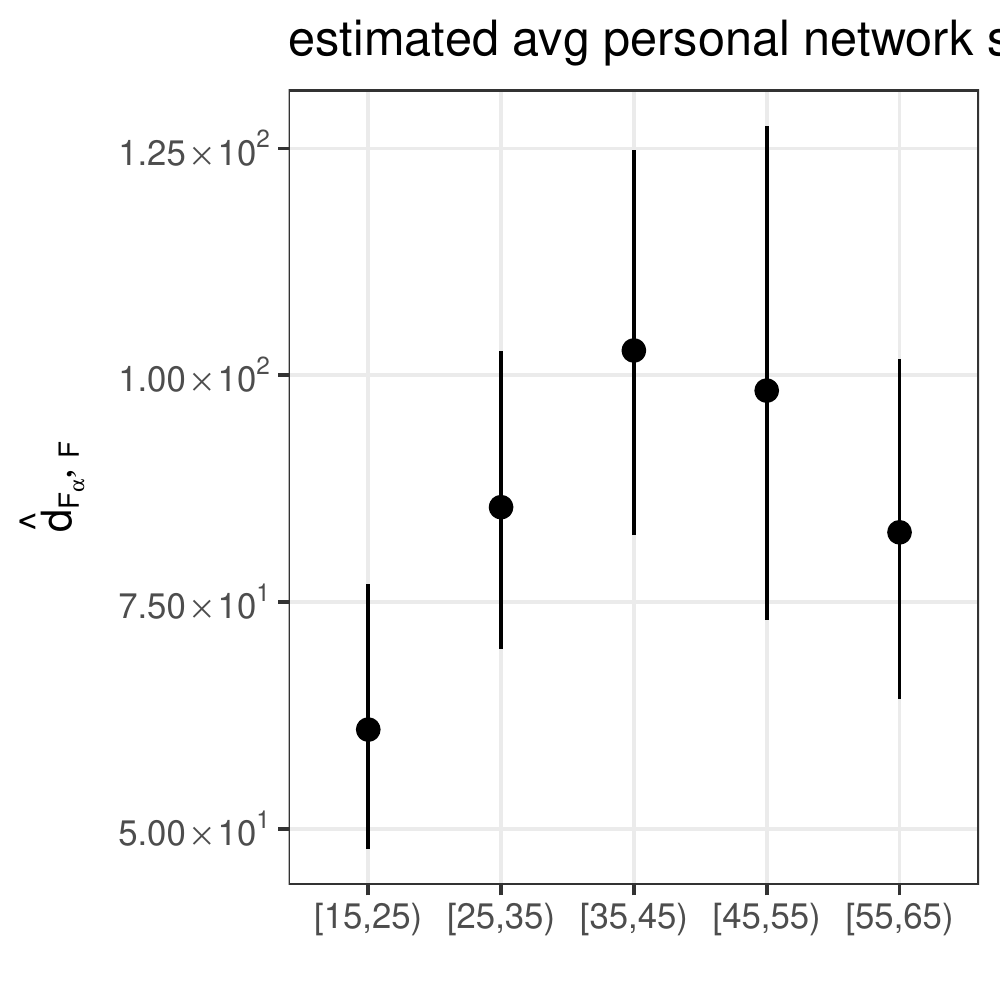}}
  \hspace{0.0in}
  \subfigure[
     $\widehat{D}_\alpha = \frac{\widehat{y}_{F, D_\alpha}}{\widehat{\bar{d}}_{F_\alpha,F}}$
  ]{%
     \label{fig:steps-meal-male-td}
     \includegraphics[keepaspectratio,width=.4\textwidth]{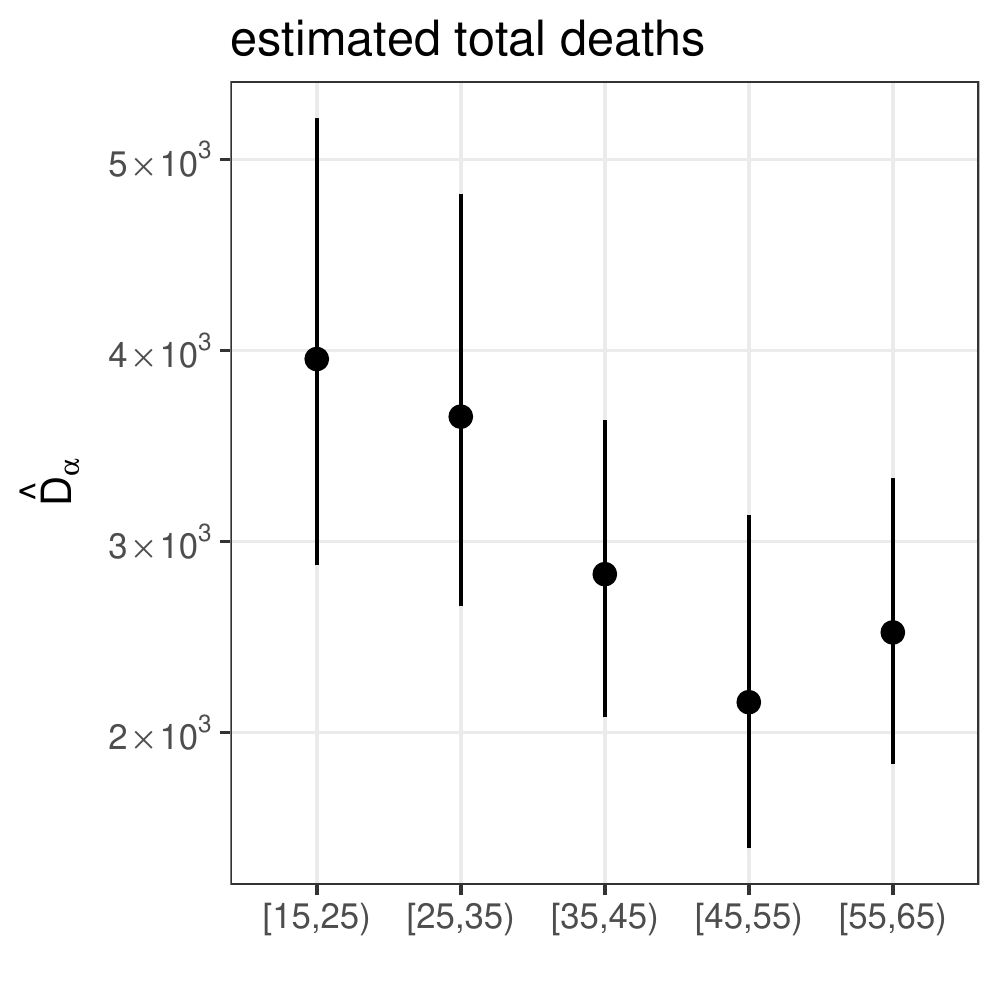}}
  \hspace{0.0in}
  \subfigure[
      $\widehat{N}_{F_\alpha}$ (an estimate of $N_\alpha$)
  ]{%
     \label{fig:steps-meal-male-exp}
     \includegraphics[keepaspectratio,width=.4\textwidth]{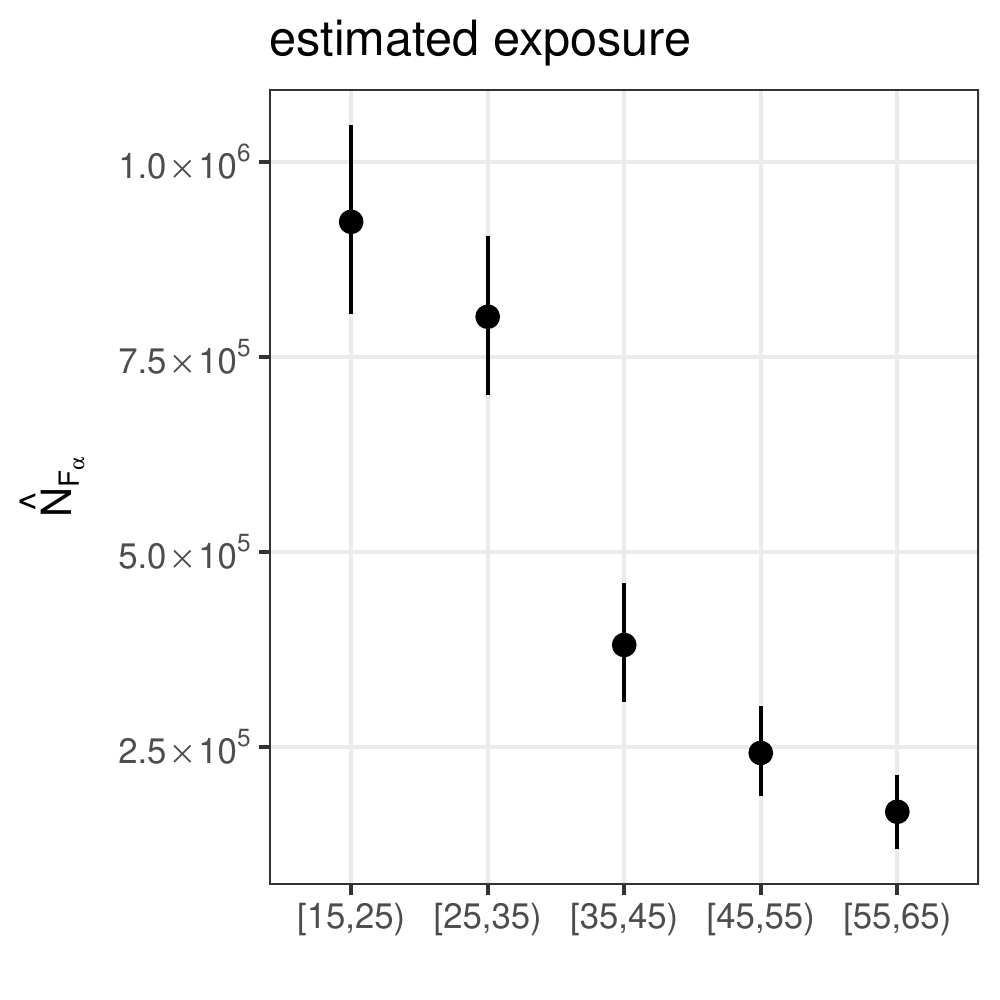}}
  \hspace{0.0in}
  \caption[
      Estimating components of age-specific death rates for Rwandan Males for 12 months prior
      to our survey using responses from the meal tie definition.  
  ]
  {
      Estimating components of age-specific death rates for Rwandan Males for 12 months prior
      to our 2011 survey using responses from the meal tie definition.  
      The average personal network size of survey respondents
      ($\widehat{\bar{d}}_{F_\alpha,F}$; Panel~\ref{fig:steps-meal-male-net}), 
      is used as an estimate of the visibility of deaths ($\bar{v}_{D_\alpha, F}$; 
      i.e., the number of times each death could be reported).
      The estimated number of deaths in the population 
      ($\widehat{D}_\alpha$; Panel~\ref{fig:steps-meal-male-td}) 
      is obtained by dividing estimated total reports about deaths 
      ($\widehat{y}_{F, D_\alpha}$; Panel~\ref{fig:steps-meal-male-dr})
      by the estimated visibility of deaths
      ($\widehat{\bar{v}}_{D_\alpha,F}$; Panel~\ref{fig:steps-meal-male-net}). 
      The estimated size of the frame population ($\widehat{N}_{F_\alpha}$) is
      used as an estimate of the population exposure $N_\alpha$.
      Estimated age-specific death rates ($\widehat{M}_\alpha$;
      Figure~\ref{fig:net-asdr-all})
      are obtained by dividing the
      estimated number of deaths ($\widehat{D}_\alpha$; Panel~\ref{fig:steps-meal-male-td}) 
      by the amount of exposure ($\widehat{N}_\alpha$; Panel~\ref{fig:steps-meal-male-exp}).
      Error bars show 95\% confidence intervals; 
      sampling uncertainty from each step is estimated using the rescaled
      bootstrap approach to account for the complex sample design
      \citep{rao_resampling_1988, rao_recent_1992}.
  } 
  \label{fig:steps-meal-male}
\end{figure}

\begin{figure}[p] 
  \captionsetup[subfigure]{justification=justified,singlelinecheck=false}
  \centering
  \subfigure[
     $\widehat{y}_{F, D_\alpha}$
  ]{%
     \label{fig:steps-meal-female-dr}
     \includegraphics[keepaspectratio,width=.4\textwidth]{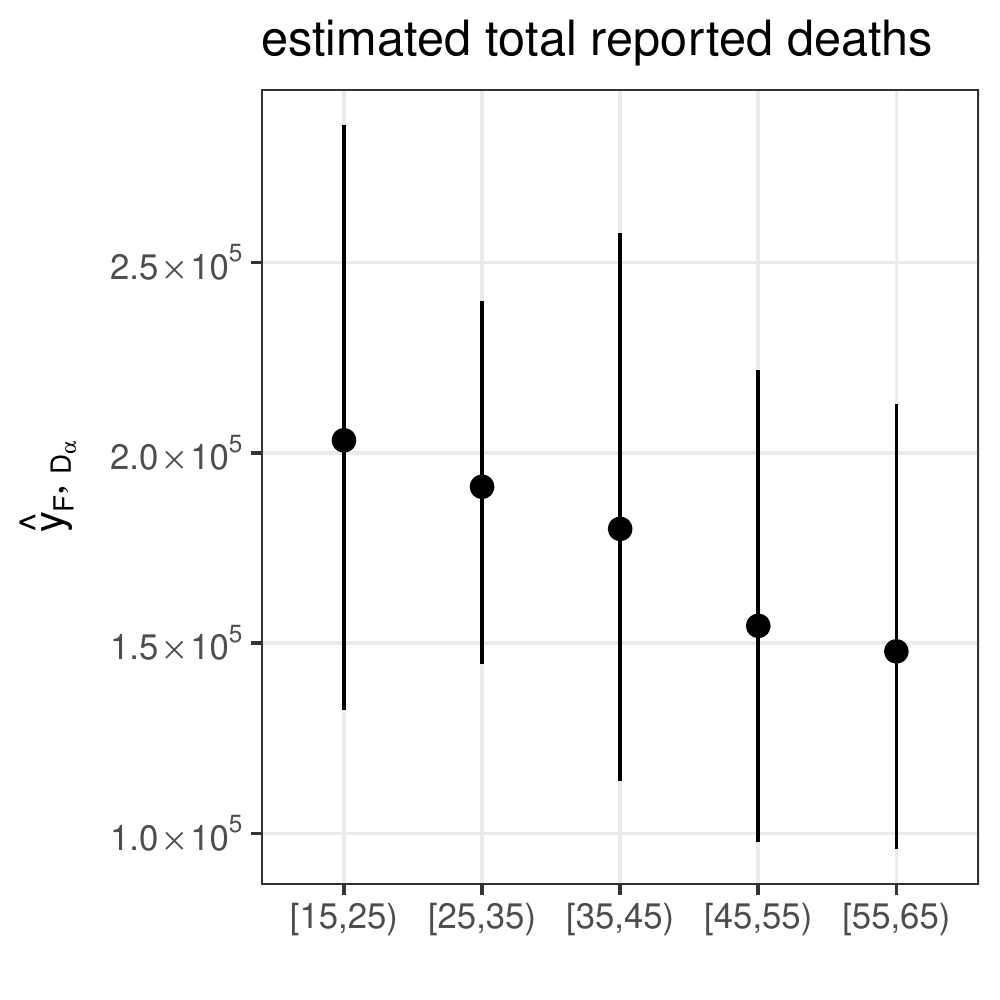}}
  \hspace{0.0in}
  \subfigure[
      $\widehat{\bar{d}}_{F_\alpha, F}$ (an estimate of $\bar{v}_{D_\alpha, F}$)
  ]{%
     \label{fig:steps-meal-female-net}
     \includegraphics[keepaspectratio,width=.4\textwidth]{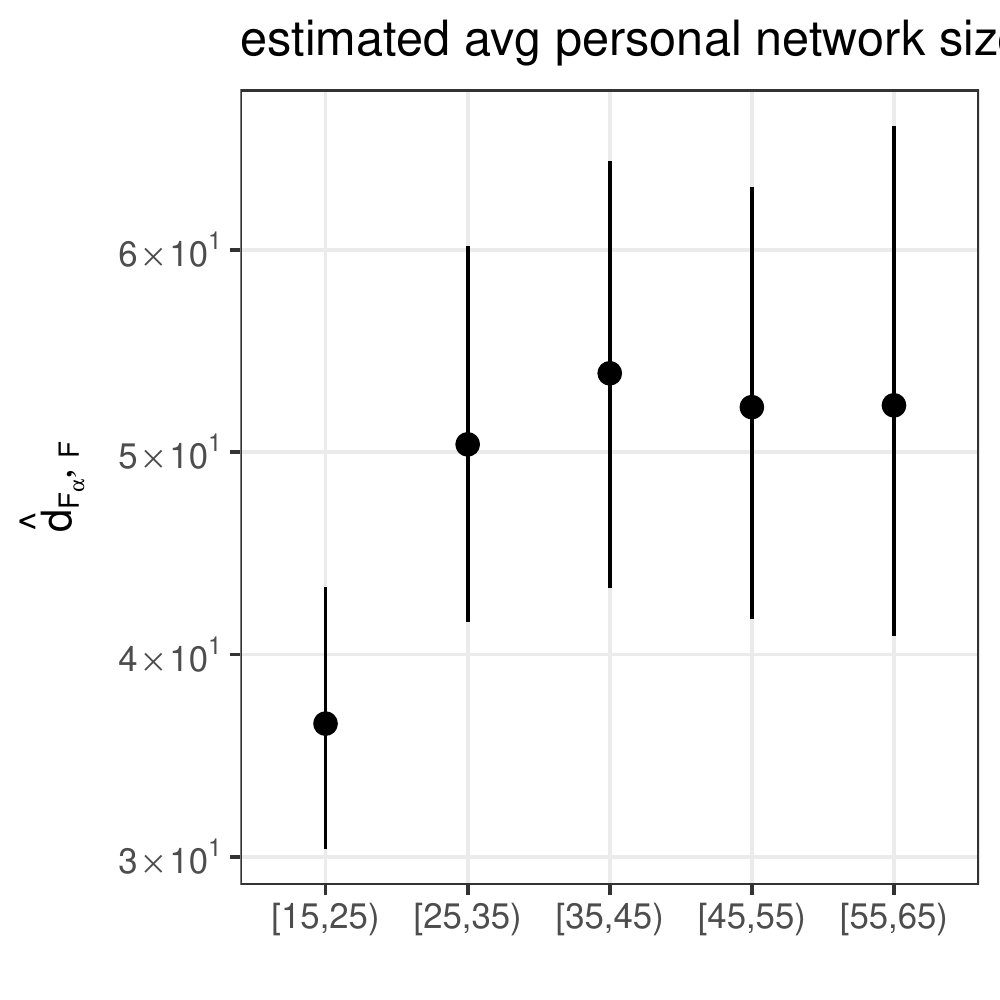}}
  \hspace{0.0in}
  \subfigure[
     $\widehat{D}_\alpha = \frac{\widehat{y}_{F, D_\alpha}}{\widehat{\bar{d}}_{F_\alpha,F}}$
  ]{%
     \label{fig:steps-meal-female-td}
     \includegraphics[keepaspectratio,width=.4\textwidth]{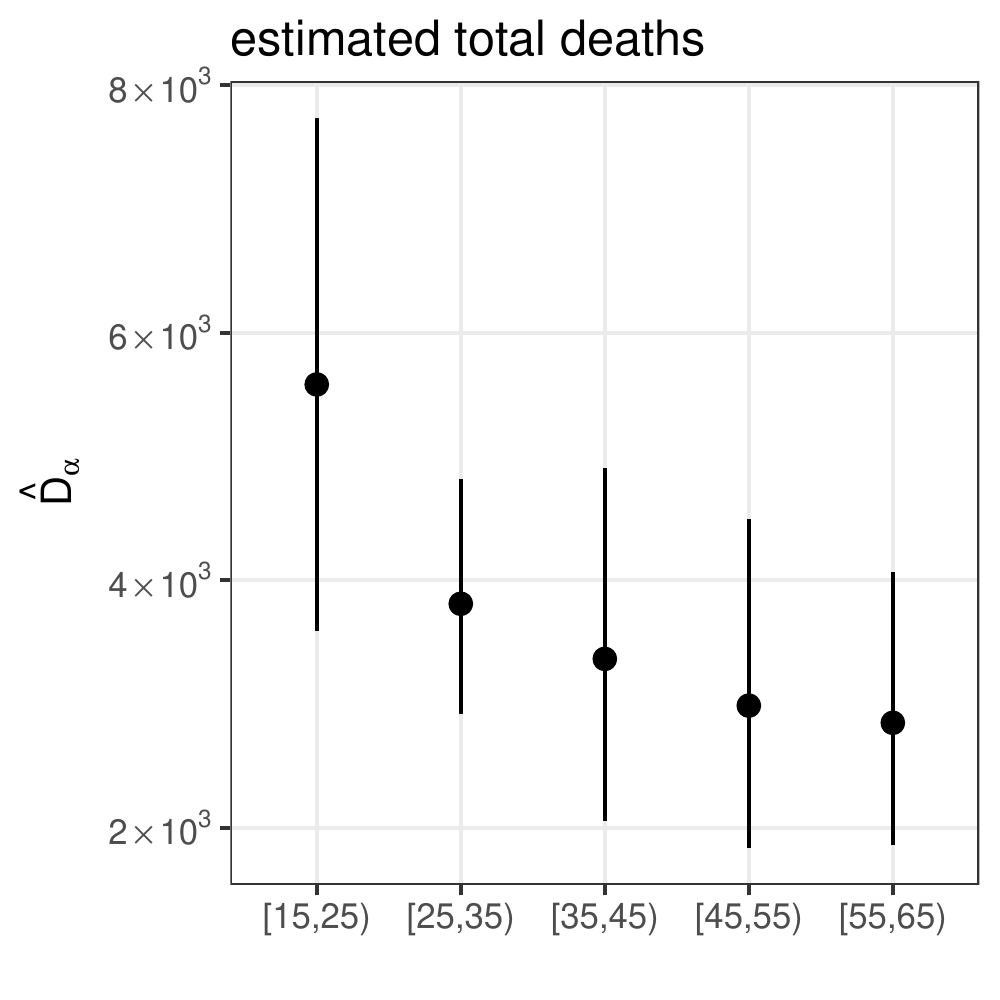}}
  \hspace{0.0in}
  \subfigure[
      $\widehat{N}_{F_\alpha}$ (an estimate of $N_\alpha$)
  ]{%
     \label{fig:steps-meal-female-exp}
     \includegraphics[keepaspectratio,width=.4\textwidth]{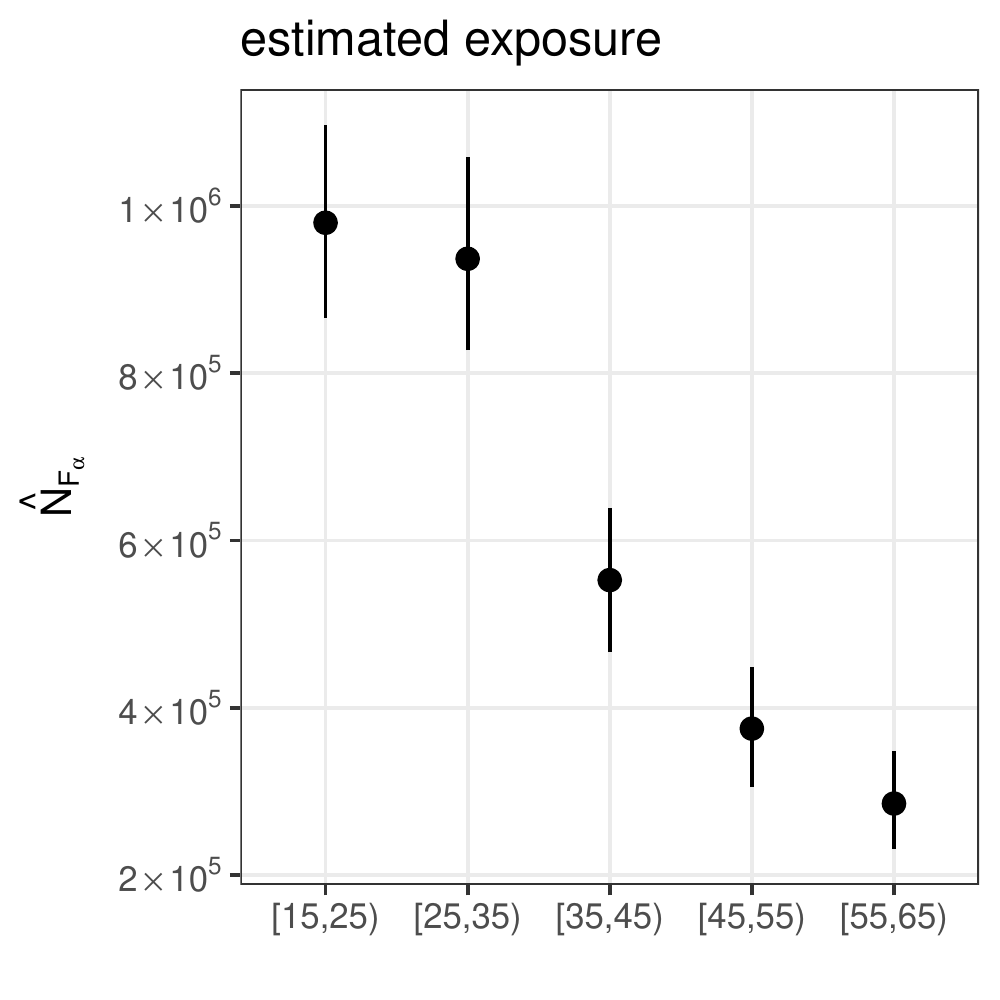}}
  \hspace{0.0in}
  \caption{
      Estimating components of age-specific death rates for Rwandan females for
      12 months prior to our survey using responses from the meal tie
      definition.  
      The interpretation of this figure is analogous to Figure~\ref{fig:steps-meal-male}.
  } 
  \label{fig:steps-meal-female}
\end{figure}

\begin{figure}[p] 
  \captionsetup[subfigure]{justification=justified,singlelinecheck=false}
  \centering
  \subfigure[
     $\widehat{y}_{F, D_\alpha}$
  ]{%
     \label{fig:steps-acq-male-dr}
     \includegraphics[keepaspectratio,width=.4\textwidth]{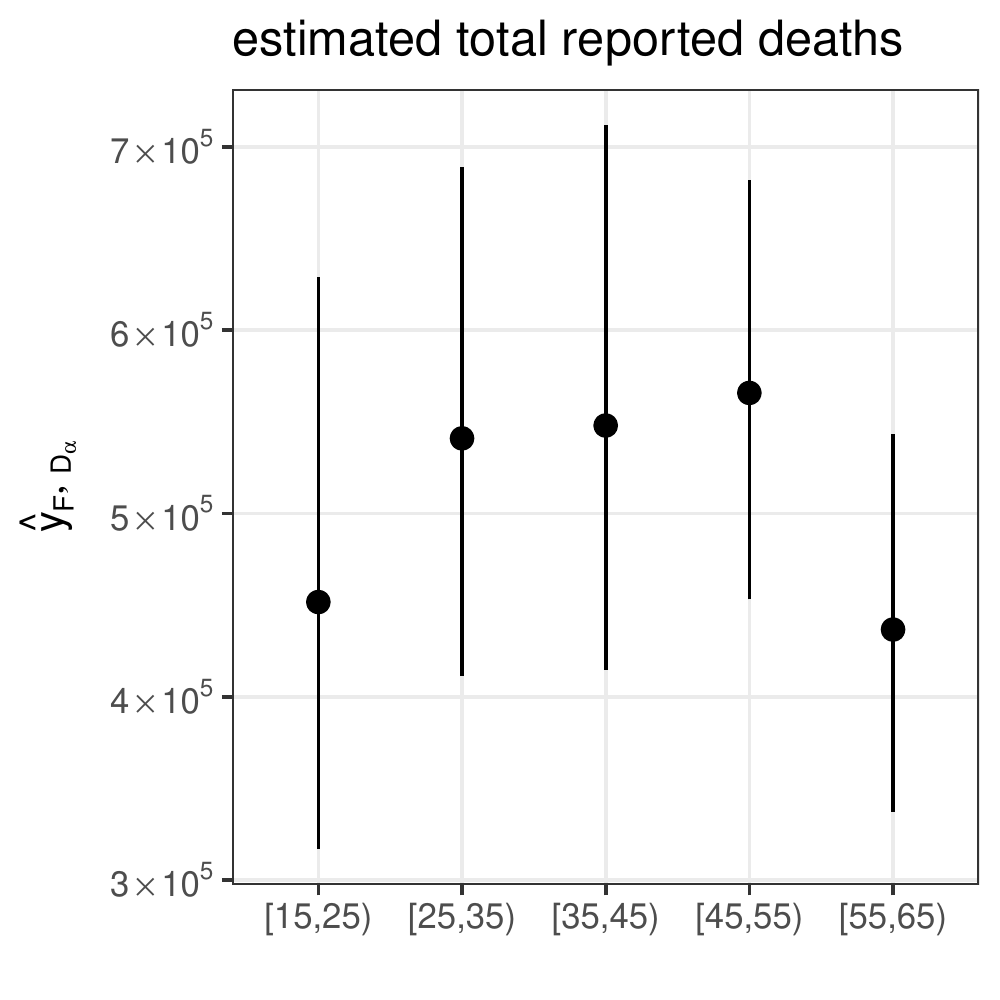}}
  \hspace{0.0in}
  \subfigure[
      $\widehat{\bar{d}}_{F_\alpha, F}$ (an estimate of $\bar{v}_{D_\alpha, F}$)
  ]{%
     \label{fig:steps-acq-male-net}
     \includegraphics[keepaspectratio,width=.4\textwidth]{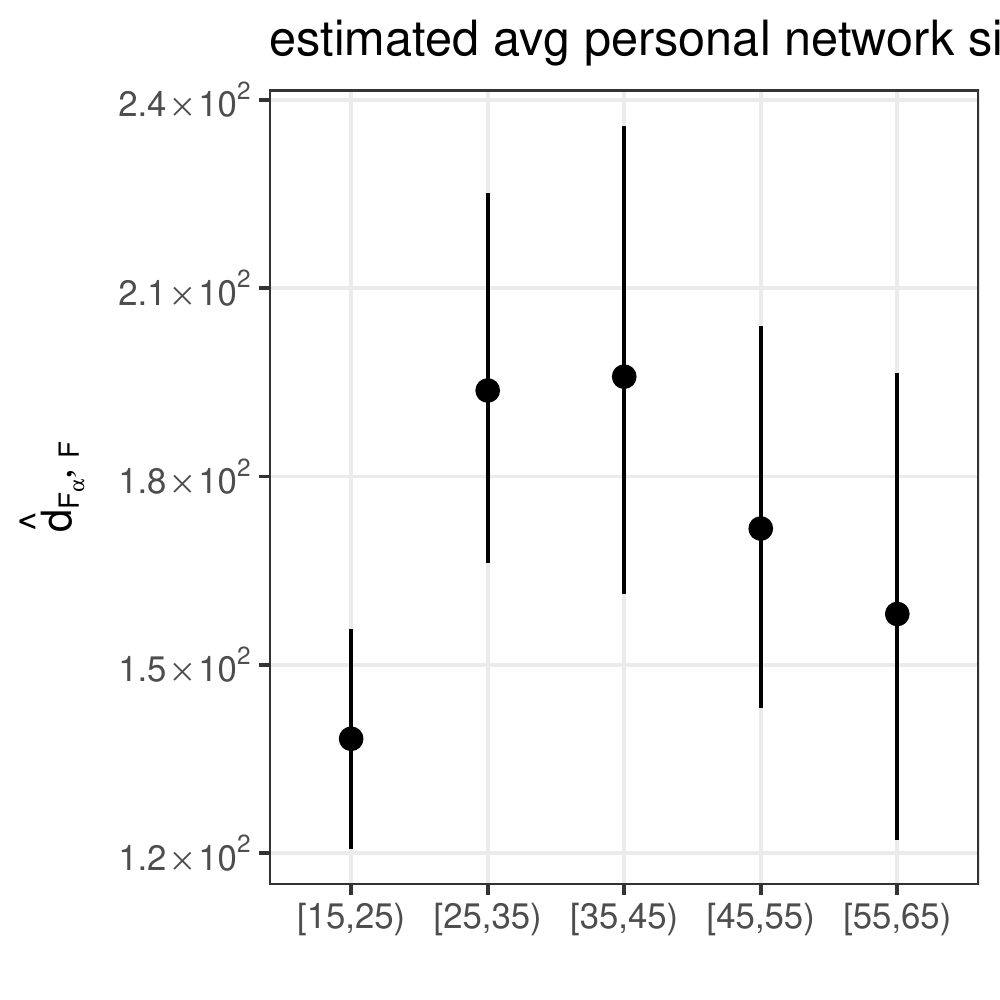}}
  \hspace{0.0in}
  \subfigure[
     $\widehat{D}_\alpha = \frac{\widehat{y}_{F, D_\alpha}}{\widehat{\bar{d}}_{F_\alpha,F}}$
  ]{%
     \label{fig:steps-acq-male-td}
     \includegraphics[keepaspectratio,width=.4\textwidth]{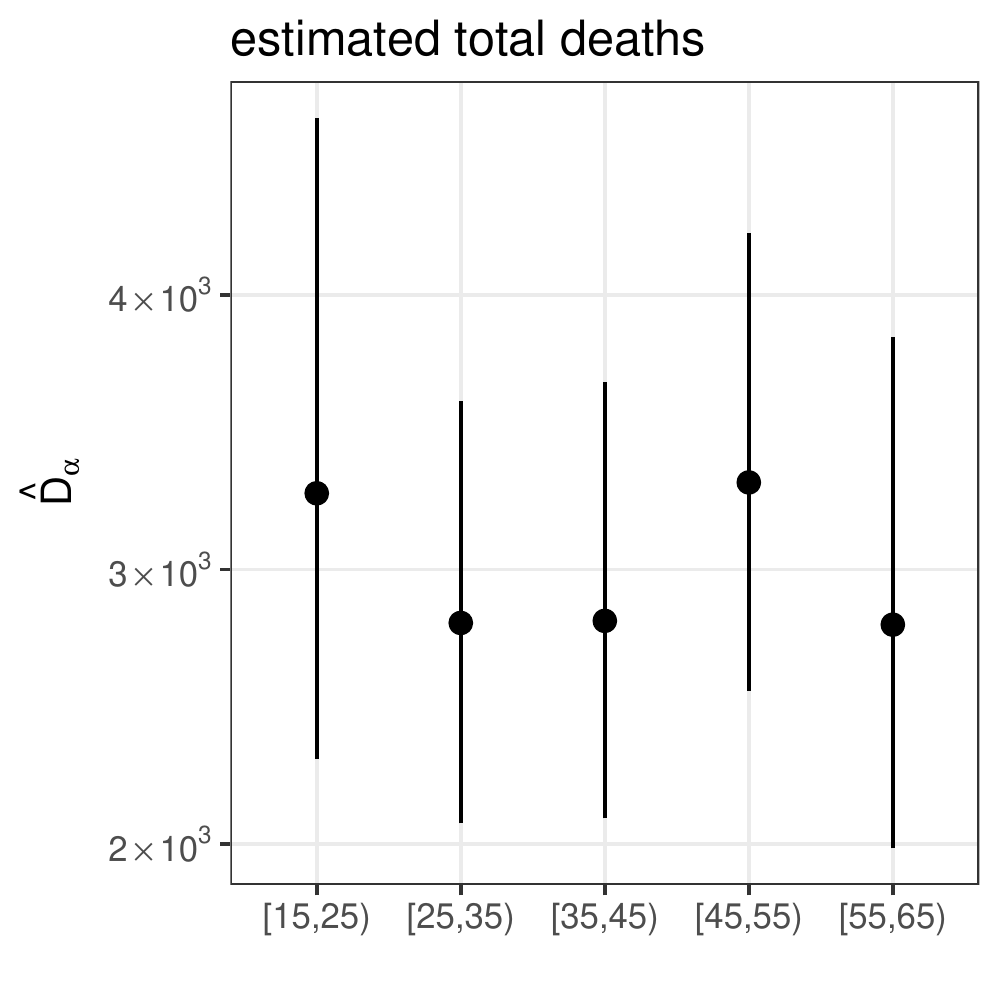}}
  \hspace{0.0in}
  \subfigure[
      $\widehat{N}_{F_\alpha}$ (an estimate of $N_\alpha$)
  ]{%
     \label{fig:steps-acq-male-exp}
     \includegraphics[keepaspectratio,width=.4\textwidth]{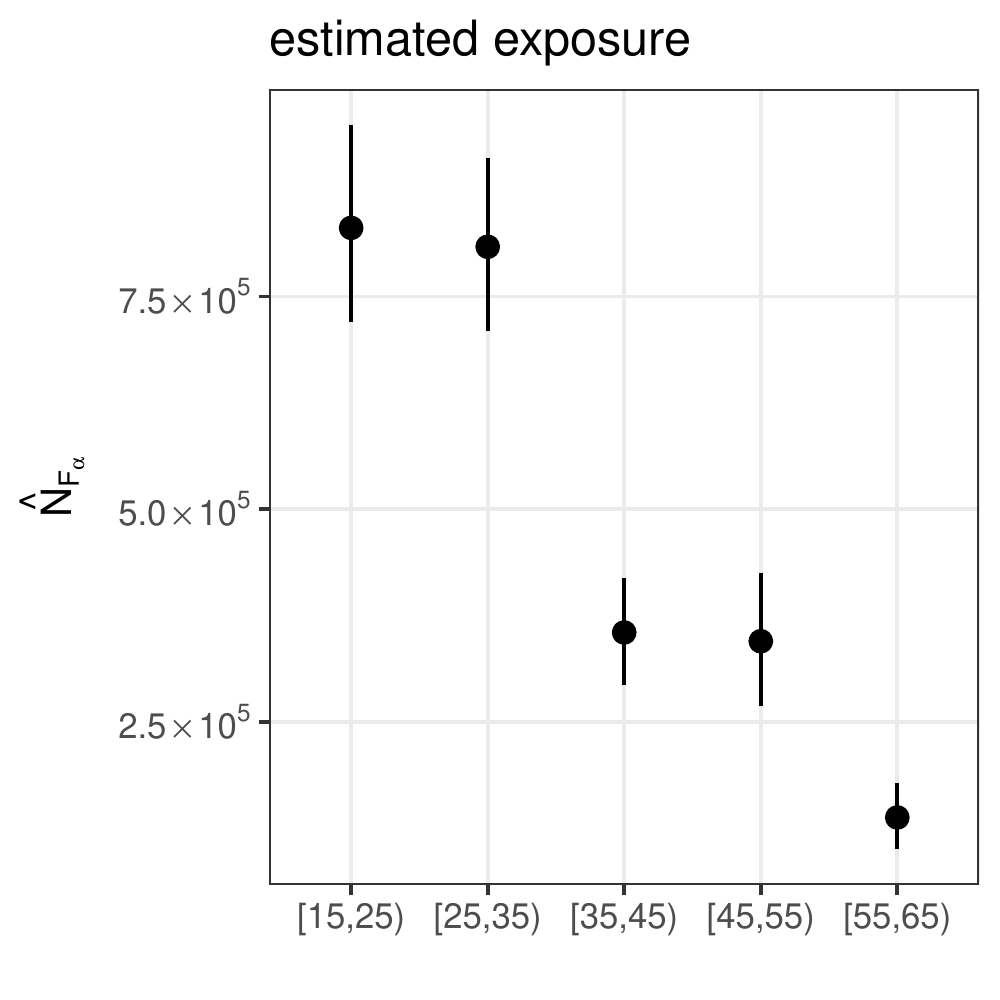}}
  \hspace{0.0in}
  \caption{
      Estimating components of age-specific death rates for Rwandan males for
      12 months prior to our survey using responses from the acquaintance tie
      definition.  
      The interpretation of this figure is analogous to Figure~\ref{fig:steps-meal-male}.
  } 
  \label{fig:steps-acq-male}
\end{figure}

\begin{figure}[p] 
  \captionsetup[subfigure]{justification=justified,singlelinecheck=false}
  \centering
  \subfigure[
     $\widehat{y}_{F, D_\alpha}$
  ]{%
     \label{fig:steps-acq-female-dr}
     \includegraphics[keepaspectratio,width=.4\textwidth]{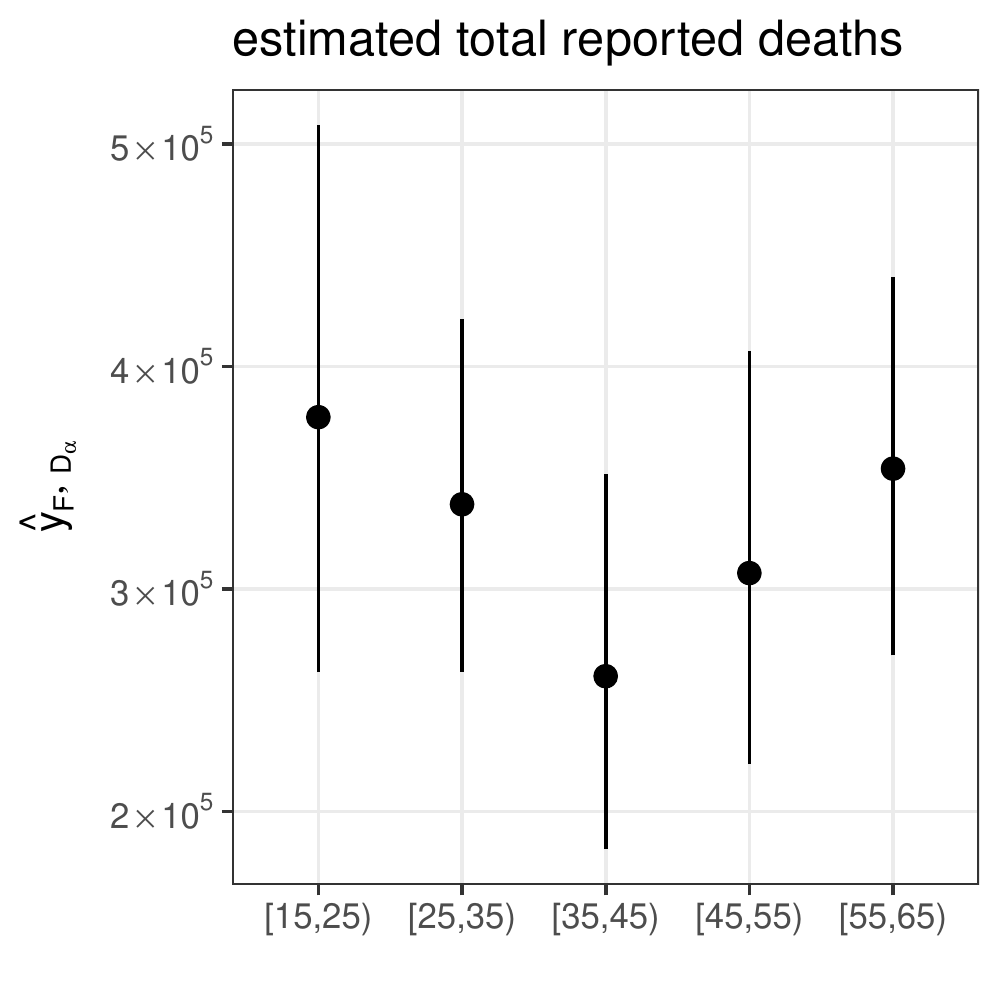}}
  \hspace{0.0in}
  \subfigure[
      $\widehat{\bar{d}}_{F_\alpha, F}$ (an estimate of $\bar{v}_{D_\alpha, F}$)
  ]{%
     \label{fig:steps-acq-female-net}
     \includegraphics[keepaspectratio,width=.4\textwidth]{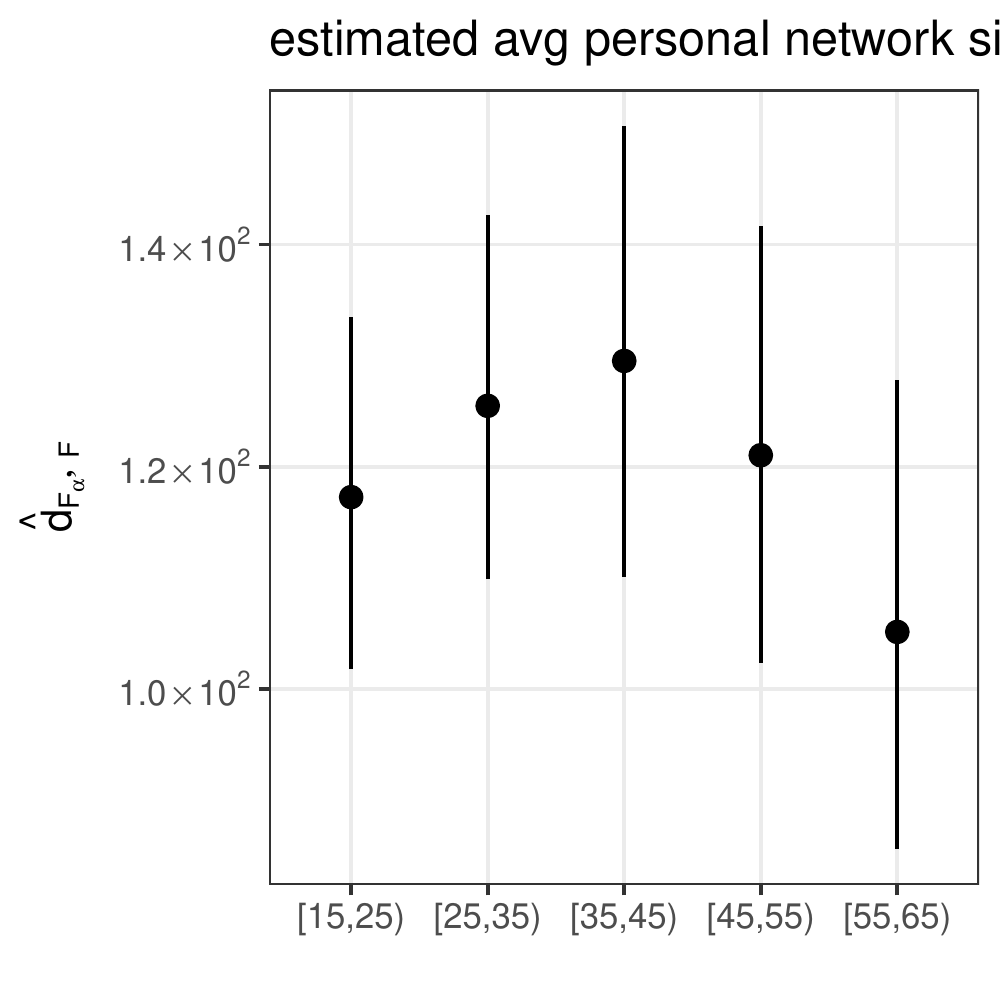}}
  \hspace{0.0in}
  \subfigure[
     $\widehat{D}_\alpha = \frac{\widehat{y}_{F, D_\alpha}}{\widehat{\bar{d}}_{F_\alpha,F}}$
  ]{%
     \label{fig:steps-acq-female-td}
     \includegraphics[keepaspectratio,width=.4\textwidth]{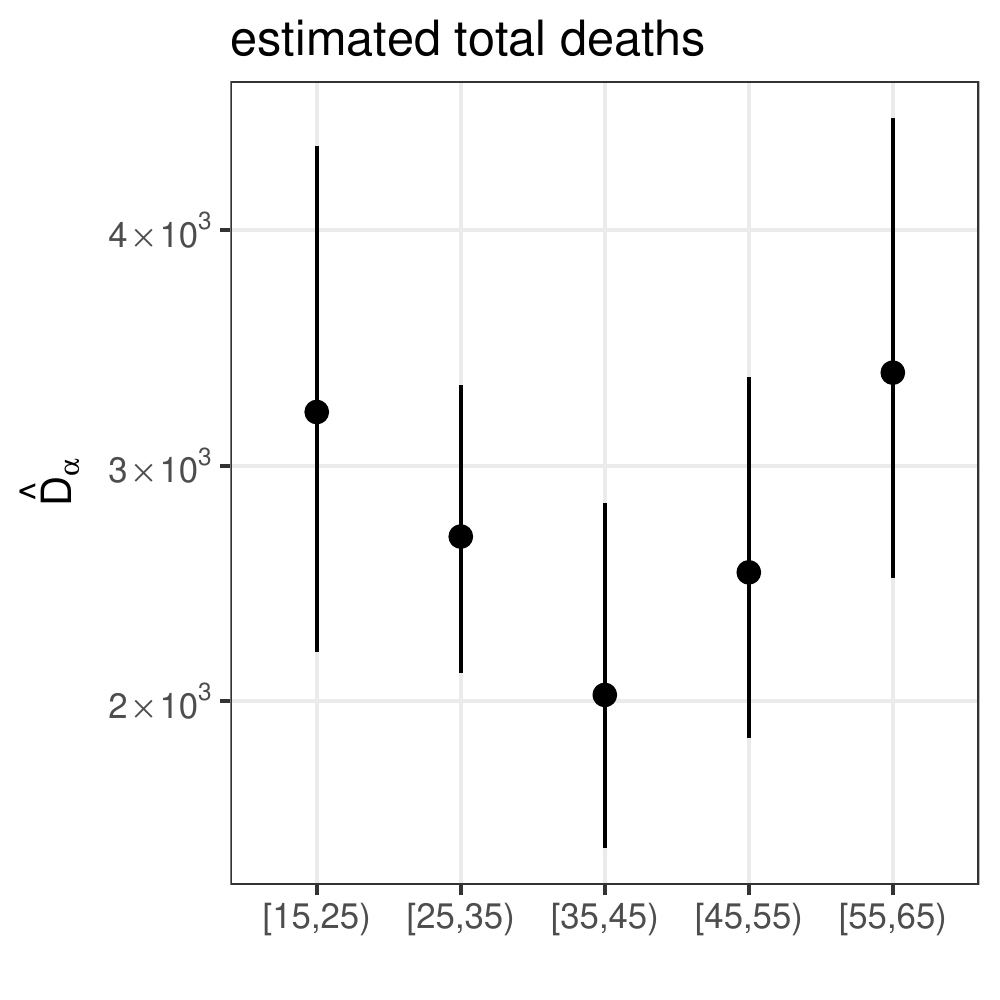}}
  \hspace{0.0in}
  \subfigure[
      $\widehat{N}_{F_\alpha}$ (an estimate of $N_\alpha$)
  ]{%
     \label{fig:steps-acq-female-exp}
     \includegraphics[keepaspectratio,width=.4\textwidth]{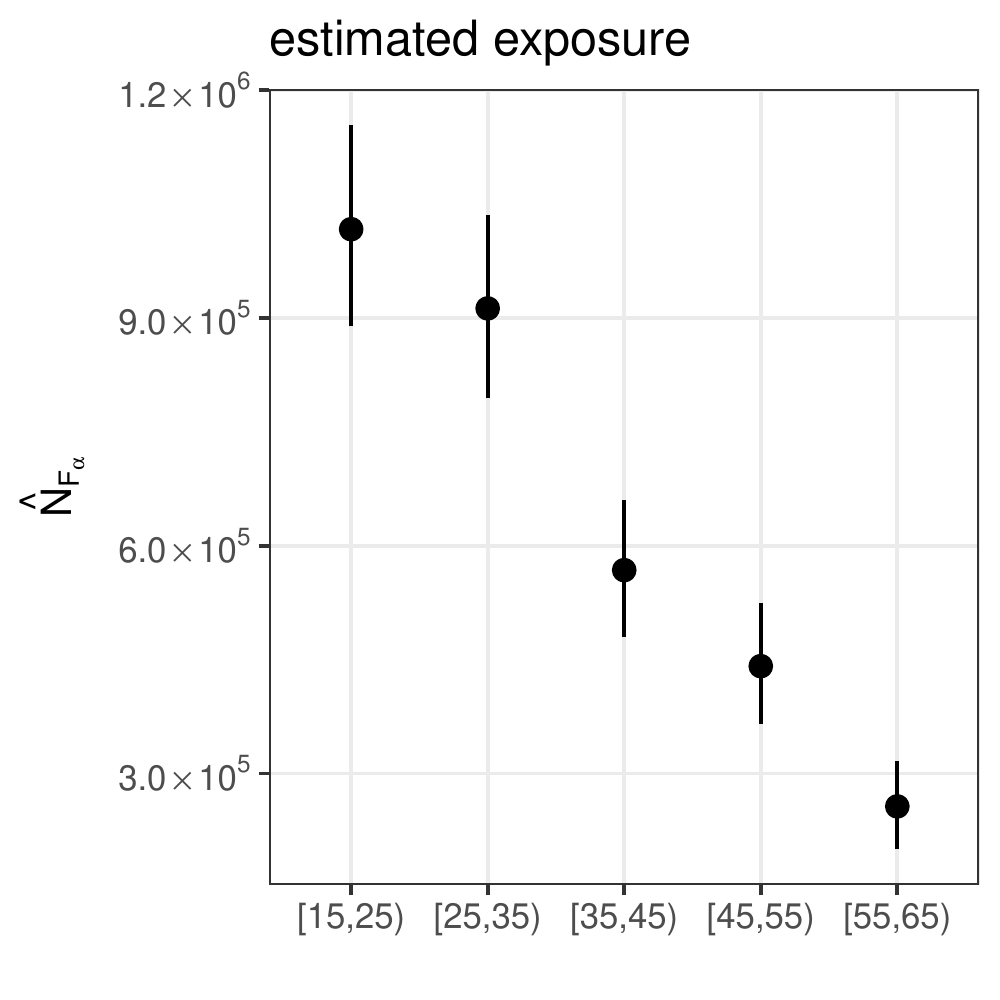}}
  \hspace{0.0in}
  \caption{
      Estimating components of age-specific death rates for Rwandan females for
      12 months prior to our survey using responses from the acquaintance tie
      definition.  
      The interpretation of this figure is analogous to Figure~\ref{fig:steps-meal-male}.
  } 
  \label{fig:steps-acq-female}
\end{figure}

\begin{figure} 
  \captionsetup[subfigure]{justification=justified,singlelinecheck=false}
  \centering
  \subfigure[
     Males, using the meal network.
  ]{%
     \label{fig:asdr-meal-male-unlogged}
     \includegraphics[keepaspectratio,width=.4\textwidth]{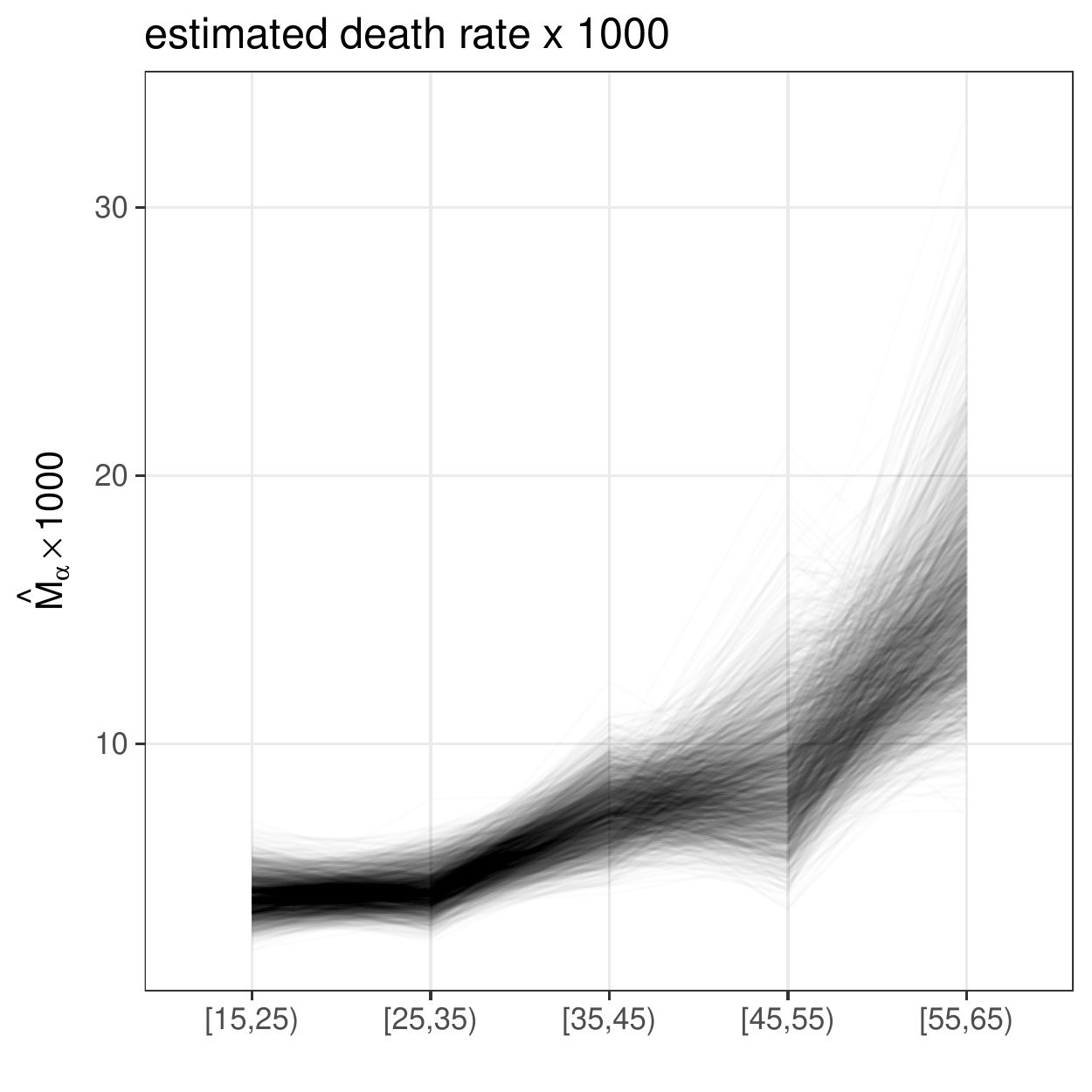}}
  \hspace{0.0in}
  \subfigure[
      Females, using the meal network.
  ]{%
     \label{fig:asdr-meal-female-unlogged}
     \includegraphics[keepaspectratio,width=.4\textwidth]{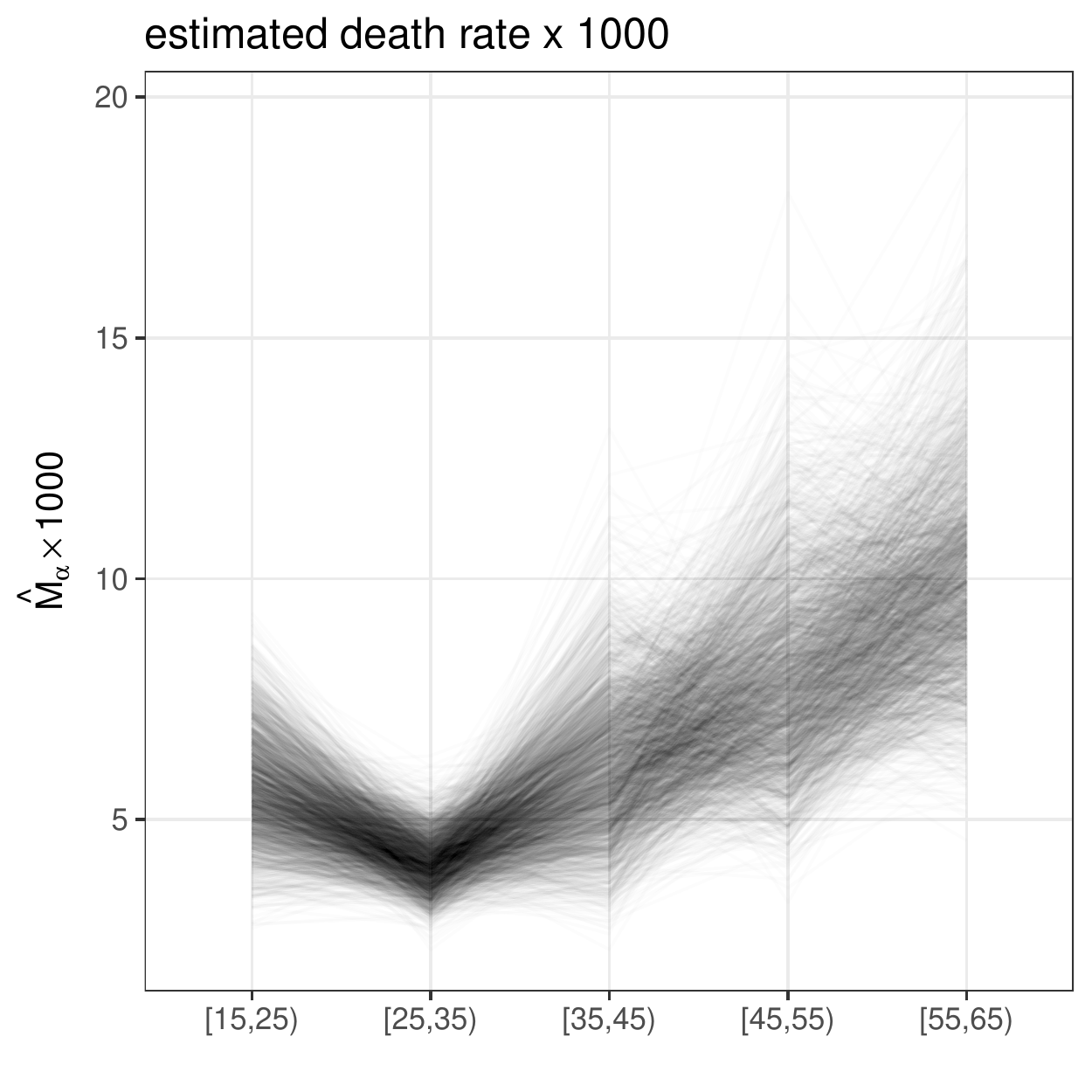}}
  \hspace{0.0in}
  \subfigure[
      Males, using the acquaintance network.
  ]{%
     \label{fig:asdr-acq-male-unlogged}
     \includegraphics[keepaspectratio,width=.4\textwidth]{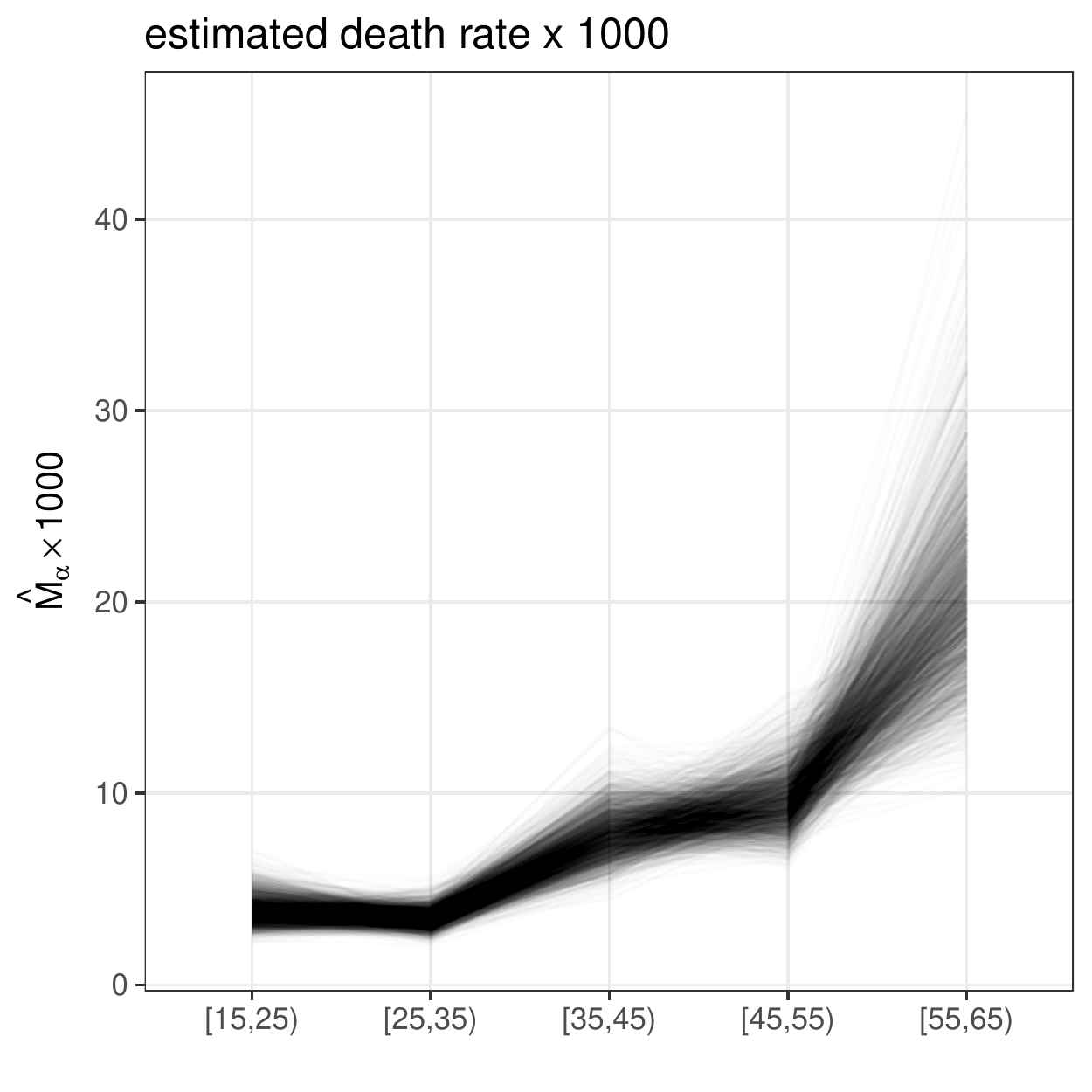}}
  \hspace{0.0in}
  \subfigure[
      Females, using the acquaintance network.
  ]{%
     \label{fig:asdr-acq-female-unlogged}
     \includegraphics[keepaspectratio,width=.4\textwidth]{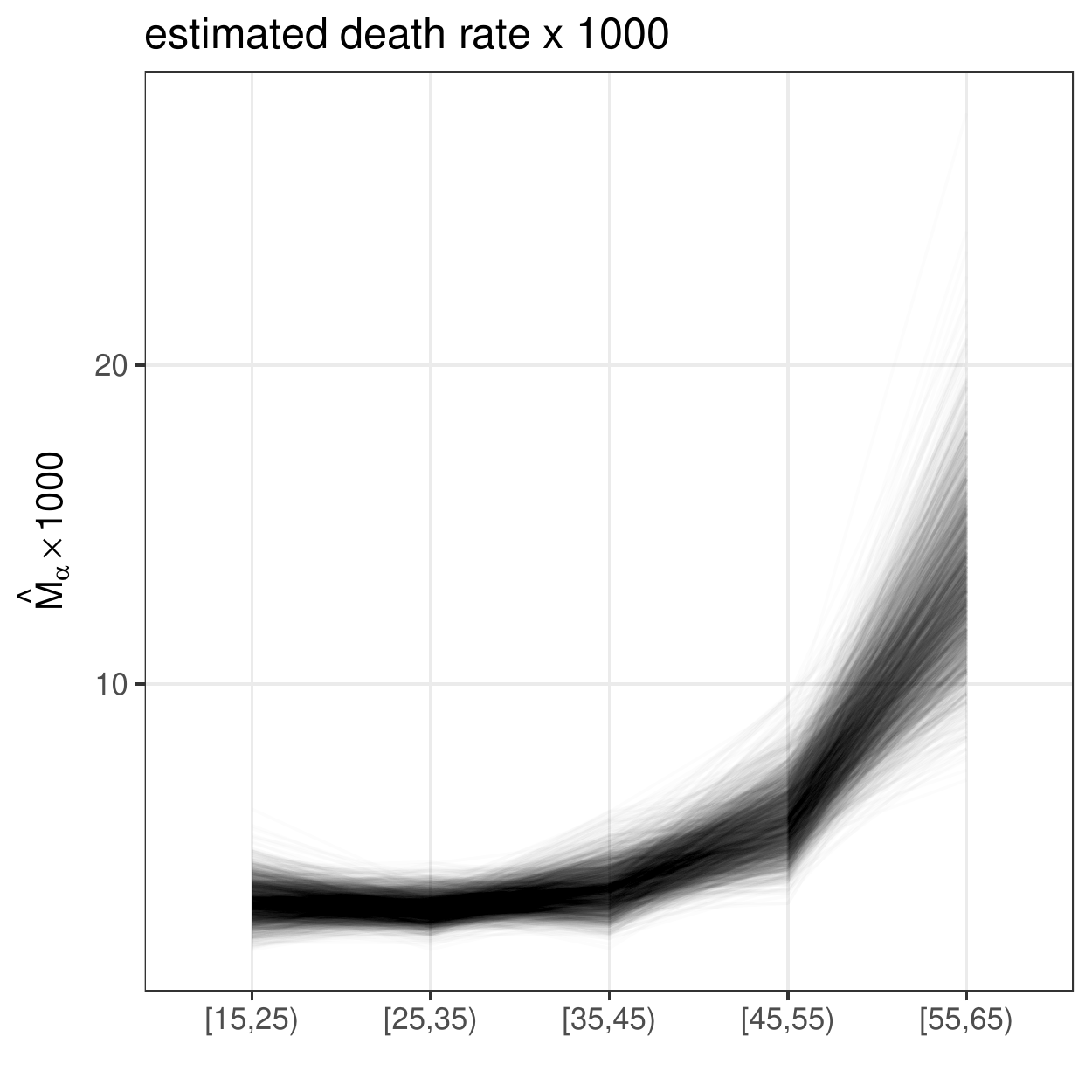}}
  \hspace{0.0in}
  \caption{
      Estimated age-specific death rates for Rwandans for 12 months prior
      to our survey using responses from the meal tie definition (top row)
      and the acquaintance tie definition (bottom row), for males (left column)
      and for females (right column).  
      These plots are not on a log scale.
      Each line shows the result of one bootstrap resample; taken together, the lines show
      the estimated sampling uncertainty for each set of death rates.
  }
  \label{fig:asdr-unlogged-all} 
\end{figure}

Our derivations in Section~\ref{sec:network_survival} and Appendix~\ref{ap:ns-estimator}
show that network survival death rate estimates are built up
from several components: the estimated number of connections to deaths; the
estimated personal network sizes; the estimated total number of deaths; and the
estimated amount of exposure. 
The first part of this appendix has figures that show each of these
components separately for all of the network survival death rate estimates from Rwanda:
male death rates from the meal network
(Figure~\ref{fig:steps-meal-male});
female death rates from the meal network
(Figure~\ref{fig:steps-meal-female}); male death rates from the
acquaintance network (Figure~\ref{fig:steps-acq-male}); and female death
rates from the acquaintance network (Figure~\ref{fig:steps-acq-female}).
The second part of this appendix has plots showing the age-specific death rates
for both sexes and tie definitions that are not on a log scale
(Figure~\ref{fig:asdr-unlogged-all}).

Figure~\ref{fig:steps-meal-male} shows detailed results for one case:
estimated Rwandan male death rates from reports about the meal tie definition. 
Panel~\ref{fig:steps-meal-male-dr} shows, for
each age group, the estimated total number of reports about deaths
($\widehat{y}_{F, D_\alpha}$, Eq.~\ref{eqn:yhat-f-oalpha}). Since each
death can be reported multiple times, this quantity on its own is not enough to
estimate the total number of deaths in the population. 
Panel~\ref{fig:steps-meal-male-net} shows, for each age group, the estimated
size of respondents' personal networks, which is used as an estimate for the
visibility of deaths ($\widehat{\bar{d}}_{F_\alpha, F}$,
Eq.~\ref{eqn:dbar-falpha-f}). 
Dividing the total estimated reports about deaths
(Panel~\ref{fig:steps-meal-male-dr}) by the estimated visibility of deaths
(Panel~\ref{fig:steps-meal-male-net}) produces the estimated total number of
deaths by age group ($\widehat{D}_{\alpha}$) shown in
Panel~\ref{fig:steps-meal-male-td}.  
Panel~\ref{fig:steps-meal-male-exp} shows the estimated number of people in
each age group ($\widehat{N}_{F_\alpha}$), which is used as an estimate of
exposure; this quantity comes from the sampling design. 
The interpretation of Figures~\ref{fig:steps-meal-female}, \ref{fig:steps-acq-male},
and \ref{fig:steps-acq-female} follow the same pattern as Figure~\ref{fig:steps-meal-male}.

\section{Comparison estimates}
\label{ap:comparison-estimates}

In this section, we provide more detail about the estimates we use to compare
with network survival estimates. First, we describe how we constructed sibling
survival estimates. Next, we give more information about the three organizations'
estimates. We also show a comparison between network survival death rates and
the death rates from the three organizations, providing a more granular comparison
than the $\ffqf$ discussed in the main text.

\subsection{Sibling survival estimates}
\label{ap:sibling}

In this section, we describe how we computed estimated adult death rates
from the sibling histories in the 2010 Rwanda DHS using the direct sibling
survival estimator.
\citet{nisr_rwanda_2012} contains detailed information about the survey, and
all of the data are freely available online through the DHS website\footnote{
    \url{http://dhsprogram.com/what-we-do/survey/survey-display-364.cfm}
}.

Section~\ref{sec:background} describes
the considerable methodological debate over how to produce estimated death
rates from DHS sibling histories. 
Our goal here was to construct the simplest direct sibling survival estimates
possible. 
We therefore follow the recommendation of the offical \emph{Guide to DHS
Statistics} \citep{rutstein_guide_2006} and the International Union for the
Scientific Study of Population's \emph{Tools for Demographic Estimation}
\citep{moultrie_tools_2013} by using the original direct sibling survival
estimator proposed by \citet{rutenberg_direct_1991}. The estimator can be
written
\begin{align}
    \label{eqn:direct-sib}
    \widehat{M}_{\alpha} &= 
    \frac{ \sum_{i \in s} \frac{1}{\pi_i} \sum_{k \in \sigma(i) } D_{k, \alpha} }
         { \sum_{i \in s} \frac{1}{\pi_i} \sum_{k \in \sigma(i) } N_{k, \alpha} },
\end{align}
\noindent where $\widehat{M}_\alpha$ is the estimated death rate in demographic
group $\alpha$; $s$ is the sample of survey respondents, $\pi_i$ is respondent
$i$'s probability of inclusion from the sampling design; $\sigma(i)$ is the set
of siblings that respondent $i$ reports about; $D_{k,\alpha}$ is an indicator
variable for whether or not $k$ died when in demographic group $\alpha$, and
$N_{k, \alpha}$ is the amount of time $k$ spent alive in demographic group
$\alpha$. 

\begin{figure}[p] 
  \centering
  \includegraphics[keepaspectratio]{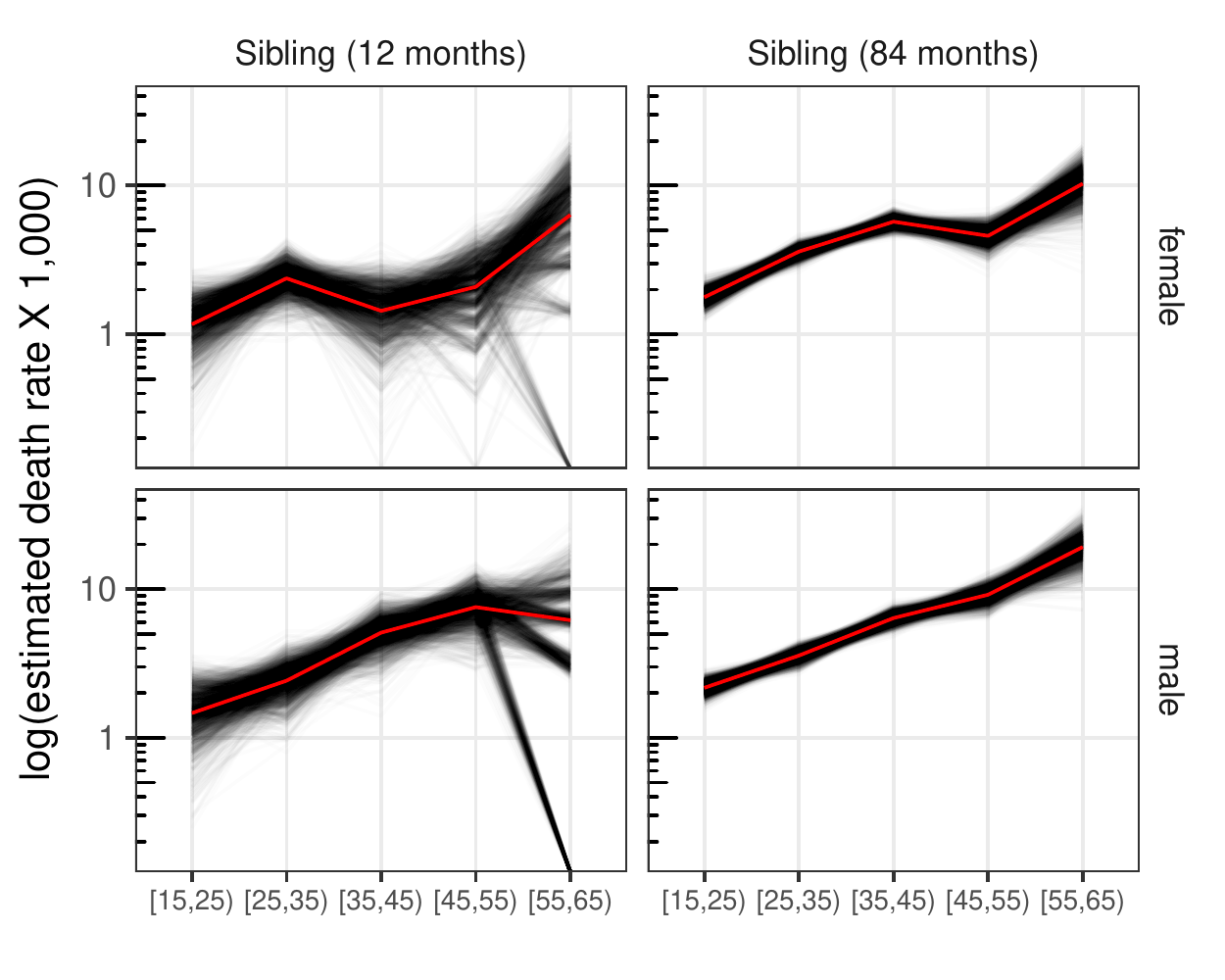}
  \caption{Comparison between sibling estimates based on deaths reported 12 months
  and 84 months before the interview. The estimates from 12 months before the
  interview are very imprecise, while the estimates from 84 months before the
  survey are much more stable. Therefore, we use the 84-month estimates when
  we compare to the network survival results in the main text.}
  \label{fig:asdr-1vs5}
\end{figure}

We wanted to compare the network survival results (based on 12 months prior to
the survey) to the sibling survival estimates. Therefore, our preference would
be to compute sibling survival estimates for the 12 months prior to the survey.
However, the left-hand panel of Figure~\ref{fig:asdr-1vs5} shows that estimates for this
time frame have too much sampling variation to be practically useful (and this
is consistent with the sibling history literature; see
Section~\ref{sec:background}). Since samples are not typically large enough to
permit estimating yearly age-specific death rates using the estimator in
Eq.~\ref{eqn:direct-sib}, in the results in the main text, we follow the
recommendation of \citet{rutstein_guide_2006} and
\citet{rutenberg_direct_1991} by producing estimates for the 84 months
(i.e., 7 years) prior to the survey. 

\subsection{Three organizations' estimates}
\label{ap:agencies}

\begin{figure}[h] 
  \centering
  \includegraphics[width=\textwidth,keepaspectratio]{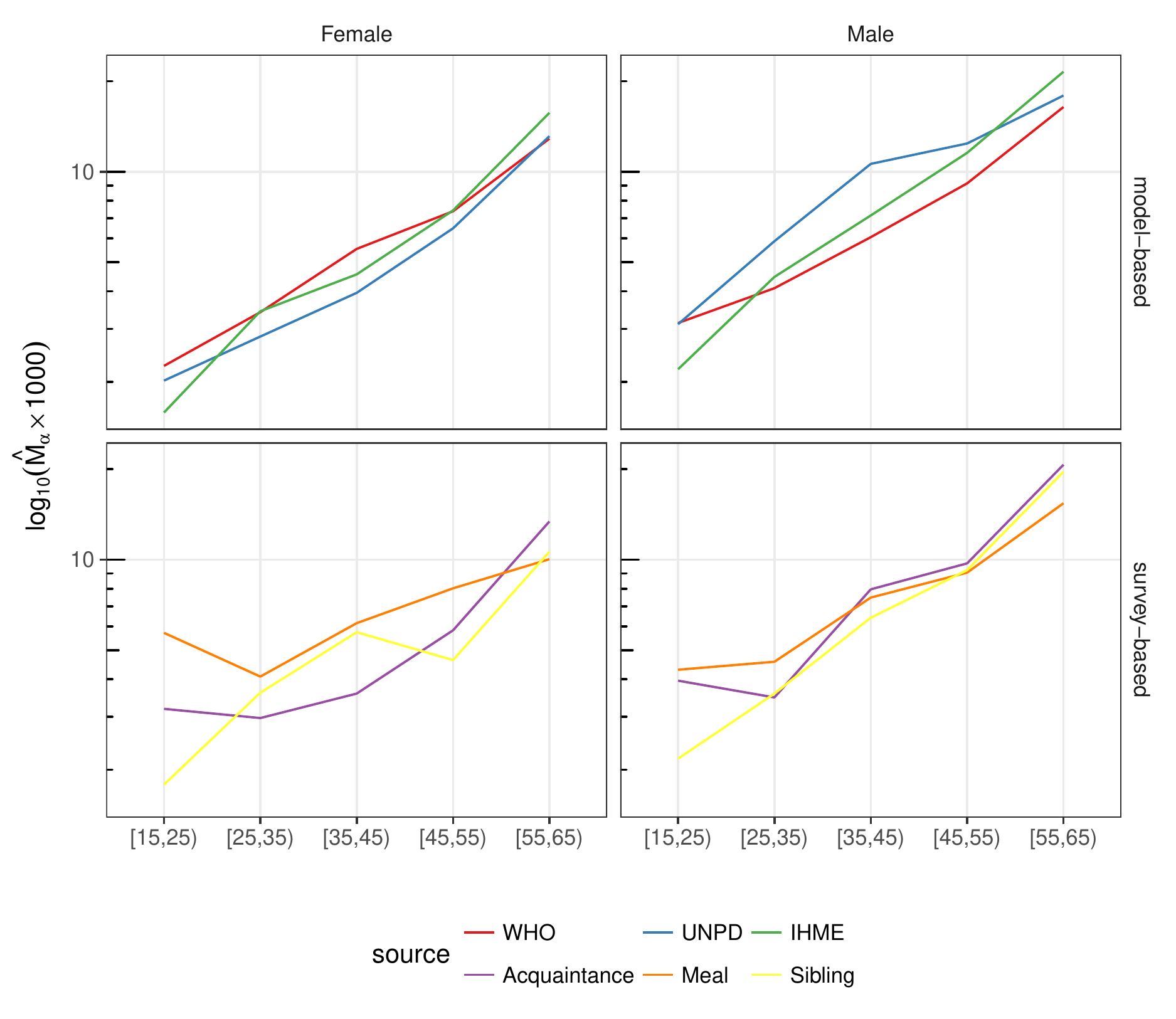}
  \caption{
    Comparison between network survival death rate estimates for two types
    of personal network, direct sibling survival death rates estimates from the
    2010 Rwanda Demographic and Health Survey, and model-based estimates for
    age-specific death rates in Rwanda from three different organization.
    Sampling uncertainty for Acquaintance, Meal, and Sibling estimates are
    shown in Figure~\ref{fig:net-asdr-all}. 
    Estimates from the WHO, UNPD, and IHME are model-based, so no comparable
    sampling-based uncertainty estimates are available.
  }
  \label{fig:ref-asdr-all}
\end{figure}

Although estimates from organizations like the WHO, UNPD, and IHME are
typically used to compare aggregate metrics of adult mortality like $\ffqf$ 
across countries, the organizations also produce age-specific death rate estimates.
Figure~\ref{fig:ref-asdr-all} shows the estimated age-specific death rates from the two network
survival estimates, the sibling survival estimates, and the age specific estimates
for each organization.

\clearpage

\section{Issues related to the frame population}
\label{ap:respondent-age}

The frame population in our study (i.e., the set of people eligible to be
interviewed) was all people age 15 and over.  Some other surveys in developing
countries, however, have different frame populations.  For example, the frame
populations in the Demographic and Health Surveys is typically women between 15
and 49 and men between 15 and 59.  The difference
between the frame population in our study and the frame populations typically used in the
Demographic and Health Surveys naturally raises questions about the ability to
embed the network reporting method as a module in other studies. Therefore, in
this appendix we describe some of the analytic and practical issues raised by
the choice of the frame population.  We also artificially truncate our
sample to match the Rwanda DHS respondents' age range (i.e., females 15-49 and males 15-59)
and show that this truncation makes very little difference in our estimate of $\tfqf$.  
Further, in Online Appendix~\ref{ap:quantity-and-quality}, we report descriptive plots
showing how the data we collected varied by the age and sex of respondents.

The network reporting identity (Eq.~\ref{eqn:text-id0}) is true for any frame
population.  When that identity is re-arranged as in
Eq.~\ref{eqn:nontext-intermsofu-id}, it reveals the key qualitative insight of
our approach: estimating the number of deaths from the number of reports of
deaths requires correctly adjusting for the visibility of deaths.  Thus, the key
issue with the network reporting method is estimating the visibility of deaths
to the frame population.  In this study, we used the average personal network
size of respondents in demographic group $\alpha$ as an estimate of the average
visibility of deaths in demographic group $\alpha$ to the frame population.
This exact approach is not possible if the frame population is more restricted;
for example, if the frame population was restricted to women between 15 and 49, we would not
have information to estimate the average personal network size of men between
15 and 29.

We see two different general approaches for the problem of estimating the
visibility of deaths when the frame population is not all people age 15 and
over.  First, researchers can make additional assumptions.  Researchers could,
for example, make assumptions about the relationship between the personal
network size of men and women or between young people and old people.
(Naturally, researchers adopting this approach would need to assess the
sensitivity of their estimates to these assumptions.)  Second, researchers can
collect additional data to directly estimate the visibility of deaths to the
frame population.  In other words, if the frame population is women between 15
and 49, then researchers could collect information to estimate the visibility
of deaths to women between 15 and 49.  We see this second approach as more
promising and some ideas in this direction might be taken from the generalized
network scale-up method, which also involves two data collections~\citep{feehan_generalizing_2016}.  

\begin{figure} 
  \centering
  \includegraphics[width=0.5\textwidth]{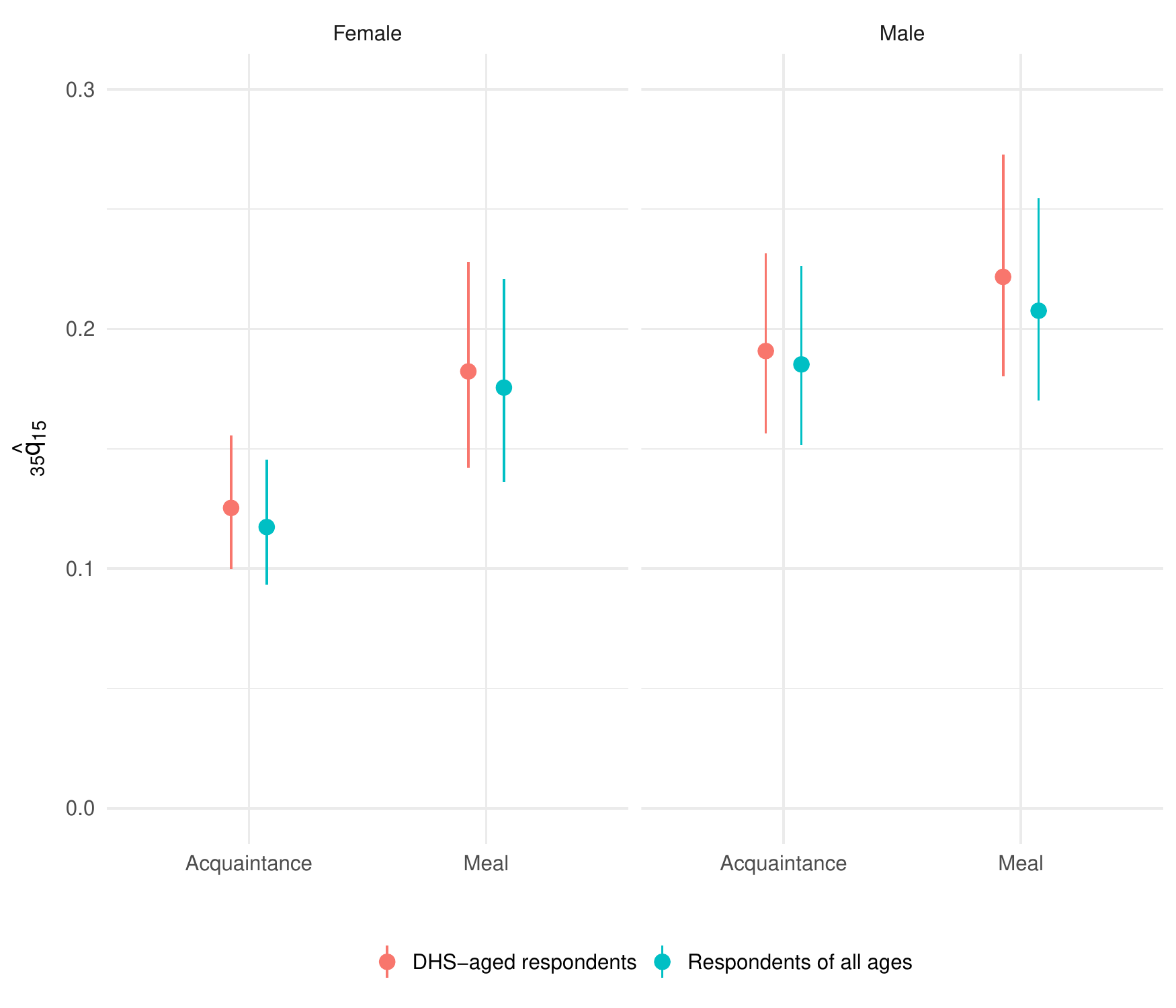}
  \caption{
    Comparison between network survival estimates of $\tfqf$ for males and for females
    using all respondents and using only DHS-aged respondents
    (women 15-49 and men 15-59).
  }
  \label{fig:tfqf-respage}
\end{figure}

Additionally, as a rough empirical check of how our results in this study might
have been impacted if we had a different frame population,  we artificially
truncate our sample to women between age 15 and 49 and men between ages 15 and
59 to match the frame population for the 2010 Rwanda DHS.  This procedure
First, Figure~\ref{fig:deaths-per-int-all} shows that the full sample and truncated
sample reported similar number of deaths per interview.  Second,
Figure~\ref{fig:tfqf-respage} shows that the full sample and the truncated
sample produce similar estimates of $\tfqf$.  Note that we estimated $\tfqf$
instead of $\ffqf$ because estimating $\ffqf$ requires information about the
visibility of deaths of people aged 50 to 65 and our study was not designed to
estimate this quantity using only the subset of respondents under age 50.

\begin{figure} 
  \centering
  \includegraphics[width=0.5\textwidth]{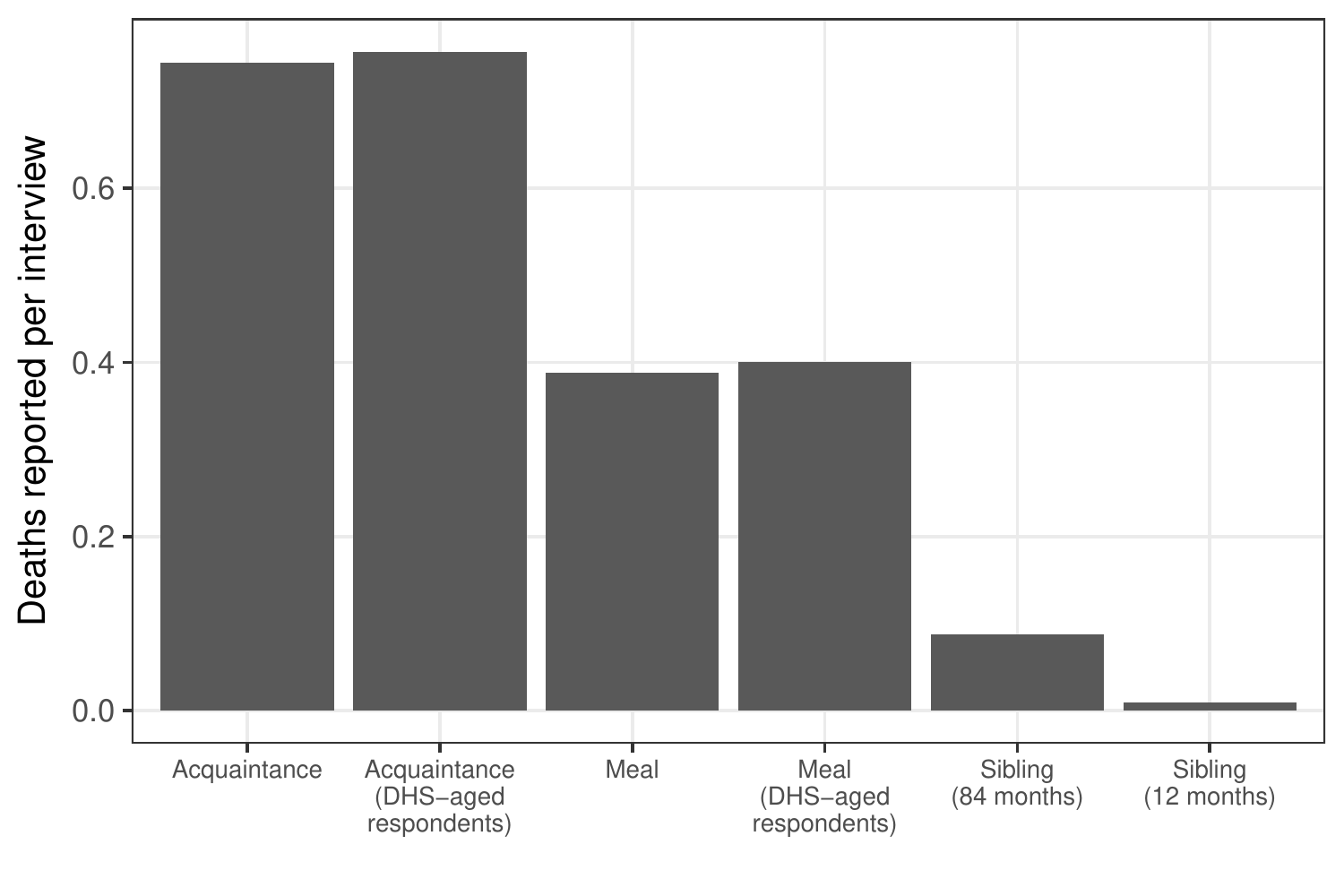}
  \caption{
    Average number of deaths reported from each interview in Rwanda using the
    acquaintance and meal tie definitions from the network survival study, and
    using the sibling history module of the DHS survey. 
    Results from the network survival study are shown for all respondents, and
    for DHS-aged respondents (women 15-49 and men 15-59).}
  \label{fig:deaths-per-int-all}
\end{figure}

\begin{figure} 
  \centering
  \includegraphics[width=\textwidth,keepaspectratio]{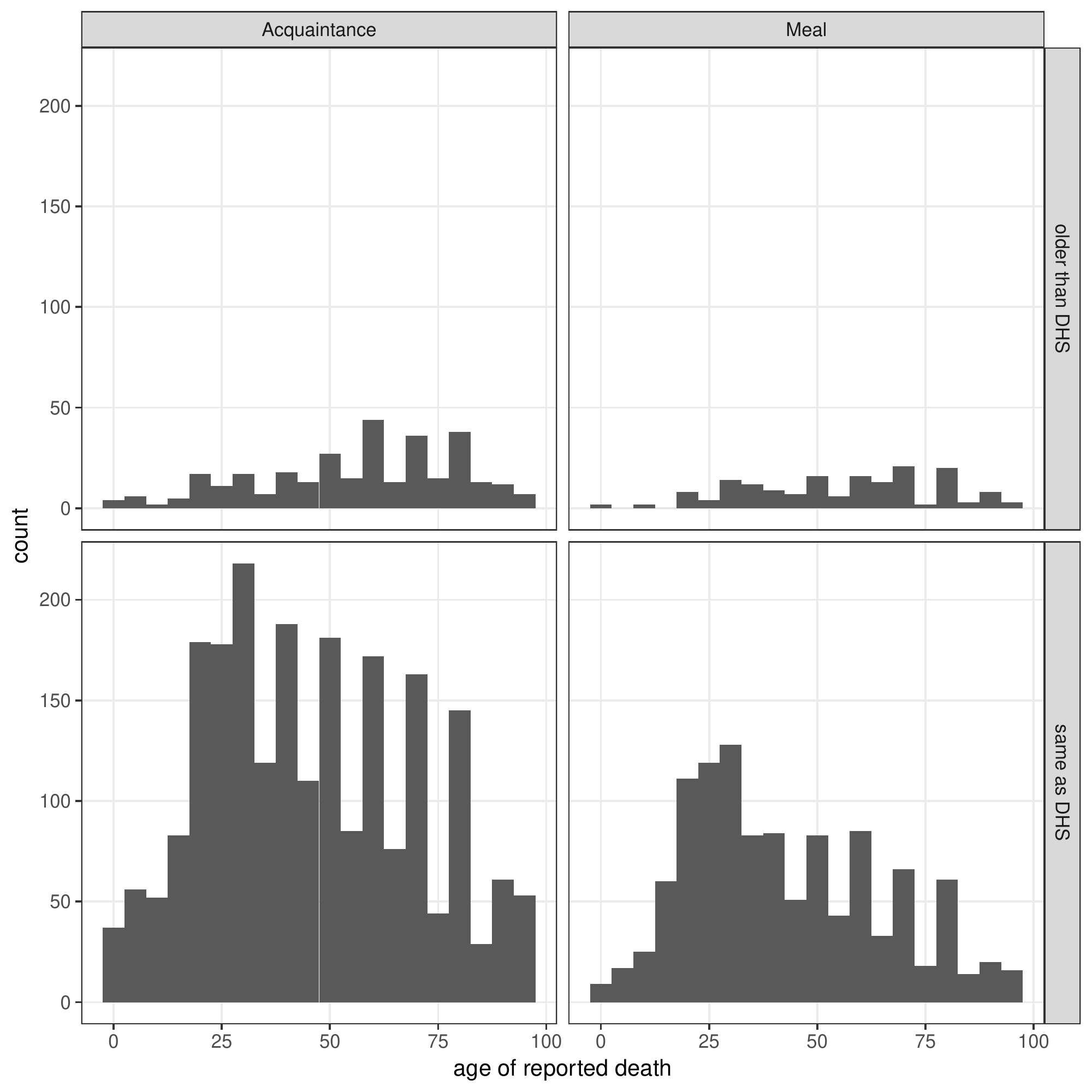}
  \caption{
	  Distribution of the ages of reported deaths by tie definition and
	  by whether or not respondents are in the age ranges typical of DHS
	  surveys (females 15-49 and males 15-59).
	  Bins have width 5 years; this figure does not use the sampling weights.
  }
  \label{fig:deaths-by-dhsage}
\end{figure}

\begin{figure} 
  \centering
  \includegraphics[width=\textwidth,keepaspectratio]{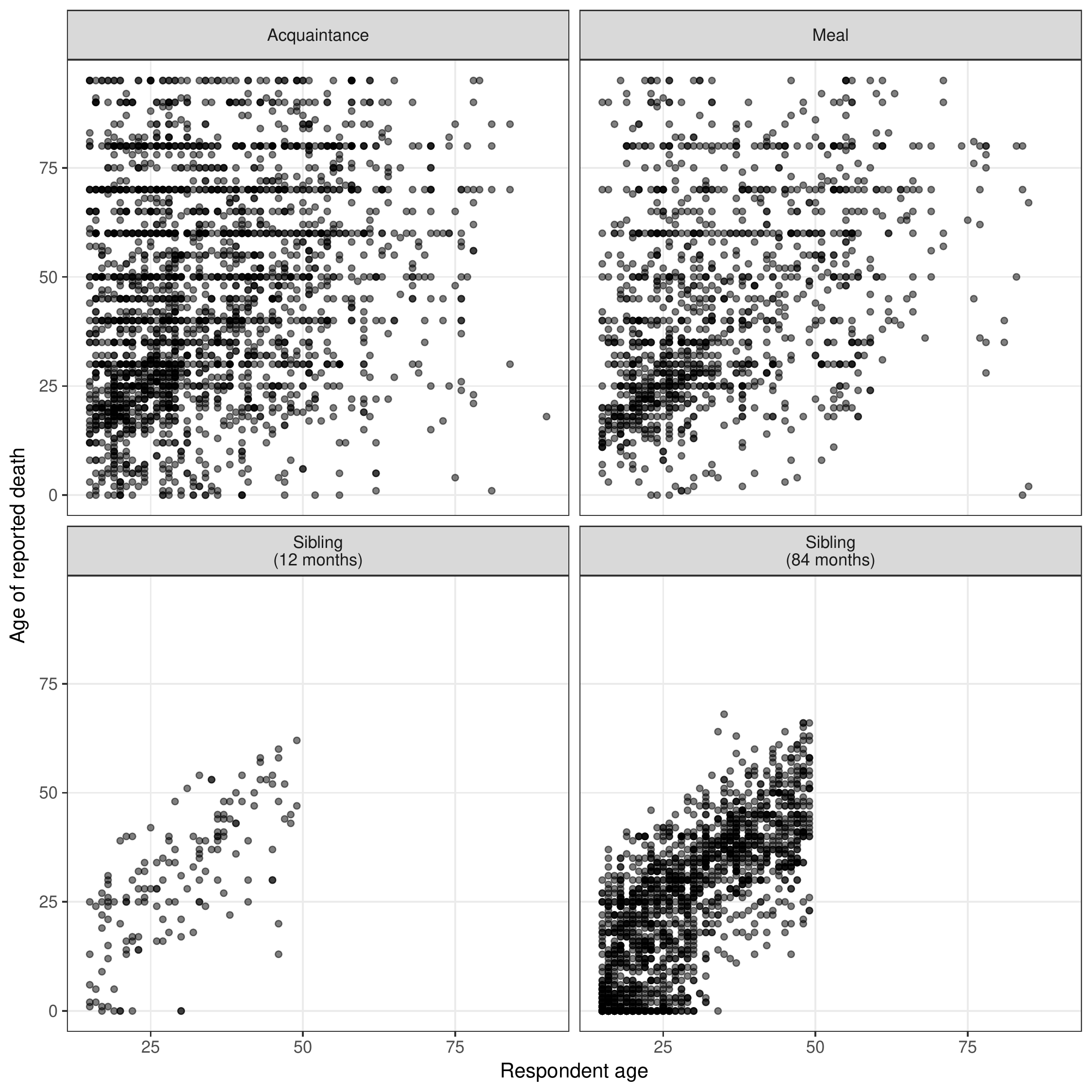}
  \caption{
	  Age of reported death versus age of survey respondent for
	  the acquaintance and meal tie definitions in our network survey, 
	  and from the sibling history of the DHS.
	  There is one point for each reported death, so survey respondents
	  who report more than one death contribute more than one point
	  to the plot.
	  The Rwanda DHS only asked the sibling histories of women,
	  so respondents for the sibling method are all under 50.
  }
  \label{fig:deathage-by-egoage}
\end{figure}

Finally, as suggested by a reviewer, we investigate the relationship between
the age of the reported deaths and the age of the respondents who reported
them.
Figure~\ref{fig:deaths-by-dhsage} shows the age distribution of reported deaths 
by the age range of respondents;
further, Table~\ref{tab:numreporteddeaths-bydeathage} shows the number of reported deaths
by tie definition, respondent age range, and death age range.
Network survival respondents who are the same age as DHS respondents
report deaths among people over 50 about one third of the time
(meal: 0.33, acquaintance: 0.38);
network survival respondents who are older than DHS respondents
report deaths among people over 50 just under two-thirds of the time
(meal: 0.57, acquaintance: 0.62).
Figure~\ref{fig:deathage-by-egoage} shows the relationship between the age
of the survey respondent and the age of the reported death, for all of the
deaths reported using both tie definitions in our survey, and using the
DHS sibling histories.
Three main conclusions emerge from Figure~\ref{fig:deaths-by-dhsage},
Table~\ref{tab:numreporteddeaths-bydeathage}, and
Figure~\ref{fig:deathage-by-egoage}:
(1) deaths over age 50 are reported both by network survival respondents who are in age ranges
typically interviewed by the DHS, and also by network survival respondents who are older than
typical DHS interviewees; (2), network survival respondents who are older than typical 
DHS interviewees report a greater fraction of deaths over age 50 than 
network survival respondents in typical DHS age ranges;
and (3), using the meal and acquaintance tie definitions, network survival 
respondents of a given age appear to report deaths across a wider range of ages
than sibling survival respondents.

\begin{table}[!htbp]  
  \caption{Number of deaths reported in Rwanda using the acquaintance and meal tie definitions from the network survival study, by age range of respondent and age of reported death.} 
  \label{tab:numreporteddeaths-bydeathage} 
\scriptsize 
\begin{tabular}{@{\extracolsep{5pt}} cccc} 
\\[-1.8ex]\hline \\[-1.8ex] 
Tie definition & Respondent age & Reported death age & Num. reported deaths \\ 
\hline \\[-1.8ex] 
Acquaintance & older than DHS & death \textless 50 & $123$ \\ 
Acquaintance & older than DHS & death 50+ & $197$ \\ 
Acquaintance & same as DHS & death \textless 50 & $1,375$ \\ 
Acquaintance & same as DHS & death 50+ & $854$ \\ 
Meal & older than DHS & death \textless 50 & $71$ \\ 
Meal & older than DHS & death 50+ & $95$ \\ 
Meal & same as DHS & death \textless 50 & $753$ \\ 
Meal & same as DHS & death 50+ & $373$ \\ 
\hline \\[-1.8ex] 
\end{tabular} 
\end{table} 

\begin{figure} 
  \centering
  \includegraphics[width=0.5\textwidth]{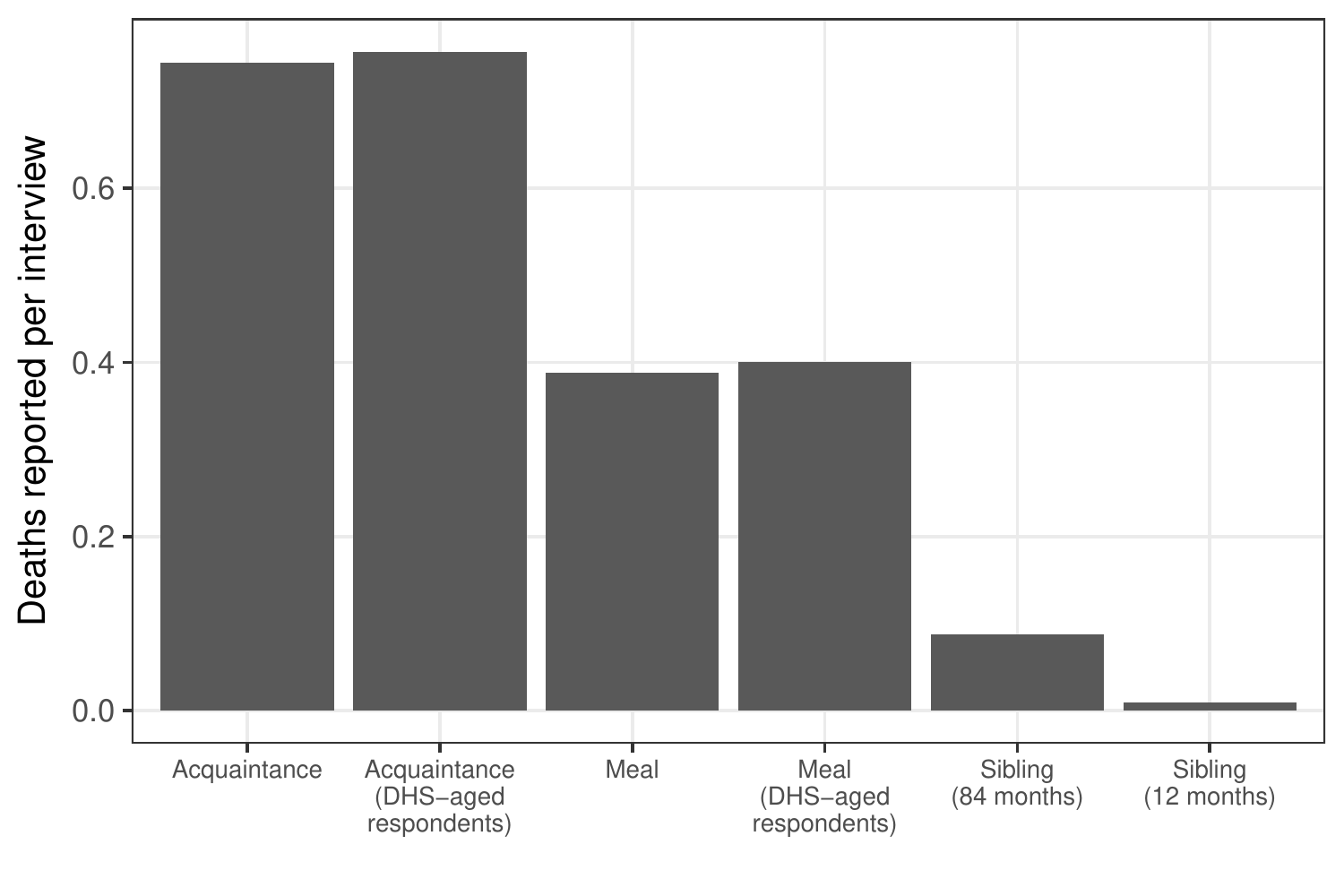}
  \caption{
    Average number of deaths reported from each interview in Rwanda using the
    acquaintance and meal tie definitions from the network survival study, and
    using the sibling history module of the DHS survey. 
    Results from the network survival study are shown for all respondents, and
    for DHS-aged respondents (women 15-49 and men 15-59).}
  \label{fig:deaths-per-int-all}
\end{figure}

In conclusion, the network reporting method can be used for any frame
population, but researchers using a frame population other than all adults
would need to make some slight modifications from the approach taken in this
paper.  We think that this represents an important area for future research.

\FloatBarrier

\section{Descriptive plots}
\label{ap:quantity-and-quality}

This appendix provides additional descriptive plots related to the network
reporting method and the sibling survival method.  In particular, we include
plots related to reports about deaths in both methods
(Sec.~\ref{sec:reports-deaths}) and reports of connections to groups of known
size in the network reporting method (Sec.~\ref{sec:reports-known}).

\subsection{Reports about deaths}
\label{sec:reports-deaths}

\begin{figure} 
  \centering
  \includegraphics[width=\textwidth,keepaspectratio]{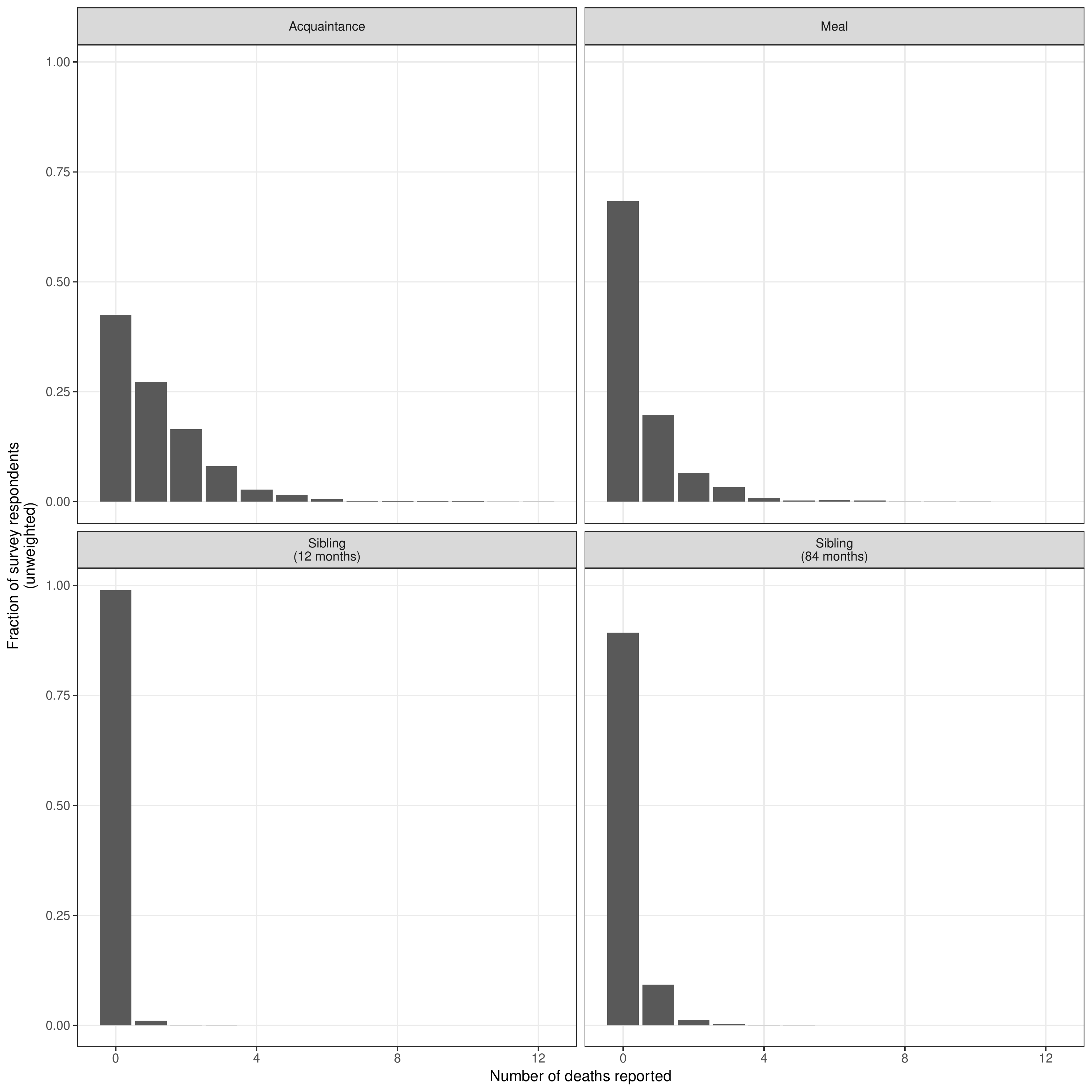}
  \caption{
      Distribution of the number of deaths reported by survey respondents to
      both types of personal network, and to the sibling histories using two time
      windows (12 months and 84 months). 
      Each panel shows the unweighted fraction of respondents who reported each possible
      number of deaths.
  }
  \label{fig:all-reported-deaths-distn}
\end{figure}

Figure~\ref{fig:all-reported-deaths-distn} shows the distribution of the number
of deaths reported by each survey respondent.  Two main findings emerge from
this plot: 1) as reported in the main paper, the network reporting method (both
tie definitions) collects more deaths per interview than the sibling method,
even when the sibling reports are taken over a 7 year time period; 2) in all
cases, the distributions seem to vary smoothly suggesting that the higher
number of reports in the network survival method are not driven by a small
number of extreme outliers. 

\begin{figure} 
  \centering
  \includegraphics[width=\textwidth,keepaspectratio]{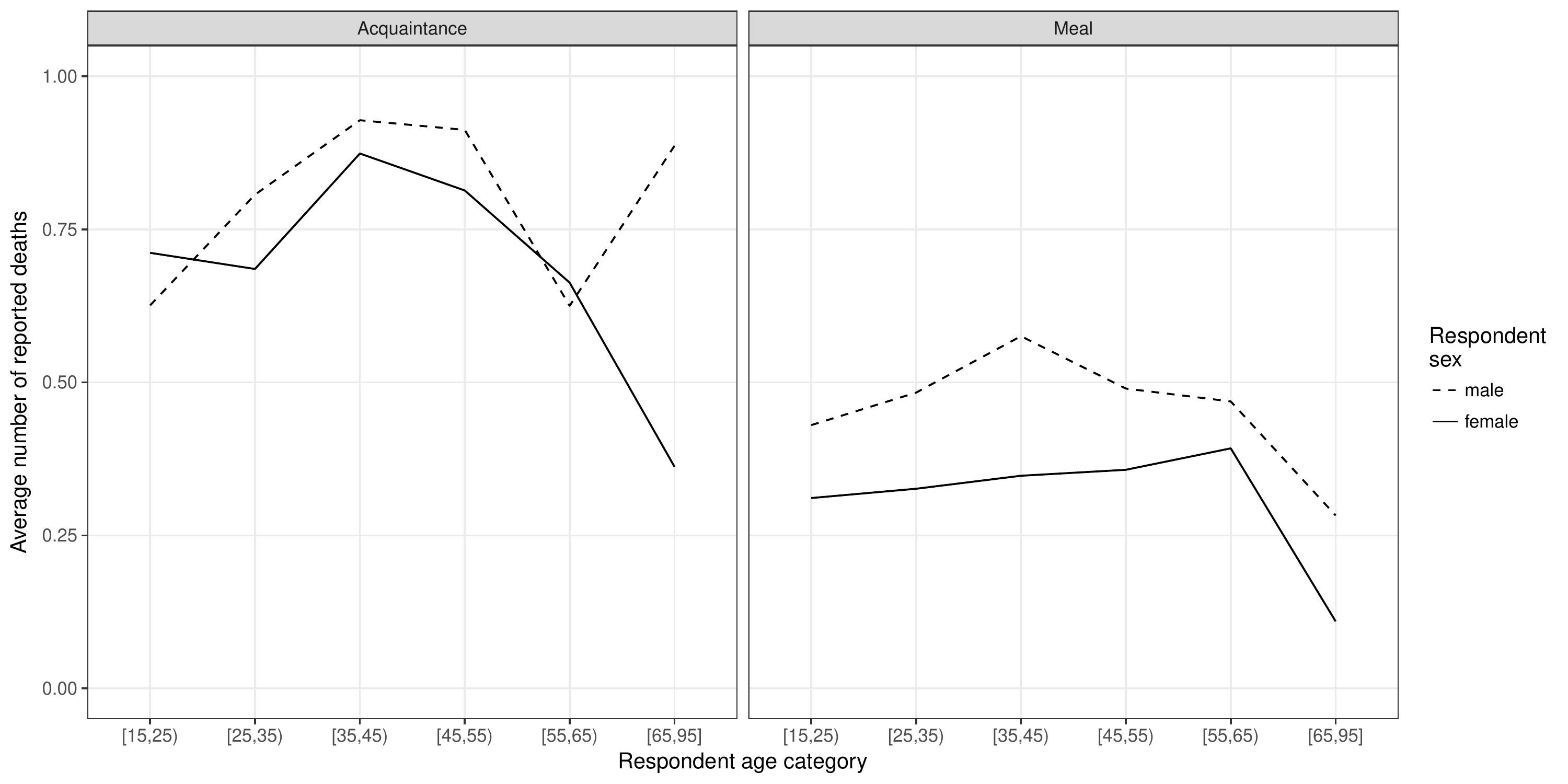}
  \caption{
    Average number of adult deaths reported for each tie definition, by age and sex of 
    survey respondents. 
  }
  \label{fig:deaths-byrespage}
\end{figure}

\begin{figure} 
  \centering
  \includegraphics[width=\textwidth,keepaspectratio]{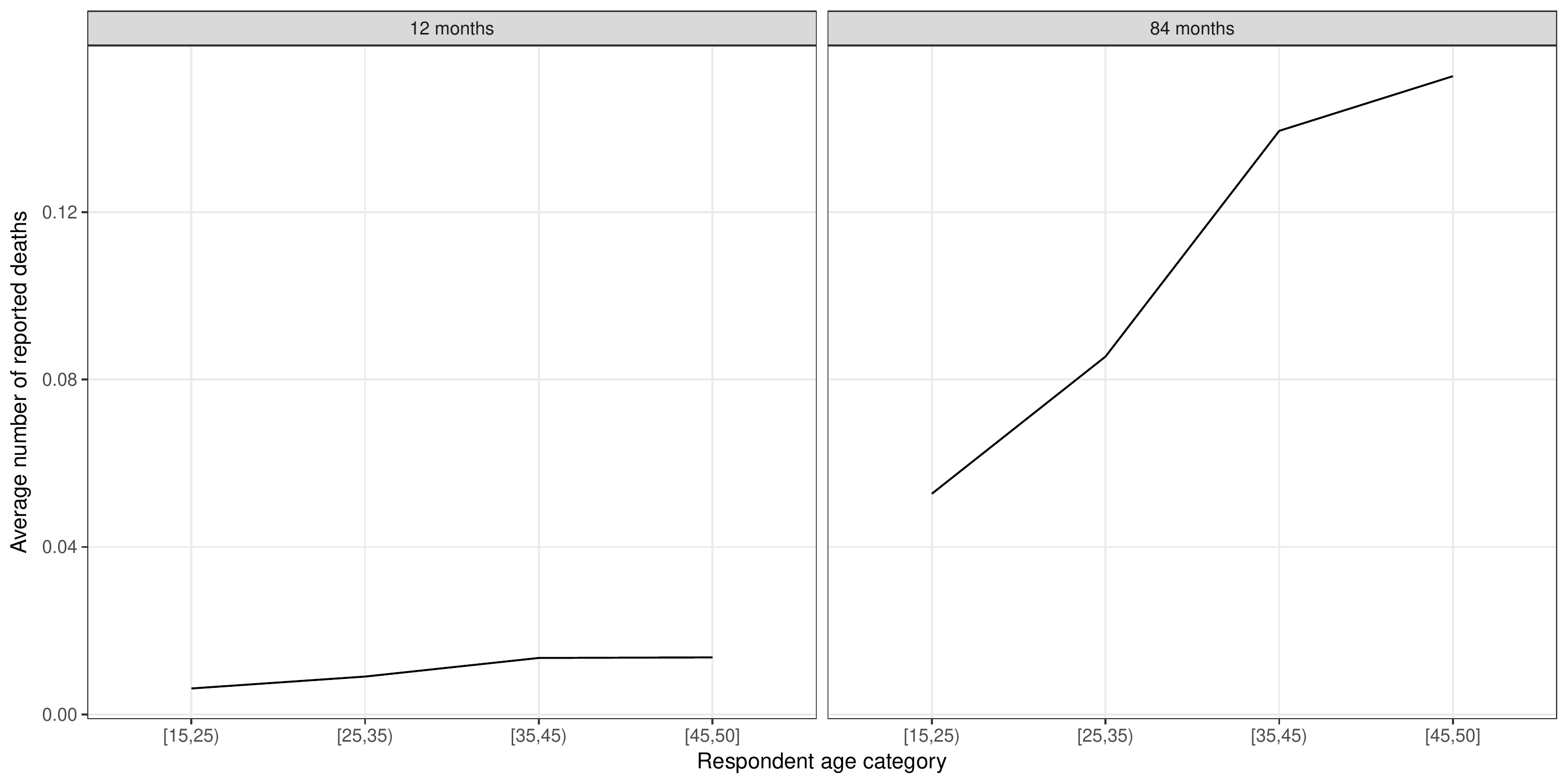}
  \caption{
    Average number of adult deaths reported for 12 months before the interview (left panel) and
    for 84 months before the interview (right panel), by age of women responding to the DHS
    sibling histories.
    Note that the last age group ends at 50, since
    the DHS only asked the sibling history module of women up to age 50.
  }
  \label{fig:sib-deaths-byrespage}
\end{figure}

Further, as described in Online Appendix~\ref{ap:respondent-age}, future
studies might use a frame population more restricted than our frame population
of all adults.  Therefore, Figure~\ref{fig:deaths-byrespage} shows the average
number of adult deaths reported by the age and sex of survey respondents.  Two
observations emerge from this figure: first, for the acquaintance network,
there appears to be a U-shaped relationship between respondent age and the
average number of deaths reported. Second, for both tie definitions, males
appear to report more deaths, on average, then females. 
Figure~\ref{fig:sib-deaths-byrespage} shows the average number of adult
deaths reported by age of women who responded to the DHS sibling history
module.  The main observation to emerge from this figure is that the number
of sibling deaths reported appears to increase with respondent age.
Taken together, one possible explanation for the difference between the
reporting patterns in sibling networks (Figure~\ref{fig:sib-deaths-byrespage})
and the reporting patterns in meal and acquaintance networks
(Figure~\ref{fig:deaths-byrespage}) is that siblings tend to be more similar to
respondents in terms of age than acquaintances or meal partners.

\begin{figure} 
  \centering
  \includegraphics[width=\textwidth,keepaspectratio]{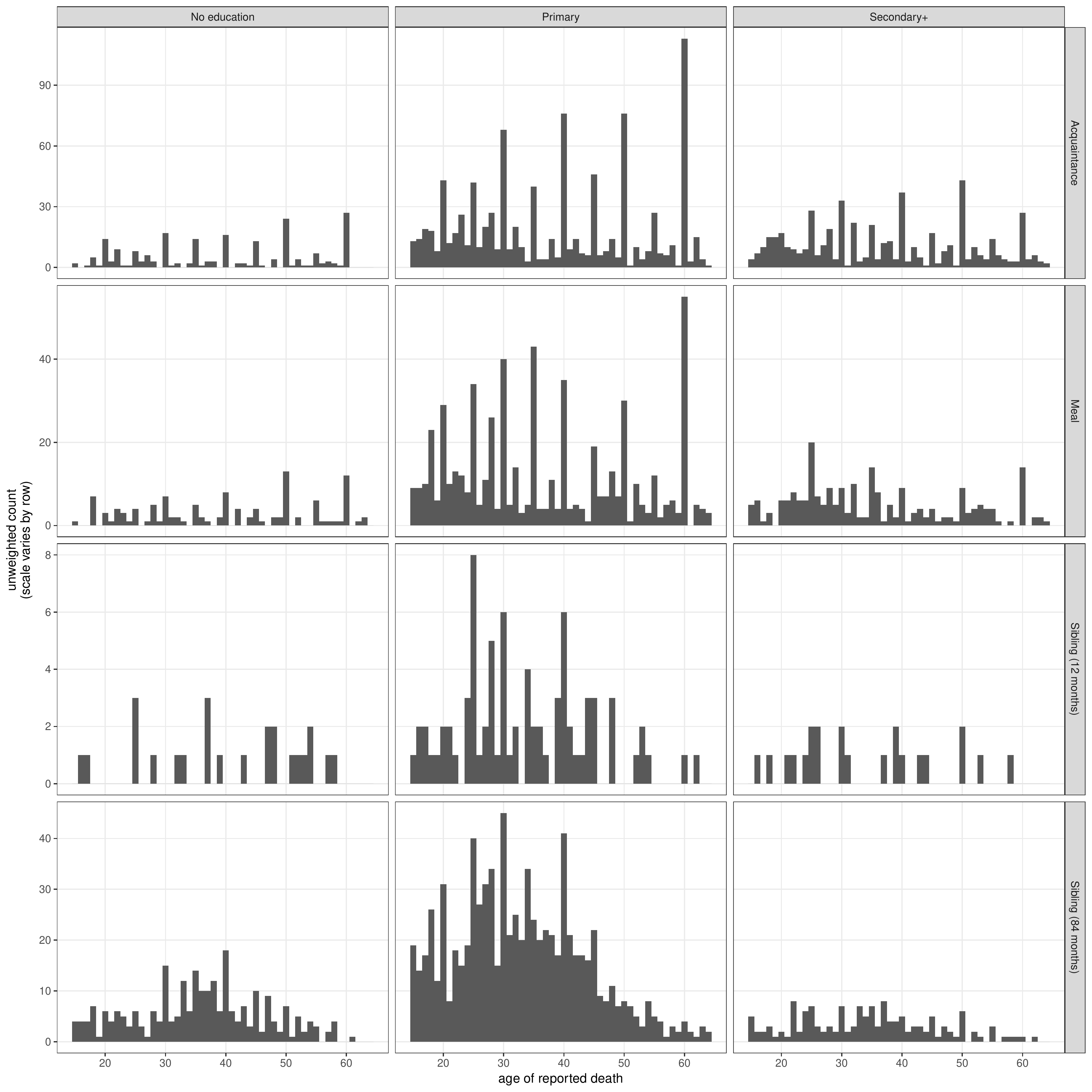}
  \caption{
      Distribution of the ages of reported deaths by single year of age from
      the two personal networks, from sibling reports 12 months prior to the survey,
      and from sibling reports 84 months prior to the survey (rows), and by
      education of survey respondent (columns).
      Note that the scale varies by row, since the total number of deaths reported
      varies considerably between the different tie definitions.
  }
  \label{fig:all-heaping}
\end{figure}

Additionally, Figure~\ref{fig:all-heaping} shows the distribution of the ages of reported deaths
from the two personal networks and from the sibling reports for two different time
periods as a function of respondent level of education. Several observations emerge from this plot: 
first, reports appear to be more heaped for less educated respondents; 
second, there appears to be considerably more heaping for the network reports, when compared to
the sibling reports over an 84 month time period. 
The small number of deaths for the sibling reports over a 12 month time period
make it very hard to draw any conclusions.

\begin{figure} 
  \centering
  \includegraphics[width=\textwidth,keepaspectratio]{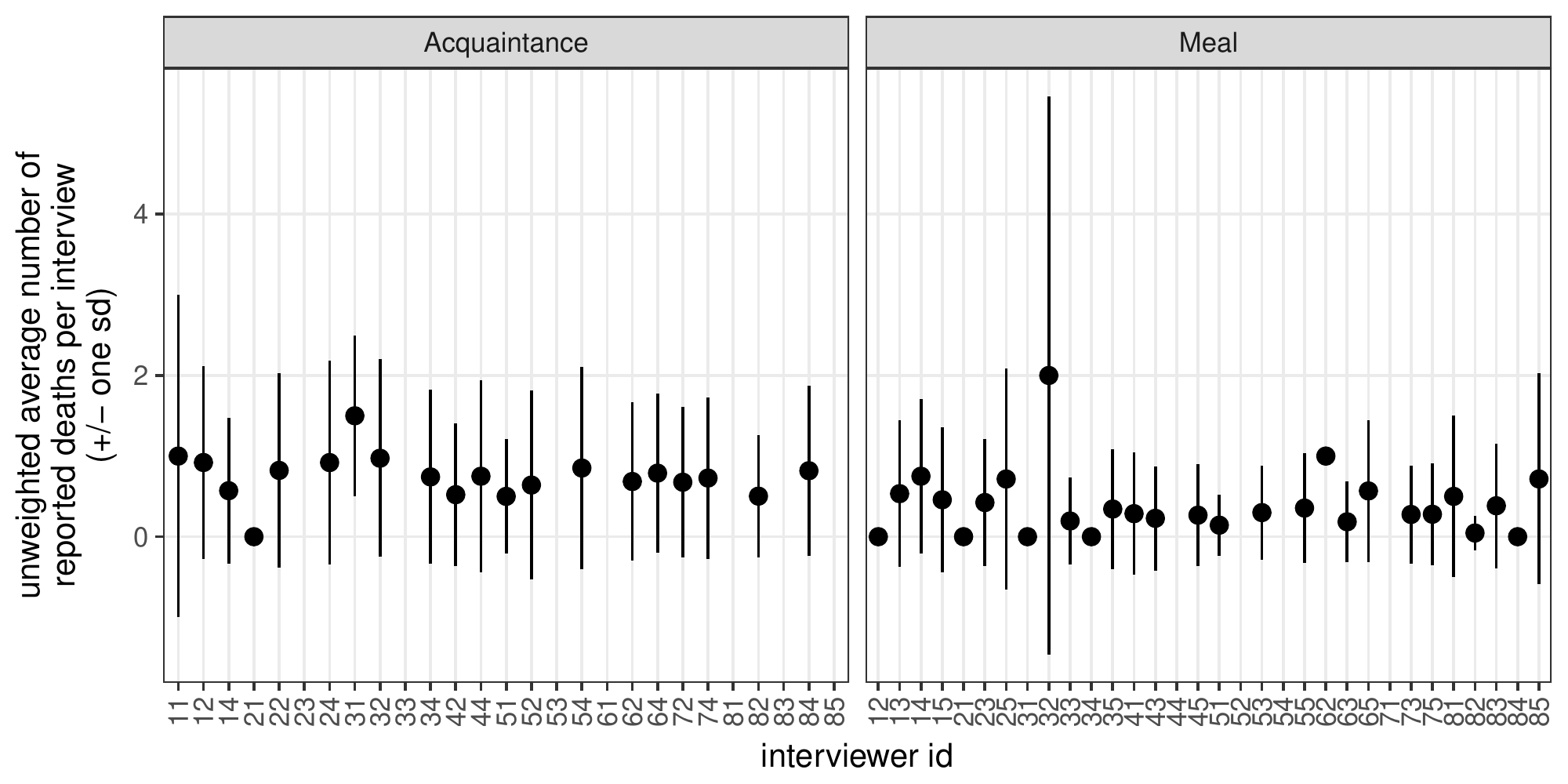}
  \caption{
      Average (+/- one s.d.) in the number of reported deaths per interview, by interviewer
      and by tie definition for the two personal networks.
      Note that interviewer id 32 only conducted 3 interviews using the meal definition,
      which may explain the large standard deviation around that observation.
  }
  \label{fig:net-interviewer}
\end{figure}

\begin{figure} 
  \centering
  \includegraphics[height=\textheight,keepaspectratio]{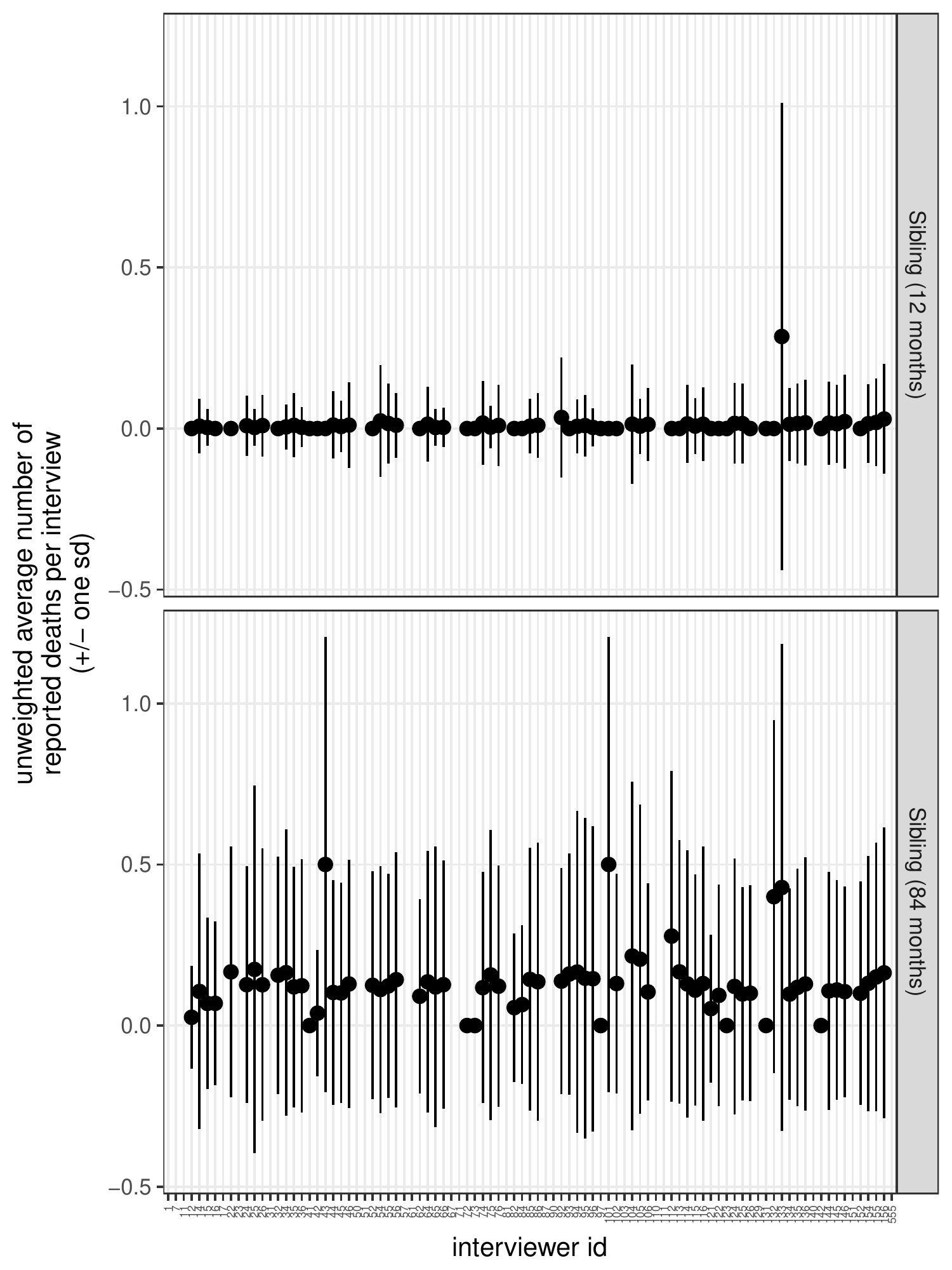}
  \caption{
      Average (+/- one s.d.) in the number of reported deaths per interview, by interviewer
      and by length of reporting interval for deaths from the Rwanda DHS sibling histories.
      Note that some interviewers conducted very few interviews, which may explain wide standard
      deviations in reports for interviewer id 43 (2 interviews), id 101 (2 interviews), and
      id 132 (5 interviews).
  }
  \label{fig:sib-interviewer}
\end{figure}

Finally, in order to explore whether the sibling survival method and the
network survival method could be impacted by interviewer effects, we plot the
number of reported deaths by interviewer.  Figure~\ref{fig:net-interviewer}
shows the average number of reported deaths per interview by interviewer and by
tie definition from our study.  And, similarly,
Figure~\ref{fig:sib-interviewer} shows the average number of reported deaths
per interview by interviewer and by time window for deaths from the 2010 Rwanda
DHS sibling histories.  These figures do not show strong evidence of
interviewer effects, but neither our survey nor the DHS were specifically
designed to measure possible interviewer effects.  We hope that this topic will
be studied in future research.

\subsection{Connections to groups of known size}
\label{sec:reports-known}

\begin{figure} 
    \thisfloatpagestyle{empty}
  \centering
  \includegraphics[width=\textwidth,keepaspectratio]{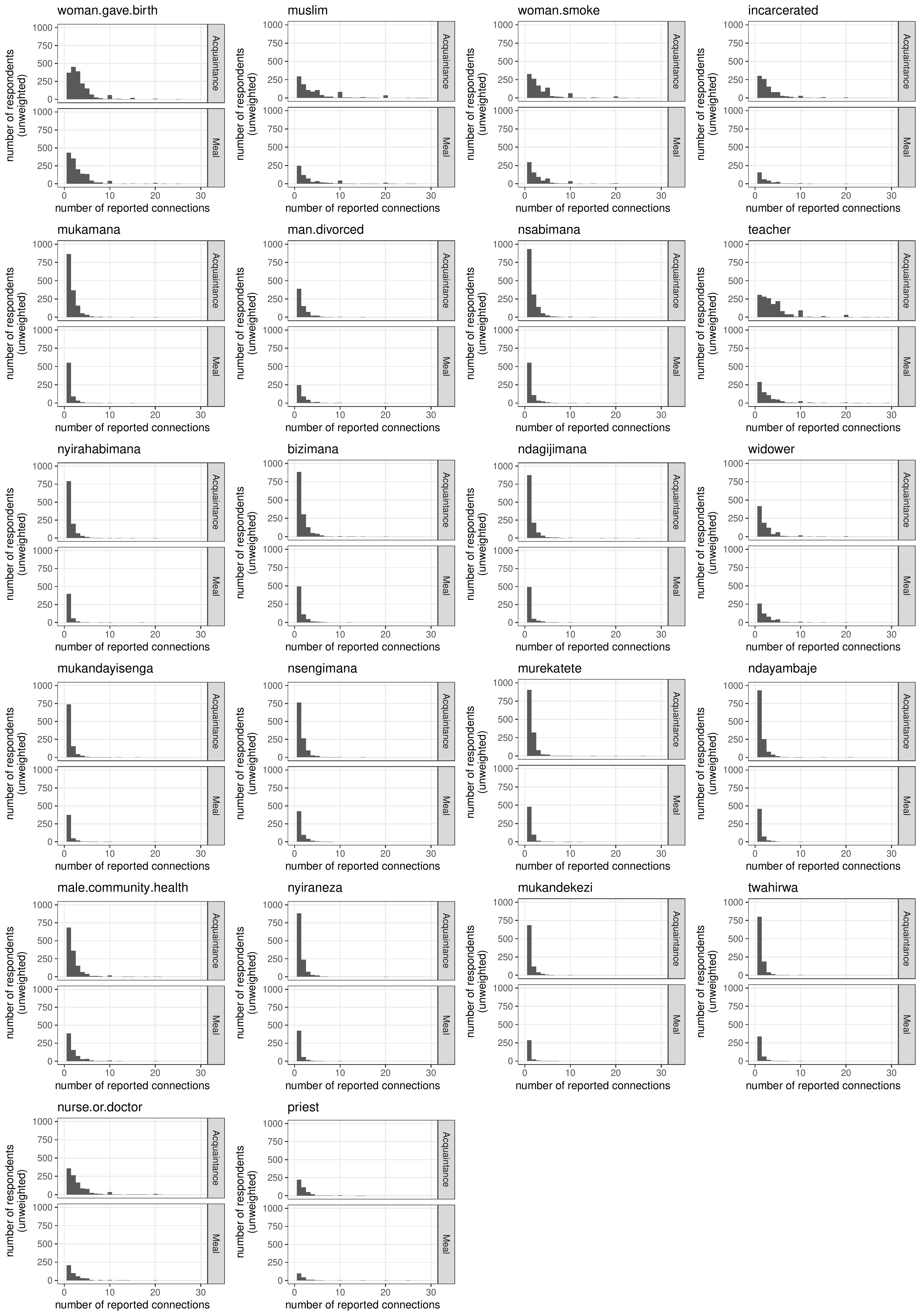}
  \caption{
      Distribution of the number of reported connections to each group of known size
      for the meal and acquaintance networks. Panels are sorted so that the largest
      known population is at the top-left and the smallest is on the bottom-right.
  }
  \label{fig:kp-marginal-hist}
\end{figure}

\begin{figure} 
    \thisfloatpagestyle{empty}
  \centering
  \includegraphics[width=\textwidth,keepaspectratio]{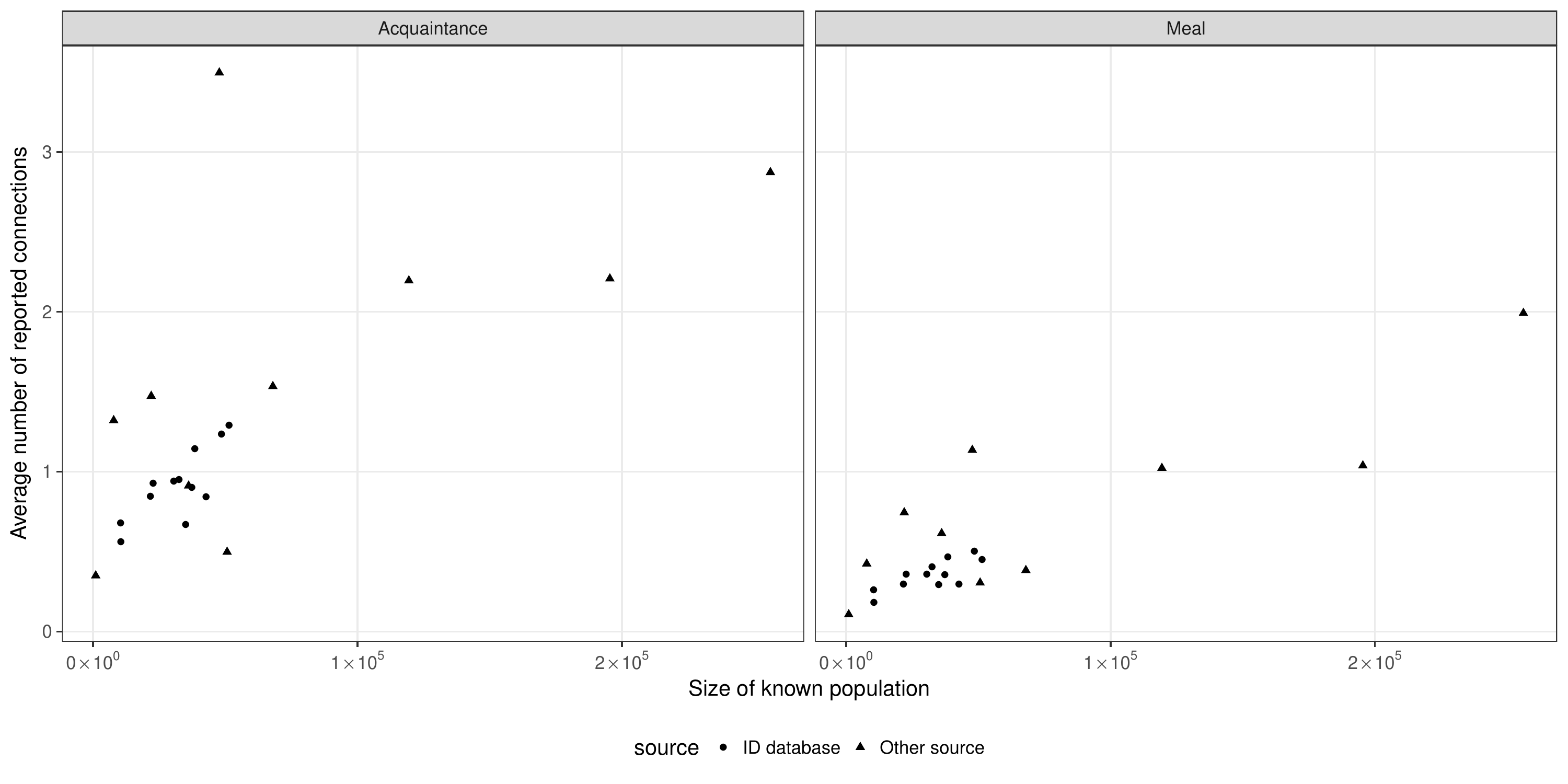}
  \caption{
      Average number of connections reported by survey respondents using the
      acquaintance network (left panel) and the meal network (right panel)
      versus the size of each known population. For both tie definitions, there
      is a strong positive relationship between the average reported
      connections and the size of known populations.
  }
  \label{fig:kp-avg-reported}
\end{figure}

The network survival method (as we operationalized it in this study) asked
respondents about their connections to groups of known size in order to
estimate their personal network size.
Figure~\ref{fig:kp-marginal-hist} shows the distribution of the number
of reported connections to each group of known size; and
Figure~\ref{fig:kp-avg-reported} and Table~\ref{tab:kp-truth} show
the relationship between the average number of  reported connections to each
known population and the size of each known population.  As expected,
respondents report more connections to larger groups, a common pattern in
studies using the network scale-up method. The correlation between the
average number of reported connections and the total size of the known populations
is 0.66 for the Acquaintance tie definition and 0.86 for Meal tie definition.
For the Acquaintance network results, Figure~\ref{fig:kp-avg-reported} shows that one
group (teachers, 3.5 average reported connections) appears to fall well above the 
pattern set by the remaining known populations.
We cannot say what causes this deviation, but one possibility is that teachers have
larger acquaintance networks than the average Rwandan. 

\begin{table}[!htbp]  
  \caption{Average number of reported connections and known group size for each of the known populations.} 
  \label{tab:kp-truth} 
\scriptsize 
\begin{tabular}{@{\extracolsep{5pt}} lccc} 
\\[-1.8ex]\hline \\[-1.8ex] 
Group & Total size & Avg. Connections (Acquaintance) & Avg. Connections (Meal) \\ 
\hline \\[-1.8ex] 
Priest & $1,004$ & $0.35$ & $0.11$ \\ 
Nurse or doctor & $7,807$ & $1.32$ & $0.42$ \\ 
Twahirwa & $10,420$ & $0.68$ & $0.26$ \\ 
Mukandekezi & $10,520$ & $0.56$ & $0.18$ \\ 
Nyiraneza & $21,705$ & $0.85$ & $0.30$ \\ 
Male community health worker & $22,000$ & $1.47$ & $0.74$ \\ 
Ndayambaje & $22,724$ & $0.93$ & $0.36$ \\ 
Murekatete & $30,531$ & $0.94$ & $0.36$ \\ 
Nsengimana & $32,528$ & $0.95$ & $0.40$ \\ 
Mukandayisenga & $35,055$ & $0.67$ & $0.29$ \\ 
Widower & $36,147$ & $0.91$ & $0.61$ \\ 
Ndagijimana & $37,375$ & $0.90$ & $0.36$ \\ 
Bizimana & $38,497$ & $1.14$ & $0.46$ \\ 
Nyirahabimana & $42,727$ & $0.84$ & $0.30$ \\ 
Teacher & $47,745$ & $3.50$ & $1.14$ \\ 
Nsabimana & $48,560$ & $1.23$ & $0.50$ \\ 
Divorced man & $50,698$ & $0.50$ & $0.31$ \\ 
Mukamana & $51,449$ & $1.29$ & $0.45$ \\ 
Incarcerated & $68,000$ & $1.53$ & $0.38$ \\ 
Woman who smokes & $119,438$ & $2.20$ & $1.02$ \\ 
Muslim & $195,449$ & $2.21$ & $1.04$ \\ 
Woman who gave birth last 12 mo. & $256,164$ & $2.87$ & $1.99$ \\ 
\hline \\[-1.8ex] 
\end{tabular} 
\end{table} %

\begin{figure} 
  \centering
  \includegraphics[width=\textwidth,keepaspectratio]{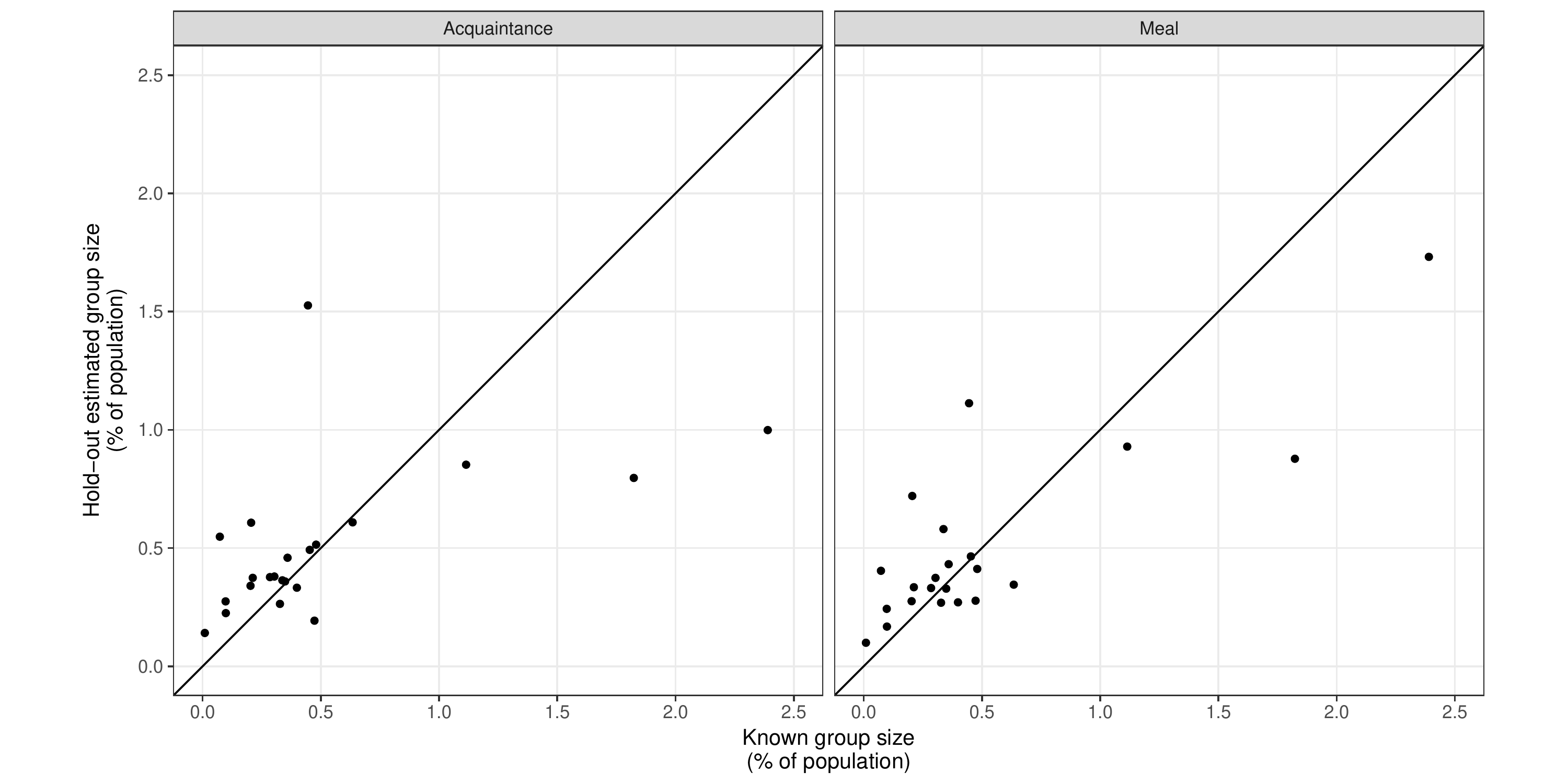}
  \caption{
	Results of internal consistency checks for the acquaintance and meal
	tie definitions in Rwanda. 
	Each point in the plot represents a single known population.
	Taking divorced men as an example, the hold-out estimate is
	calculated by (1) estimating personal network size using all
	known populations \emph{except} divorced men;
	(2) using number of reported connections to divorced men
	together with the hold-out estimates of personal network size
	to estimate the number of divorced men;
	and (3) comparing the hold-out estimate for the number of divorced
	men to the known size of that group.
	This exercise is repeated once for each group of known size, and for
	each tie definition.
	If these hold-out estimates were perfectly accurate, then all of the points
	in the two panels would lie along the diagonal lines.
  }
  \label{fig:ic-checks}
\end{figure}

Figure~\ref{fig:ic-checks} shows the results of internal consistency checks
that provide further evidence about the plausibility of the reported
connections to groups of known size. 
These internal consistency checks are based on taking each
known population, pretending its size is not known, estimating network
size using the remaining known populations, and then using those
estimated network sizes to predict the size of the held-out known population
(see~\citet{feehan_quantity_2016} for more details).
Almost all of the hold-out estimates shown in Figure~\ref{fig:ic-checks}
lie close to the diagonal line, suggesting that reported connections to
the groups of known size are internally consistent;
however, two groups (women who gave birth in the past 12 months and Muslims)
both of appear to be underestimated in the hold-out checks.

\begin{figure} 
  \centering
  \includegraphics[width=.85\textwidth,keepaspectratio]{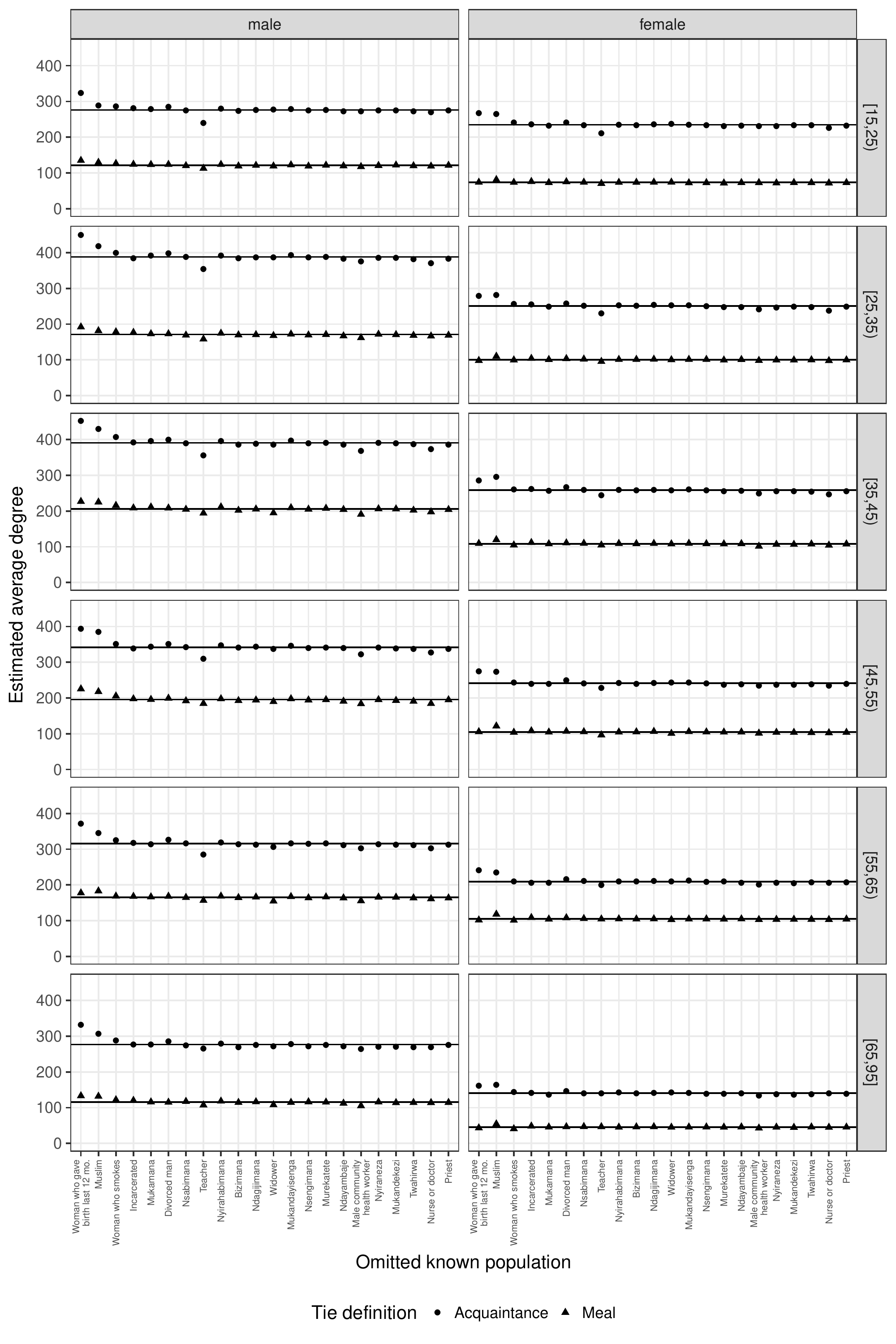}
  \caption{
	Impact of each known population on estimating average personal network size,
	by sex, age group, and tie definition.
	The horizontal line shows the estimated average personal network size using 
	all of the known populations, and each point shows the estimated personal network
	size calculated using all of the known populations except for the one listed
	on the x axis.
	The distance between each point and the horizontal line shows how different the
	estimated personal network size would be if the corresponding known population
	was not used.
	The groups are shown on the x axis in order of their total size
	from largest to smallest.
  }
  \label{fig:loo-cell-degree-estimates}
\end{figure}

Finally, Figure~\ref{fig:loo-cell-degree-estimates} plots, for each age group, sex,
and tie definition, how the estimated average personal network size would change
if each known population was not used. Figure~\ref{fig:loo-cell-degree-estimates} shows
that estimated average personal network size appears not to be dramatically 
affected by the decision to include any particular group of known size.
To be clear, we consider Figure~\ref{fig:loo-cell-degree-estimates} to be a
heuristically useful diagnostic plot.
However, it is important to note that a desirable set of known populations is
one that satisfies the conditions required by the adapted known population
estimator (Result~\ref{res:adapted-kp}).  Such a set of known populations could
include individual groups whose removal appreciably impacts estimated average
personal network size.

\FloatBarrier

\newpage

\section{Network survival survey instrument}
\label{ap:ns-instrument}

In this appendix, we reproduce an excerpt of the English translation of the survey instrument
that we used for the meal tie definition, and we comment on its
design. All of the survey materials---including the original Kinyarwanda
instruments for both the meal and tie definition, as well as their English
translations---are freely available from the DHS website\footnote{
    \url{http://dhsprogram.com/what-we-do/survey/survey-display-422.cfm}
}.

We had to pay careful attention to constructing the wording of the question
that asked respondents to report about deaths (Q226). Both tie definitions in our
study were based on interactions (Table~\ref{tab:tiedefns})---either
contact (for the acquaintance definition) or sharing a meal or drink (for the
meal definition). Of course, people who have died cannot continue to interact
with others.  Therefore, in this section, we generalize the framework introduced in the main text to 
account for tie definitions where people's degree could change daily (e.g., tie definitions that are based on interactions).  Without loss of generality, we will consider the meal definition.

When asking respondents about connections to people in the groups of known size, we ask about people who the respondent has shared a meal with in the 12 months before the interview.  When asking about people who have died, we asked about people where: (i) the person died in
the 12 months before the interview; and (ii) the person shared a meal with the
respondent \emph{in the 12 months before death} (see Q226).  In this situation, the decedent network condition needs to be generalized into the
\emph{dynamic decedent network condition}. 

The decedent network condition discussed in main text and in Result~\ref{res:vbar-oalpha-f} says that:
\begin{equation}
  \label{eqn:static-condition}
  \bar{d}_{D_\alpha, F} = \bar{d}_{F_\alpha, F},
\end{equation}

\noindent where  $\bar{d}_{D_\alpha, F}$ is the average degree of people who have died in group $\alpha$ and $\bar{d}_{F_\alpha, F}$ is the average degree of frame population members in group $\alpha$.

The analogous dynamic decedent network condition says that:
\begin{equation}
    \label{eqn:dynamic-condition}
    \frac{1}{D_\alpha} \sum_{i \in D_\alpha} \Delta_{i,F}^{\delta(i)} = \frac{1}{N_{F_\alpha}}\sum_{i \in F_\alpha} \Delta_{i,F}^{\omega},
\end{equation}

\noindent where $\Delta_{i,F}^{t}$ is the number of personal network connections from $i$ to the frame population $F$ at time $t$; 
$\delta(i)$ is the day in which $i$ died (for $i \in D_\alpha$); 
and $\omega$ be the date of the survey (we will assume all of the interviews take place on the same date).  For example, the dynamic decedent network connection says that the average
number of meals shared by men 35-44 in the 12 months before the interview is
equal to the average number of meals shared by dead men aged 35-44 in the 12
months before they died.
If the size of people's networks is fixed over time, then Equation~\ref{eqn:dynamic-condition}
is equivalent to \ref{eqn:static-condition}, which we discuss throughout the paper.

We expect that the most common reason for the dynamic decent network condition
to fail is that people who are going to die share fewer meals than otherwise
similar people who are not about to die (perhaps due to poor health).  Ideally,
future research would attempt to measure this directly, but even if this
measurement does not take place researchers can use the degree ratio parameter
in the sensitivity framework ($\delta_{F, \alpha}$) to assess the impact that
violating the dynamic decedent network condition would have on death rate estimates
(see Appendix~\ref{ap:ns-decomposition-framework}).

A second possible reason for the dynamic decent network condition to fail is a
societal change in the frequency of meal sharing.  This issue arises because we
learn about meal sharing over two different time periods: for the people
who die, we learn about meal sharing in the 12 months before their death and
for the respondents, we learn about meal sharing in the 12 months before the
interview.  For example, suppose an interview was conducted on January 1, 2010
in a country where meal sharing was common in 2009 but there was no meal
sharing at all in 2008.  We would use the known population method to estimate
the respondents' meal sharing during 2009.  Now imagine a women who died in the
middle of 2009.  Half of the year before her death was in the time period where
meal sharing never happened.  Therefore, the number of meals she shared in the
12 months before she died (i.e., her degree) will be lower than a women
who lived during the entire period.  Just as the previous possible concern with
the dynamic decedent network assumption, we hope that future work would attempt
to measure this possibility directly. But, even if this
measurement does not take place, researchers can use the degree ratio parameter
in the sensitivity framework ($\delta_{F, \alpha}$) to assess the impact that
violating the dynamic decedent network condition would have on death rate estimates
(see Appendix~\ref{ap:ns-decomposition-framework}).

The need to use the dynamic decedent network condition is caused by the tie
definition we chose; it is not a property of the network survival estimator
generally.  If we had used a tie definition that was fixed over time---for
example, ties based on a kinship relation (e.g., siblings or cousins) or ties
based on mutual attendance at some fixed event---then only the decedent network
condition would be needed.  Therefore, we consider the trade-off between the
decedent network condition and the dynamic decedent network condition to be one
of the trade-offs researchers will need to make when considering different tie
definitions.

Finally, we note that we designed this specific instrument for our study in
Rwanda. Researchers who are interested in applying the network survival method
in the future should consider modifying it to account for the context in which
they will work. For example, researchers should considering adjusting tie
definitions to be more appropriate for their context.  Further, if network
survival data are collected in a conflict setting, where some respondents may
have many connections to people who died, researchers should allow respondents
to report more than 12 deaths.



\section*{Supplementary material}

| NO. | QUESTIONS AND FILTERS | CODING CATEGORIES | SKIP |
|---|---|---|---|
| 200 | Now I am going to ask you some questions about people that you know.  These questions will help us count the number of people who may be in need of certain health services  These people should be: <br>- people you know by sight AND name, and who also know you by sight and name. In other words, you should not consider famous people that you know about, but who do not know about you. <br>- people you have shared a meal or drink with in the past 12 months. These could be family members, friends, co-workers, or neighbors. You should include meals or drinks taken at any location, such as at home, at work, or in a restaurant. <br>- people of all ages who live in Rwanda. | | |
| 201 | How many men have you shared a meal or drink with whose wife has died and they have not remarried? <br>IF DOES NOT KNOW ANY, RECORD '00' <br>IF KNOWS 95 OR MORE, RECORD '95 | NUMBER OF MEN WHOSE <br>WIFE HAS DIED        . . . . . . . . | |
| 202 | How many people have you shared a meal or drink with who are currently nurses or doctors? <br>IF DOES NOT KNOW ANY, RECORD '00' <br>IF KNOWS 95 OR MORE, RECORD '95 | NUMBER OF <br>NURSES/DOCTORS        . . . . . . . . . | |
| 203 | How many people have you shared a meal or drink with who are currently male community health workers in 2010? <br>IF DOES NOT KNOW ANY, RECORD '00' <br>IF KNOWS 95 OR MORE, RECORD '95 | NUMBER OF MALE <br>COM. HEALTH WORKERS  . . . | |
| 204 | How many people have you shared a meal or drink with who are currently primary or secondary teachers? <br>IF DOES NOT KNOW ANY, RECORD '00' <br>IF KNOWS 95 OR MORE, RECORD '95 | NUMBER OF <br>TEACHERS        . . . . .  . . . . . . . | |
| 205 | How many women have you shared a meal or drink with who currently smoke a pipe or cigarettes? <br>IF DOES NOT KNOW ANY, RECORD '00' <br>IF KNOWS 95 OR MORE, RECORD '95 | NUMBER OF WOMEN <br>WHO SMOKE  . . . . . . . . . . . . . . | |
| 206 | How many men have you shared a meal or drink with who are currently catholic priests? <br>IF DOES NOT KNOW ANY, RECORD '00' <br>IF KNOWS 95 OR MORE, RECORD '95 | NUMBER OF PRIEST  . . . . . . . . | |
| 207 | How many people have you shared a meal or drink with who are currently civil servants? <br>IF DOES NOT KNOW ANY, RECORD '00' <br>IF KNOWS 95 OR MORE, RECORD '95 | NUMBER OF <br>CIVIL SERVANTS        . . . . . . . . . | |
| 208 | How many women have you shared a meal or drink with who gave birth in the last 12 months? <br>IF DOES NOT KNOW ANY, RECORD '00' <br>IF KNOWS 95 OR MORE, RECORD '95 | NUMBER OF WOMEN <br>WHO GAVE BIRTH        . . . . . . . . | |
| 209 | How many people have you shared a meal or drink with who are Muslims? <br>IF DOES NOT KNOW ANY, RECORD '00' <br>IF KNOWS 95 OR MORE, RECORD '95 | NUMBER OF <br>MUSLIMS        . . . . . . . . . . . . . . | |
| 210 | How many people have you shared a meal or drink with who are currently incarcerated? <br>IF DOES NOT KNOW ANY, RECORD '00' <br>IF KNOWS 95 OR MORE, RECORD '95 | NUMBER OF PEOPLE <br>INCARCERATED  . . . . . . . . . . | |
| 211 | How many people have you shared a meal or drink with who were Gacaca judges in 2010? <br>IF DOES NOT KNOW ANY, RECORD '00' <br>IF KNOWS 95 OR MORE, RECORD '95 | NUMBER OF <br>GACACA JUDGES        . . . . . . . . | |
| 212 | How many men have you shared a meal or drink with who are divorced or separated and not remarried? <br>IF DOES NOT KNOW ANY, RECORD '00' <br>IF KNOWS 95 OR MORE, RECORD '95 | NUMBER OF MEN <br>DIVORCED/SEPARATED    . . . | |

| NO. | QUESTIONS AND FILTERS | CODING CATEGORIES | SKIP |
|-----|----------------------|-------------------|------|
| 213 | How many people have you shared a meal or drink with who are being treated for TB?<br>IF DOES NOT KNOW ANY, RECORD '00'<br>IF KNOWS 95 OR MORE, RECORD '95 | NUMBER OF PEOPLE<br>TREATED FOR TB ........ ▢▢ | |
| | Just as a reminder I am only interested in<br>- people you shared a meal or drink with in the past 12 months<br>- People of all ages who live in Rwanda. | | |
| 214 | How many people have you shared a meal or drink with are named NSENGIMANA?<br>IF DOES NOT KNOW ANY, RECORD '00'<br>IF KNOWS 95 OR MORE, RECORD '95 | NUMBER OF<br>NSENGIMANA ..... ......... ▢▢ | |
| 215 | How many people have you shared a meal or drink with named MUREKATETE?<br>IF DOES NOT KNOW ANY, RECORD '00'<br>IF KNOWS 95 OR MORE, RECORD '95 | NUMBER OF<br>MUREKATETE ..... ......... ▢▢ | |
| 216 | How many people have you shared a meal or drink with are named TWAHIRWA?<br>IF DOES NOT KNOW ANY, RECORD '00'<br>IF KNOWS 95 OR MORE, RECORD '95 | NUMBER OF<br>TWAHIRWA ..... ......... ▢▢ | |
| 217 | How many people have you shared a meal or drink with are named MUKANDEKEZI?<br>IF DOES NOT KNOW ANY, RECORD '00'<br>IF KNOWS 95 OR MORE, RECORD '95 | NUMBER OF<br>MUKANDEKEZI ............ ▢▢ | |
| 218 | How many people have you shared a meal or drink with are named NSABIMANA?<br>IF DOES NOT KNOW ANY, RECORD '00'<br>IF KNOWS 95 OR MORE, RECORD '95 | NUMBER OF<br>NSABIMANA ............ ▢▢ | |
| 219 | How many people have you shared a meal or drink with are named MUKAMANA?<br>IF DOES NOT KNOW ANY, RECORD '00'<br>IF KNOWS 95 OR MORE, RECORD '95 | NUMBER OF<br>MUKAMANA ............... ▢▢ | |
| 220 | How many people have you shared a meal or drink with are named NDAYAMBAJE?<br>IF DOES NOT KNOW ANY, RECORD '00'<br>IF KNOWS 95 OR MORE, RECORD '95 | NUMBER OF<br>NDAYAMBAJE ........ ▢▢ | |
| 221 | How many people have you shared a meal or drink with are named NYIRANEZA?<br>IF DOES NOT KNOW ANY, RECORD '00'<br>IF KNOWS 95 OR MORE, RECORD '95 | NUMBER OF<br>NYIRANEZA ............... ▢▢ | |
| 222 | How many people have you shared a meal or drink with are named BIZIMANA?<br>IF DOES NOT KNOW ANY, RECORD '00'<br>IF KNOWS 95 OR MORE, RECORD '95 | NUMBER OF<br>BIZIMANA ............... ▢▢ | |
| 223 | How many people have you shared a meal or drink with are named NYIRAHABIMANA?<br>IF DOES NOT KNOW ANY, RECORD '00'<br>IF KNOWS 95 OR MORE, RECORD '95 | NUMBER OF<br>NYIRAHABIMANA ........ ▢▢ | |
| 224 | How many people have you shared a meal or drink with are named NDAGIJIMANA?<br>IF DOES NOT KNOW ANY, RECORD '00'<br>IF KNOWS 95 OR MORE, RECORD '95 | NUMBER OF<br>NDAGIJIMANA ........ ▢▢ | |
| 225 | How many people have you shared a meal or drink with are named MUKANDAYISENGA?<br>IF DOES NOT KNOW ANY, RECORD '00'<br>IF KNOWS 95 OR MORE, RECORD '95 | NUMBER OF<br>MUKANDAYISENGA ........ ▢▢ | |

| NO. | QUESTIONS AND FILTERS | CODING CATEGORIES | | SKIP |
|---|---|---|---|---|
| 226 | Now I would like to ask you a few questions about people who have died.<br><br>Similar to the previous questions only tell me about<br>- people you shared a meal or drink with in the past 12 months before they died.<br>- These should be people of all ages living in Rwanda.<br><br>How many people have you shared a meal or drink with who have died in the past 12 months? | NUMBER OF DEATHS . . . . . . . . ☐☐<br><br>NONE . . . . . . . . . . . . . . . . . . . . . . . . 00 | | → 301 |
| 227 | I would like to ask a couple of questions about each of these people who died. To keep track of the different people we are discussing, could you tell me the first name of each person you know who died in the past 12 months?<br><br>RECORD THE FIRST NAME OF EACH PERSON WHO HAS DIED AND ASK Q.228 AND 229<br><br>IF AGE IS NOT KNOWN, GET THE BEST POSSIBLE ESTIMATE IF AGE 95 OR MORE, RECORD '95' | **228**<br>Was (NAME) male or female? | **229**<br>How old was (NAME)? | |
| | NAME 1 _____ | MALE . . . . . . . . 1<br>FEMALE . . . . . 2 | ☐☐ | |
| | NAME 2 _____ | MALE . . . . . . . . 1<br>FEMALE . . . . . 2 | ☐☐ | |
| | NAME 3 _____ | MALE . . . . . . . . 1<br>FEMALE . . . . . 2 | ☐☐ | |
| | NAME 4 _____ | MALE . . . . . . . . 1<br>FEMALE . . . . . 2 | ☐☐ | |
| | NAME 5 _____ | MALE . . . . . . . . 1<br>FEMALE . . . . . 2 | ☐☐ | |
| | NAME 6 _____ | MALE . . . . . . . . 1<br>FEMALE . . . . . 2 | ☐☐ | |
| | NAME 7 _____ | MALE . . . . . . . . 1<br>FEMALE . . . . . 2 | ☐☐ | |
| | NAME 8 _____ | MALE . . . . . . . . 1<br>FEMALE . . . . . 2 | ☐☐ | |
| | NAME 9 _____ | MALE . . . . . . . . 1<br>FEMALE . . . . . 2 | ☐☐ | |
| | NAME 10 _____ | MALE . . . . . . . . 1<br>FEMALE . . . . . 2 | ☐☐ | |
| | NAME 11 _____ | MALE . . . . . . . . 1<br>FEMALE . . . . . 2 | ☐☐ | |
| | NAME 12 _____ | MALE . . . . . . . . 1<br>FEMALE . . . . . 2 | ☐☐ | |



\end{document}